\documentclass[journal]{IEEEtran}
\usepackage[figuresright]{rotating}
\usepackage{amssymb}
\usepackage{amsthm}
\usepackage{multicol}
\usepackage{subfigure}
\usepackage{graphicx}
\usepackage{epstopdf}
\usepackage{fullpage}
\usepackage{latexsym,amsmath}
\usepackage{stmaryrd}
\usepackage{algorithm,algorithmic}
\usepackage{subfigure}
\usepackage{amsfonts}
\usepackage{xcolor,multirow}
\usepackage{cite}
\usepackage{bm}
\usepackage{url}
\usepackage{diagbox}
\usepackage{array}
\usepackage{booktabs}
\usepackage{footmisc} 
\usepackage{pifont}

\newtheorem{lemma}{Lemma}
\newtheorem{theorem}{Theorem}
\newtheorem{definition}{Definition}

\ifCLASSINFOpdf
\else
\fi
\begin{document}
	
\title{Quaternion-based bilinear factor matrix norm minimization for color image inpainting}

\author{Jifei~Miao and Kit~Ian~Kou

\thanks{The authors are with the Department of Mathematics, Faculty of Science
	and Technology, University of Macau, Macau 999078, China  (e-mail: jifmiao@163.com;
	kikou@umac.mo)}}

\markboth{Journal of \LaTeX\ Class Files,~Vol.~14, No.~8, August~2019}%
{Shell \MakeLowercase{\textit{et al.}}: Bare Demo of IEEEtran.cls for IEEE Journals}

\maketitle

\begin{abstract}
As a new color image representation tool, quaternion has achieved excellent results in the color image processing, 	
because it treats the color image as a whole rather than as a separate color space component, thus it can make full use of the high correlation among RGB channels. Recently, low-rank quaternion matrix completion (LRQMC) methods have proven very useful for color image inpainting. In this paper, we propose three novel LRQMC methods based on three quaternion-based bilinear factor (QBF) matrix norm minimization models. Specifically, we define quaternion double Frobenius norm (Q-DFN), quaternion double nuclear norm (Q-DNN) and quaternion Frobenius/nuclear norm (Q-FNN), and then show their relationship with quaternion-based matrix Schatten-$p$ (Q-Schatten-$p$ ) norm for certain  $p$ values. The proposed methods can avoid computing quaternion singular value decompositions (QSVD) for large quaternion matrices, and thus can effectively reduce the calculation time compared with existing (LRQMC) methods.
The experimental results demonstrate the superior performance of the proposed methods over some state-of-the-art low-rank (quaternion) matrix completion methods. 
\end{abstract}

\IEEEpeerreviewmaketitle

\section{Introduction}
\IEEEPARstart{L}ow-rank matrix completion (LRMC)-based techniques have made a great success in the application of image inpainting. Most of the existing LRMC methods generally can be divided into two categories \cite{DBLP:journals/csur/ZhouYZY14}: matrix rank minimization and matrix factorization. Matrix rank minimization is generally achieved by sorts of rank approximation regularizers. For example, nuclear norm \cite{DBLP:journals/jacm/CandesLMW11, DBLP:journals/siamjo/CaiCS10} which is the tightest convex relaxation of the NP-hard rank minimization function \cite{DBLP:journals/focm/CandesR09}. Assigning different weights with different singular values, the authors in \cite{DBLP:conf/cvpr/GuZZF14,DBLP:journals/ijcv/GuXMZFZ17} proposed the weighted nuclear norm minimization (WNNM) algorithm, which can better approximate the rank function. Besides, as a generalization of nuclear norm  minimization (NNM) and WNNM, the authors in \cite{DBLP:conf/aaai/NieHD12, DBLP:journals/jcam/LiuHC14} proposed the Schatten-p norm  minimization. Combining WNNM  and the Schatten-p norm, the authors in \cite{DBLP:journals/tip/XieGLZZZ16} proposed  the weighted Schatten-p norm minimization. However, the main solution strategy for these kinds of methods require computing singular value decompositions (SVD) which requires increasingly cost as matrix sizes and rank increase. On the other hand, matrix factorization generally factorizes the original larger matrix into at least two much smaller matrices \cite{DBLP:journals/mpc/WenYZ12,DBLP:conf/cvpr/KimLO15, DBLP:conf/cvpr/ZhengLSYO12, DBLP:journals/pami/ShangCLLL18}. These approaches benefit from fast numerical methods for optimization and easy kernelization \cite{DBLP:conf/iccv/CabralTCB13}. The main issue of these kinds of methods is the lack of the rank values in many cases. To tackle this problem, the authors in \cite{DBLP:journals/mpc/WenYZ12} proposed a rank estimation strategy, the effectiveness of which has been proven by many applications and pieces of literature \cite{DBLP:journals/pami/ShangCLLL18, DBLP:journals/sigpro/FanLYLN19}.

Most of the LRMC algorithms mentioned above have obtained excellent performance for grayscale images. When handling color images, these algorithms usually processes each color channel independently using the monochromatic model or processes the concatenation of three color channels using the concatenation model \cite{DBLP:journals/tip/MairalES08, DBLP:conf/iccv/XuZ0F17}. However, these two schemes may not make full use of the high correlation among RGB channels, thus they may cause unsatisfactory results.

Recently, quaternion, as an elegant color image representation tool, has attracted much attention in the field of color image processing. For instance, it has achieved excellent results in the following applications: color image filtering \cite{DBLP:journals/iet-ipr/ChenLSLS14}, color image edge detection \cite{DBLP:journals/imst/XuYL10}, color image denoising \cite{DBLP:journals/mssp/GaiYW015,DBLP:journals/ijon/YuZY19},  color image watermarking \cite{1199526_P2003}, color face recognition \cite{DBLP:journals/tip/ZouKW16}, color image inpainting \cite{DBLP:journals/tip/ChenXZ20} and so on. By using quaternion algebra, a color image is encoded as a pure quaternion matrix, that means it processes a color image holistically as a vector field and handles the coupling between the color channels naturally \cite{DBLP:journals/iet-ipr/ChenLSLS14, DBLP:journals/ijcv/SubakanV11, DBLP:journals/sigpro/ChenSZCTDL12}, and thus color information of source image is fully used. More recently, the authors in \cite{DBLP:journals/tip/ChenXZ20} extended traditional LRMC methods to quaternion field, and proposed a general low-rank quaternion matrix completion (LRQMC) model based on several nonconvex rank functions including quaternion nuclear norm (QNN), Laplace function and German function. These kinds of methods show promising performance for color image inpainting. Nonetheless, these methods need to compute the quaternion singular value decompositions (QSVD) for large quaternion matrices in each iteration, which are calculated by their equivalent complex matrices with twice sizes and thus suffer from high computational complexity and time-consuming.  Therefore, it is necessary to design more efficient algorithms for LRQMC.

In this paper, we develop three novel LRQMC methods based on three quaternion-based bilinear factor (QBF) matrix norm minimization models for color image inpainting.
The contributions of this paper can be summarized as
\begin{itemize}
	\item We propose three novel LRQMC methods based on three QBF matrix norm minimization models including  quaternion double Frobenius norm (Q-DFN)-based, quaternion double nuclear norm (Q-DNN)-based and quaternion Frobenius/nuclear norm (Q-FNN)-based models. Compared with traditional LRMC-based methods, the proposed models process three RGB channels information in a parallel way and thus can preserve the correlation among the color channels well. Compared with existing LRQMC methods, the proposed models only need to handle two much smaller factor quaternion matrices, thus it can effectively reduce the time consumption caused by the calculation of QSVD.
	\item We show the relationship among quaternion-based matrix Schatten-$p$ (Q-Schatten-$p$ ) norm for certain  $p$ values with the defined three norms. Then, the three models are optimized by applying the alternating direction method of multipliers (ADMM) framework. An effective rank-estimation method is used to estimate the quaternion rank adaptively as the number of iterations increases. 
	\item The experimental results on real color images show their empirical convergence and illustrate their competitive performance over several state-of-the-art methods. 
\end{itemize}

The remainder of this paper is organized as follows. Section \ref{sec2} introduces some notations and preliminaries for quaternion algebra. Section \ref{sec3} revisits the matrix completion theory and related works, then gives our three quaternion-based matrix completion methods. Section \ref{sec4} provides some experiments to illustrate the performance of our algorithms,
and compare it with several state-of-the-art methods. Finally, some conclusions are drawn in Section \ref{sec5}.

\section{Notations and preliminaries}
\label{sec2}
In this section, we first summarize some main notations and
then introduce some basic knowledge of quaternion algebra.

\subsection{Notations}
In this paper, $\mathbb{R}$, $\mathbb{C}$, and $\mathbb{H}$ respectively denote the real space, complex space, and quaternion space. A scalar, a vector, and a matrix are written as $a$, $\mathbf{a}$, and $\mathbf{A}$, respectively. $\dot{a}$,  $\dot{\mathbf{a}}$, and $\dot{\mathbf{A}}$
respectively represent a quaternion scalar, a quaternion vector, and a quaternion matrix. $(\cdot)^{\ast}$, $(\cdot)^{-1}$, $(\cdot)^{T}$, and $(\cdot)^{H}$ denote the
conjugation,  inverse, transpose, and  conjugate transpose, respectively. $|\cdot|$, $\|\cdot\|_{F}$, and $\|\cdot\|_{\ast}$ are respectively the absolute value or modulus, the Frobenius norm, and the nuclear norm. $\langle\cdot,\cdot\rangle$ denotes the inner product operation.
${\rm{tr}}\{\cdot\}$ and ${\rm{rank}}(\cdot)$ denote the trace and rank operators respectively. $\mathfrak{R}(\cdot)$ denotes the real part of quaternion (scalar, vector, and matrix). $\mathbf{I}$ represents the identity matrix with appropriate size.

\subsection{Basic knowledge of quaternion algebras}
Quaternion algebras have attracted extensive attention in recent years, especially in the field of signal and image processing \cite{8611251,DBLP:journals/tip/ChenXZ20}, \emph{etc.}. Some basic knowledge of quaternion algebras can be found in the Appendix \ref{a_sec1}. In the following, we give two theorems about quaternion matrix.
\begin{theorem}
\label{Th1}
(Quaternion singular value decomposition (QSVD) \cite{10029950538}): For any quaternion matrix $\dot{\mathbf{Q}}\in\mathbb{H}^{M\times N}$ with rank $r$,  there exist unitary quaternion matrices $\dot{\mathbf{U}}\in\mathbb{H}^{M\times M}$ and $\dot{\mathbf{V}}\in\mathbb{H}^{N\times N}$ such that
\begin{equation}\small
\dot{\mathbf{Q}}=\dot{\mathbf{U}}\left(\begin{array}{cc}
\mathbf{D}_{r}&  \mathbf{0} \\ 
\mathbf{0}&  \mathbf{0}  
\end{array} \right)\dot{\mathbf{V}}^{H},
\end{equation}
where $\mathbf{D}_{r}={\rm{diag}}\{d_{1},\ldots,d_{r}\}\in\mathbb{R}^{r\times r}$ and the $d's$ are the positive singular values of $\dot{\mathbf{Q}}$.
\end{theorem}
The way that to obtain the QSVD of a quaternion matrix can be found in the Appendix \ref{a_sec1}. Following the \textbf{Theorem \ref{Th1}}, we can see that the rank of a quaternion matrix $\dot{\mathbf{Q}}\in\mathbb{H}^{M\times N}$ is equal to the number of its positive singular values  \cite{10029950538}.	
\begin{theorem}
\label{Th2} Let $\dot{\mathbf{X}}\in\mathbb{H}^{M\times N}$ be an arbitrary quaternion matrix. 
If ${\rm{rank}}(\dot{\mathbf{X}})=r\leq d$, then $\dot{\mathbf{X}}$ can be written into a quaternion matrix product form $\dot{\mathbf{X}}=\dot{\mathbf{U}}\dot{\mathbf{V}}^{H}$, where
	 $\dot{\mathbf{U}}\in\mathbb{H}^{M\times d}$ and  $\dot{\mathbf{V}}\in\mathbb{H}^{N\times d}$ are two quaternion matrices of smaller sizes and they meet ${\rm{rank}}(\dot{\mathbf{U}})={\rm{rank}}(\dot{\mathbf{V}})=r$.
\end{theorem}
The proof of \textbf{Theorem \ref{Th2}} can be found in the Appendix \ref{a_sec2}.

\section{Problem formulation}
\label{sec3}
This section firstly revisits the matrix completion theory and related works, then gives our three quaternion-based matrix completion methods.
\subsection{Matrix completion}
The matrix completion problem consists of recovering a data matrix from its partial entries by the well-known rank optimization model \cite{Cand2009Exact}:
\begin{equation}\small
\label{e1}
\mathop{{\rm{min}}}\limits_{\mathbf{X}}\  \lambda{\rm{rank}}(\mathbf{X}),\ \text{s.t.},\ \mathcal{P}_{\Omega}(\mathbf{X}-\mathbf{T})=\mathbf{0},	
\end{equation}
where $\lambda$ is a nonnegative parameter, $\mathbf{X}\in\mathbb{R}^{M\times N}$ is a completed output matrix, $\mathbf{T}\in\mathbb{R}^{M\times N}$ is an incomplete input
matrix and $\mathcal{P}_{\Omega}$ is the unitary projection onto the linear space of matrices supported on the entries set $\Omega$, defined as
\begin{equation*}\small
(\mathcal{P}_{\Omega}(\mathbf{X}))_{mn}=\left\{
\begin{array}{c}
\!\!\!x_{mn},\qquad (m,n)\in \Omega, \\
0,\qquad\quad\:  (m,n)\notin\Omega.
\end{array}
\right.
\end{equation*}
Directly solve the problem (\ref{e1}) is difficult as the rank minimization problem is known as NP-hard \cite{DBLP:journals/siammax/GillisG11}. Various heuristics approaches have been developed to solve this problem. These methods usually adopt the convex or non-convex surrogates to
replace the rank function in (\ref{e1}), and formulate this problem into the following general form
\begin{equation}\small
\label{e2}
\mathop{{\rm{min}}}\limits_{\mathbf{X}}\ \lambda\|\mathbf{X}\|_{S_{p}}^{p}, \ \text{s.t.},\ \mathcal{P}_{\Omega}(\mathbf{X}-\mathbf{T})=\mathbf{0},	
\end{equation}\small
where $\|\mathbf{X}\|_{S_{p}}$ is the Schatten-$p$ norm $(0<p<\infty)$ of matrix $\mathbf{X}$, and defined as 
\begin{equation}\small
\label{e3}
\|\mathbf{X}\|_{S_{p}}:=\bigg(\sum_{k}^{{\rm{min}}(M,N)}\sigma_{k}^{p}(\mathbf{X})\bigg)^{1/p},
\end{equation}
where $\sigma_{k}(\mathbf{X})$ denotes the $k$-th singular value of $\mathbf{X}$. When $p=1$ the Schatten-$1$ norm is the well-known nuclear norm $\|\mathbf{X}\|_{\ast}$ which has been widely applied in low-rank matrix approximation problems. For example, there are many matrix completion approaches \cite{DBLP:journals/focm/CandesR09, DBLP:journals/pami/HuZYLH13, DBLP:conf/cvpr/LuTYL14, DBLP:journals/ijcv/GuXMZFZ17} by using nuclear norm $\|\mathbf{X}\|_{\ast}$ or weighted nuclear norm $\|\mathbf{X}\|_{w,\ast}$ to replace the first term in (\ref{e2}). In addition, as the non-convex surrogate for the rank function, the Schatten-$p$ quasi-norm with $0 < p < 1$ makes a closer approximation to the rank function than the nuclear norm \cite{DBLP:journals/tit/ZhangHZ13}. Therefore, the
Schatten-$p$ quasi-norm has attracted a great deal of attention in low-rank matrix approximation problems \cite{DBLP:conf/aaai/NieHD12, DBLP:journals/tip/XieGLZZZ16, DBLP:journals/tip/ZhangQZYGW20}. Besides, the authors in \cite{DBLP:journals/pami/ShangCLLL18} adopted the following two bilinear factor (BF) matrix norms (double nuclear norm  $\|\mathbf{X}\|_{\rm{D-N}}$ and Frobenius/nuclear norm $\|\mathbf{X}\|_{\rm{F-N}}$) to replace the Schatten-$p$ norm and obtained an impressive performance
\begin{equation}\small
\label{e4}
\|\mathbf{X}\|_{\rm{D-N}}=\mathop{{\rm{min}}}\limits_{\mathbf{U},\mathbf{V}\atop \mathbf{X}=\mathbf{U}\mathbf{V}^{T}} \frac{1}{4}(\|\mathbf{U}\|_{\ast}+\|\mathbf{V}\|_{\ast})^{2},
\end{equation}
\begin{equation}
\label{e5}
\|\mathbf{X}\|_{\rm{F-N}}=\mathop{{\rm{min}}}\limits_{\mathbf{U},\mathbf{V}\atop \mathbf{X}=\mathbf{U}\mathbf{V}^{T}} \big[(\|\mathbf{U}\|_{F}^{2}+2\|\mathbf{V}\|_{\ast})/3\big]^{3/2}.
\end{equation}
\subsection{Quaternion matrix completion}
The traditional low-rank approximation based matrix completion methods are inherently designed in the real settings for grayscale image inpainting and thus may
suffer from performance degradation of color image case. Quaternion matrix completion can be regarded as the generalization of the traditional matrix completion in the quaternion number field, which, for color image inpainting, allows different color channels to talk to each other rather than each channel being independently manipulated. We define the quaternion-based matrix Schatten-$p$ norm (Q-Schatten-$p$) as follows.
\begin{definition}(Q-Schatten-$p$):
Given a quaternion matrix $\dot{\mathbf{X}}\in\mathbb{H}^{M\times N}$, the Q-Schatten-$p$ norm $(0<p<\infty)$ is defined as
\begin{equation}\small
\label{e6}
\|\dot{\mathbf{X}}\|_{Q-S_{p}}:=\bigg(\sum_{k}^{{\rm{min}}(M,N)}\sigma_{k}^{p}(\dot{\mathbf{X}})\bigg)^{1/p},
\end{equation}
\end{definition}
where $\sigma_{k}(\dot{\mathbf{X}})$ denotes the $k$-th singular value of quaternion matrix $\dot{\mathbf{X}}$.
When $p=1$ the above quaternion-based matrix Schatten-$1$ norm is quaternion nuclear norm (QNN) denoted as $\|\dot{\mathbf{X}}\|_{\ast}$ \cite{DBLP:journals/ijon/YuZY19, DBLP:journals/tip/ChenXZ20}. However, analogous to the traditional nuclear norm based approaches, the QNN-based methods involve computing the QSVD of a large quaternion matrix in each iteration, and thus suffer from high computational complexity and time-consuming. Hence, it is necessary to design more efficient algorithms.
\subsection{Proposed quaternion-based matrix completion models}
The \textbf{Theorem \ref{Th2}} allows us to make bilinear factorization of larger quaternion matrices. Thus, in the following,
we first define the quaternion double Frobenius norm (Q-DFN). Then, motivated by the definition of double nuclear norm and Frobenius/nuclear norm in \cite{DBLP:journals/pami/ShangCLLL18}, we also define the quaternion double nuclear norm (Q-DNN) and the quaternion Frobenius/nuclear norm (Q-FNN) in this section.
\begin{definition}\label{def0}(Q-DFN):
	Given a quaternion matrix $\dot{\mathbf{X}}\in\mathbb{H}^{M\times N}$ with $rank(\dot{\mathbf{X}})=r\leq d$, we decompose it into two much smaller quaternion factor matrices $\dot{\mathbf{U}}\in\mathbb{H}^{M\times d}$ and $\dot{\mathbf{V}}\in\mathbb{H}^{N\times d}$ such that $\dot{\mathbf{X}}=\dot{\mathbf{U}}\dot{\mathbf{V}}^{H}$. Then the quaternion double Frobenius norm  is defined as
	\begin{equation}\small
	\label{en1}
	\|\dot{\mathbf{X}}\|_{\rm{Q-DFN}}:=\mathop{{\rm{min}}}\limits_{\dot{\mathbf{U}},\dot{\mathbf{V}}\atop \dot{\mathbf{X}}=\dot{\mathbf{U}}\dot{\mathbf{V}}^{H}} \frac{1}{2}\|\dot{\mathbf{U}}\|_{F}^{2}+\frac{1}{2}\|\dot{\mathbf{V}}\|_{F}^{2}.
	\end{equation}
\end{definition}
\begin{definition}\label{def1}(Q-DNN):
Given a quaternion matrix $\dot{\mathbf{X}}\in\mathbb{H}^{M\times N}$ with $rank(\dot{\mathbf{X}})=r\leq d$, we decompose it into two much smaller quaternion factor matrices $\dot{\mathbf{U}}\in\mathbb{H}^{M\times d}$ and $\dot{\mathbf{V}}\in\mathbb{H}^{N\times d}$ such that $\dot{\mathbf{X}}=\dot{\mathbf{U}}\dot{\mathbf{V}}^{H}$. Then the quaternion double nuclear norm is defined as
\begin{equation}\small
\label{e7}
\|\dot{\mathbf{X}}\|_{\rm{Q-DNN}}:=\mathop{{\rm{min}}}\limits_{\dot{\mathbf{U}},\dot{\mathbf{V}}\atop \dot{\mathbf{X}}=\dot{\mathbf{U}}\dot{\mathbf{V}}^{H}} \frac{1}{4}(\|\dot{\mathbf{U}}\|_{\ast}+\|\dot{\mathbf{V}}\|_{\ast})^{2}.
\end{equation}
\end{definition}
\begin{definition}\label{def2}(Q-FNN):
	Given a quaternion matrix $\dot{\mathbf{X}}\in\mathbb{H}^{M\times N}$ with $rank(\dot{\mathbf{X}})=r\leq d$, we decompose it into two much smaller quaternion factor matrices $\dot{\mathbf{U}}\in\mathbb{H}^{M\times d}$ and $\dot{\mathbf{V}}\in\mathbb{H}^{N\times d}$ such that $\dot{\mathbf{X}}=\dot{\mathbf{U}}\dot{\mathbf{V}}^{H}$. Then the quaternion Frobenius/nuclear norm is defined as
	\begin{equation}\small
	\label{e8}
	\|\dot{\mathbf{X}}\|_{\rm{Q-FNN}}:=\mathop{{\rm{min}}}\limits_{\dot{\mathbf{U}},\dot{\mathbf{V}}\atop \dot{\mathbf{X}}=\dot{\mathbf{U}}\dot{\mathbf{V}}^{H}} \big[(\|\dot{\mathbf{U}}\|_{F}^{2}+2\|\dot{\mathbf{V}}\|_{\ast})/3\big]^{3/2}.
	\end{equation}
\end{definition}
In the following three theorems, we give the relationships among the Q-Schatten-$p$ norm, the quaternion double Frobenius norm, the quaternion double nuclear norm, and the quaternion Frobenius/nuclear norm.
\begin{theorem}
	\label{th0}
	The quaternion double Frobenius norm is in essence the Q-Schatten-$1$ norm (or QNN), i.e.,
	\begin{equation}\small
	\label{en2}
	\|\dot{\mathbf{X}}\|_{\rm{Q-DFN}}=\|\dot{\mathbf{X}}\|_{\ast}.
	\end{equation}
\end{theorem}
\begin{theorem}
\label{th1}
The quaternion double nuclear norm is in essence the Q-Schatten-$1/2$ norm, i.e.,
\begin{equation}\small
\label{e9}
\|\dot{\mathbf{X}}\|_{\rm{Q-DNN}}=\|\dot{\mathbf{X}}\|_{Q-S_{1/2}}.
\end{equation}
\end{theorem}
\begin{theorem}
\label{th2}
The quaternion Frobenius/nuclear norm is in essence the Q-Schatten-$1/2$ norm, i.e.,
\begin{equation}\small
\label{e10}
\|\dot{\mathbf{X}}\|_{\rm{Q-FNN}}=\|\dot{\mathbf{X}}\|_{Q-S_{2/3}}.
\end{equation}
\end{theorem}
The proofs of \textbf{Theorem \ref{th0}}, \textbf{Theorem \ref{th1}}  and \textbf{Theorem \ref{th2}} can be found in
the Appendix \ref{a_sec3}.

Based on the \textbf{Definition \ref{def0}}, \textbf{Definition \ref{def1}} and the \textbf{Definition \ref{def2}}, we present the following three quaternion-based matrix completion models
\begin{equation}\small
\label{en3}
\begin{split}
&\mathop{{\rm{min}}}\limits_{\dot{\mathbf{U}},\dot{\mathbf{V}},\dot{\mathbf{X}}}\ \frac{\lambda}{2}\big(\|\dot{\mathbf{U}}\|_{F}^{2}+\|\dot{\mathbf{V}}\|_{F}^{2}\big),\\ 
&\text{s.t.},\ \mathcal{P}_{\Omega}(\dot{\mathbf{X}}-\dot{\mathbf{T}})=\mathbf{0}, \dot{\mathbf{X}}=\dot{\mathbf{U}}\dot{\mathbf{V}}^{H}. 	
\end{split}
\end{equation}
\begin{equation}\small
\label{e11}
\begin{split}
&\mathop{{\rm{min}}}\limits_{\dot{\mathbf{U}},\dot{\mathbf{V}},\dot{\mathbf{X}}}\ \frac{\lambda}{2}\big(\|\dot{\mathbf{U}}\|_{\ast}+\|\dot{\mathbf{V}}\|_{\ast}\big),\\ 
&\text{s.t.},\ \mathcal{P}_{\Omega}(\dot{\mathbf{X}}-\dot{\mathbf{T}})=\mathbf{0}, \dot{\mathbf{X}}=\dot{\mathbf{U}}\dot{\mathbf{V}}^{H}. 	
\end{split}
\end{equation}
\begin{equation}\small
\label{e12}
\begin{split}
&\mathop{{\rm{min}}}\limits_{\dot{\mathbf{U}},\dot{\mathbf{V}},\dot{\mathbf{X}}}\ \frac{\lambda}{3}\big(\|\dot{\mathbf{U}}\|_{F}^{2}+2\|\dot{\mathbf{V}}\|_{\ast}\big),\\ 
&\text{s.t.},\ \mathcal{P}_{\Omega}(\dot{\mathbf{X}}-\dot{\mathbf{T}})=\mathbf{0}, \dot{\mathbf{X}}=\dot{\mathbf{U}}\dot{\mathbf{V}}^{H}. 	
\end{split}
\end{equation}
The above models have at least two obvious advantages. First of all, the above models are built in the quaternion field, which, compared with the traditional real matrix-based method, can directly deal with the problem of color image inpainting, and can make full use of the connection between color channels. 
In addition, compared with the existing quaternion matrix completion methods, the above models only need to handle two much smaller factor quaternion matrices, thus it can effectively reduce the time consumption caused by the calculation of QSVD.

\subsection{Optimization process}
The problem  (\ref{en3}) can be solved by minimizing the following augmented Lagrangian function
\begin{equation}\small
\label{en4}
\begin{split}
&\mathcal{L}_{\mu}(\dot{\mathbf{U}},\dot{\mathbf{V}},\dot{\mathbf{X}},\dot{\mathbf{F}})\\
&=\frac{\lambda}{2}\big(\|\dot{\mathbf{U}}\|_{F}^{2}+\|\dot{\mathbf{V}}\|_{F}^{2}\big)+\langle\dot{\mathbf{F}},\dot{\mathbf{X}}-\dot{\mathbf{U}}\dot{\mathbf{V}}^{H}\rangle\\
&\quad+\frac{\mu}{2}\|\dot{\mathbf{X}}-\dot{\mathbf{U}}\dot{\mathbf{V}}^{H}\|_{F}^{2}+\frac{1}{2}\|\mathcal{P}_{\Omega}(\dot{\mathbf{X}}-\dot{\mathbf{T}})\|_{F}^{2},
\end{split}
\end{equation}
where $\mu>0$ is the penalty parameter, $\dot{\mathbf{F}}$ is Lagrange multiplier.

\textbf{Updating $\dot{\mathbf{U}}$ and $\dot{\mathbf{V}}$}: In the $\tau+1$-th iteration, fixing the other variables at their latest values, $\dot{\mathbf{U}}^{\tau+1}$ and $\dot{\mathbf{V}}^{\tau+1}$ are respectively the optimal solutions of the following problems
\begin{equation}\small
\label{en5}
\begin{split}
\dot{\mathbf{U}}^{\tau+1}=&\mathop{{\rm{arg\, min}}}\limits_{\dot{\mathbf{U}}}\ \frac{1}{2}\|\dot{\mathbf{X}}^{\tau}-\dot{\mathbf{U}}(\dot{\mathbf{V}}^{\tau})^{H}+\dot{\mathbf{F}}^{\tau}/\mu^{\tau}\|_{F}^{2}\\
&+\frac{\lambda}{2\mu^{\tau}}\|\dot{\mathbf{U}}\|_{F}^{2},
\end{split}
\end{equation}
\begin{equation}\small
\label{en6}
\begin{split}
\dot{\mathbf{V}}^{\tau+1}=&\mathop{{\rm{arg\, min}}}\limits_{\dot{\mathbf{V}}}\ \frac{1}{2}\|\dot{\mathbf{X}}^{\tau}-\dot{\mathbf{U}}^{\tau+1}\dot{\mathbf{V}}^{H}+\dot{\mathbf{F}}^{\tau}/\mu^{\tau}\|_{F}^{2}\\
&+\frac{\lambda}{2\mu^{\tau}}\|\dot{\mathbf{V}}\|_{F}^{2}.
\end{split}
\end{equation}
Let $\mathcal{Q}(\dot{\mathbf{U}}):=\frac{1}{2}\|\dot{\mathbf{X}}^{\tau}-\dot{\mathbf{U}}(\dot{\mathbf{V}}^{\tau})^{H}+\dot{\mathbf{F}}^{\tau}/\mu^{\tau}\|_{F}^{2}+\frac{\lambda}{2\mu^{\tau}}\|\dot{\mathbf{U}}\|_{F}^{2}$, and $\mathcal{G}(\dot{\mathbf{V}}):=\frac{1}{2}\|\dot{\mathbf{X}}^{\tau}-\dot{\mathbf{U}}^{\tau+1}\dot{\mathbf{V}}^{H}+\dot{\mathbf{F}}^{\tau}/\mu^{\tau}\|_{F}^{2}+\frac{\lambda}{2\mu^{\tau}}\|\dot{\mathbf{V}}\|_{F}^{2}$. We see that $\mathcal{Q}(\dot{\mathbf{U}})$ and $\mathcal{G}(\dot{\mathbf{V}})$ are real functions of quaternion variable, thus the left and right generalized HR (GHR) derivatives are identical \cite{DBLP:journals/tsp/XuM15}. Using the related theories of quaternion matrix derivatives in \cite{DBLP:journals/tsp/XuM15}, the gradient of $\mathcal{Q}(\dot{\mathbf{U}})$ can be calculated as
\begin{equation}\small
\label{en7}
\begin{split}
\frac{\partial\mathcal{Q}(\dot{\mathbf{U}})}{\partial\dot{\mathbf{U}}^{\ast}}=&\frac{1}{2}\frac{\partial {\rm{Tr}}\{(\dot{\mathbf{C}}_{1}-\dot{\mathbf{U}}(\dot{\mathbf{V}}^{\tau})^{H})^{H}(\dot{\mathbf{C}}_{1}-\dot{\mathbf{U}}(\dot{\mathbf{V}}^{\tau})^{H})\} }{\partial\dot{\mathbf{U}}^{\ast}}\\
&+\frac{\lambda}{2\mu^{\tau}}\frac{\partial {\rm{Tr}}\{\dot{\mathbf{U}}^{H}\dot{\mathbf{U}}\} }{\partial\dot{\mathbf{U}}^{\ast}}\\
=&\frac{1}{2}\left(\frac{1}{2}\dot{\mathbf{C}}_{1}^{\ast}\dot{\mathbf{V}}^{\tau}-\mathfrak{R}(\dot{\mathbf{C}}_{1})\dot{\mathbf{V}}^{\tau}+\mathfrak{R}(\dot{\mathbf{U}}(\dot{\mathbf{V}}^{\tau})^{H})\dot{\mathbf{V}}^{\tau}\right.\\
&\left.-\frac{1}{2}\dot{\mathbf{U}}^{\ast}(\dot{\mathbf{V}}^{\tau})^{T}\dot{\mathbf{V}}^{\tau}\right)
+\frac{\lambda}{2\mu^{\tau}}(\mathfrak{R}(\dot{\mathbf{U}})-\frac{1}{2}\dot{\mathbf{U}}^{\ast})\\
=&\frac{1}{4}\left(\dot{\mathbf{U}}(\dot{\mathbf{V}}^{\tau})^{H}\dot{\mathbf{V}}^{\tau}-\dot{\mathbf{C}}_{1}\dot{\mathbf{V}}^{\tau}+\frac{\lambda}{\mu^{\tau}}\dot{\mathbf{U}}\right),
\end{split}
\end{equation}
where $\dot{\mathbf{C}}_{1}=\dot{\mathbf{X}}^{\tau}+\dot{\mathbf{F}}^{\tau}/\mu^{\tau}$. By the similar approach, we can obtain  the gradient of $\mathcal{G}(\dot{\mathbf{V}})$ as
\begin{equation}\small
\label{en8}
\frac{\partial\mathcal{G}(\dot{\mathbf{V}})}{\partial\dot{\mathbf{V}}^{\ast}}=\frac{1}{4}\left(\dot{\mathbf{V}}(\dot{\mathbf{U}}^{\tau+1})^{H}\dot{\mathbf{U}}^{\tau+1}-\dot{\mathbf{C}}_{1}^{H}\dot{\mathbf{U}}^{\tau+1}+\frac{\lambda}{\mu^{\tau}}\dot{\mathbf{V}}\right).
\end{equation}
Setting (\ref{en7}) and (\ref{en8}) to zero, we can obtain the unique solutions of $\dot{\mathbf{U}}^{\tau+1}$ and $\dot{\mathbf{V}}^{\tau+1}$ separately as
\begin{equation}\small
\label{en9}
\dot{\mathbf{U}}^{\tau+1}=\dot{\mathbf{C}}_{1}\dot{\mathbf{V}}^{\tau}\left((\dot{\mathbf{V}}^{\tau})^{H}\dot{\mathbf{V}}^{\tau}+\frac{\lambda}{\mu^{\tau}}\mathbf{I}\right)^{-1}
\end{equation}
and
\begin{equation}\small
\label{en10}
\dot{\mathbf{V}}^{\tau+1}=\dot{\mathbf{C}}_{1}^{H}\dot{\mathbf{U}}^{\tau+1}\left((\dot{\mathbf{U}}^{\tau+1})^{H}\dot{\mathbf{U}}^{\tau+1}+\frac{\lambda}{\mu^{\tau}}\mathbf{I}\right)^{-1}.
\end{equation}
\textbf{Updating $\dot{\mathbf{X}}$}: In the $\tau+1$-th iteration, fixing the other variables at their latest values,  $\dot{\mathbf{X}}^{\tau+1}$ is the optimal solution of the following problem
\begin{equation}\small
\begin{split}
\dot{\mathbf{X}}^{\tau+1}=&\mathop{{\rm{arg\, min}}}\limits_{\dot{\mathbf{X}}}\ \frac{1}{2}\|\mathcal{P}_{\Omega}(\dot{\mathbf{X}}-\dot{\mathbf{T}})\|_{F}^{2}\\
&+\frac{\mu^{\tau}}{2}\|\dot{\mathbf{X}}-\big(\dot{\mathbf{U}}^{\tau+1}(\dot{\mathbf{V}}^{\tau+1})^{H}-\frac{\dot{\mathbf{F}}^{\tau}}{\mu^{\tau}}\big)\|_{F}^{2}.
\end{split}
\end{equation}
Then, we can directly obtain the optimal $\dot{\mathbf{X}}^{\tau+1}$ as
\begin{equation}\small
\label{en11}
\begin{split}
\dot{\mathbf{X}}^{\tau+1}=&\mathcal{P}_{\Omega^{c}}\big(\dot{\mathbf{U}}^{\tau+1}(\dot{\mathbf{V}}^{\tau+1})^{H}-\dot{\mathbf{F}}^{\tau}/\mu^{\tau}\big)\\
&+\mathcal{P}_{\Omega}\bigg(\frac{\mu^{\tau}\dot{\mathbf{U}}^{\tau+1}(\dot{\mathbf{V}}^{\tau+1})^{H}-\dot{\mathbf{F}}^{\tau}+\dot{\mathbf{T}}}{1+\mu^{\tau}}\bigg),
\end{split}
\end{equation}
where $\Omega^{c}$ is the complement of $\Omega$, and we have used the
fact that $\mathcal{P}_{\Omega^{c}}(\dot{\mathbf{T}})=\mathbf{0}$ in (\ref{en11}).

\textbf{Updating $\dot{\mathbf{F}}$ and $\mu$}:
	The multiplier $\dot{\mathbf{F}}^{\tau+1}$ is directly obtained by
	\begin{equation}\small
	\label{en12}
	\dot{\mathbf{F}}^{\tau+1}=\dot{\mathbf{F}}^{\tau}+\mu^{\tau}(\dot{\mathbf{X}}^{\tau+1}-\dot{\mathbf{U}}^{\tau+1}(\dot{\mathbf{V}}^{\tau+1})^{H}).
	\end{equation}
	And the penalty parameter $\mu^{\tau+1}$ is
	\begin{equation}\small
	\label{en13}
	\mu^{\tau+1}={\rm{min}}(\beta\mu^{\tau}, \mu_{max}),
	\end{equation}
	where $\mu_{max}$ is the  default maximum value of the penalty parameter $\mu$, and $\beta\geq 1$ is a constant parameter.

Introducing the auxiliary variables $\dot{\mathbf{A}}_{U}$ and $\dot{\mathbf{A}}_{V}$, the problems (\ref{e11}) and (\ref{e12}) can be reformulated into the following equivalent formulations
\begin{equation}\small
\label{e13}
\begin{split}
&\mathop{{\rm{min}}}\limits_{\dot{\mathbf{U}},\dot{\mathbf{V}},\dot{\mathbf{A}}_{U},\dot{\mathbf{A}}_{V},\dot{\mathbf{X}}}\ \frac{\lambda}{2}\big(\|\dot{\mathbf{A}}_{U}\|_{\ast}+\|\dot{\mathbf{A}}_{V}\|_{\ast}\big),\\ 
&\text{s.t.},\ \dot{\mathbf{A}}_{U}=\dot{\mathbf{U}}, \dot{\mathbf{A}}_{V}=\dot{\mathbf{V}},
\mathcal{P}_{\Omega}(\dot{\mathbf{X}}-\dot{\mathbf{T}})=\mathbf{0},\\ &\qquad\dot{\mathbf{X}}=\dot{\mathbf{U}}\dot{\mathbf{V}}^{H}. 	
\end{split}
\end{equation}

\begin{equation}\small
\label{e14}
\begin{split}
&\mathop{{\rm{min}}}\limits_{\dot{\mathbf{U}},\dot{\mathbf{V}},\dot{\mathbf{A}}_{U},\dot{\mathbf{A}}_{V},\dot{\mathbf{X}}}\ \frac{\lambda}{3}\big(\|\dot{\mathbf{U}}\|_{F}^{2}+2\|\dot{\mathbf{A}}_{V}\|_{\ast}\big),\\ 
&\text{s.t.},\ \dot{\mathbf{A}}_{V}=\dot{\mathbf{V}},
\mathcal{P}_{\Omega}(\dot{\mathbf{X}}-\dot{\mathbf{T}})=\mathbf{0}, \dot{\mathbf{X}}=\dot{\mathbf{U}}\dot{\mathbf{V}}^{H}. 	
\end{split}
\end{equation}
Note that, (\ref{e13}) and (\ref{e14}) split the interdependent terms such that they can be solved independently. Then, the problems (\ref{e13}) and (\ref{e14}) can be solved by the ADMM framework.

 We first solve the problem (\ref{e13}) by minimizing the following augmented Lagrangian function
\begin{equation}\small
\label{e15}
\begin{split}
&\mathcal{L}_{\mu}(\dot{\mathbf{U}},\dot{\mathbf{V}},\dot{\mathbf{A}}_{U},\dot{\mathbf{A}}_{V},\dot{\mathbf{X}},\dot{\mathbf{F}}_{1},\dot{\mathbf{F}}_{2},\dot{\mathbf{F}}_{3})\\
&=\frac{\lambda}{2}\big(\|\dot{\mathbf{A}}_{U}\|_{\ast}+\|\dot{\mathbf{A}}_{V}\|_{\ast}\big)+\langle\dot{\mathbf{F}}_{1},\dot{\mathbf{U}}-\dot{\mathbf{A}}_{U}\rangle\\
&\quad +\langle\dot{\mathbf{F}}_{2},\dot{\mathbf{V}}-\dot{\mathbf{A}}_{V}\rangle+\langle\dot{\mathbf{F}}_{3},\dot{\mathbf{X}}-\dot{\mathbf{U}}\dot{\mathbf{V}}^{H}\rangle\\
&\quad+\frac{\mu}{2}\left(\|\dot{\mathbf{U}}-\dot{\mathbf{A}}_{U}\|_{F}^{2}+\|\dot{\mathbf{V}}-\dot{\mathbf{A}}_{V}\|_{F}^{2}\right.\\
&\quad+\left.\|\dot{\mathbf{X}}-\dot{\mathbf{U}}\dot{\mathbf{V}}^{H}\|_{F}^{2}\right)+\frac{1}{2}\|\mathcal{P}_{\Omega}(\dot{\mathbf{X}}-\dot{\mathbf{T}})\|_{F}^{2},
\end{split}
\end{equation}
where $\dot{\mathbf{F}}_{1}$, $\dot{\mathbf{F}}_{2}$ and $\dot{\mathbf{F}}_{3}$ are Lagrange multipliers.

\textbf{Updating $\dot{\mathbf{U}}$ and $\dot{\mathbf{V}}$}: In the $\tau+1$-th iteration, fixing the other variables at their latest values, $\dot{\mathbf{U}}^{\tau+1}$ and $\dot{\mathbf{V}}^{\tau+1}$ are respectively the optimal solutions of the following problems
\begin{equation}\small
\label{e16}
\begin{split}
\dot{\mathbf{U}}^{\tau+1}=&\mathop{{\rm{arg\, min}}}\limits_{\dot{\mathbf{U}}}\ \langle\dot{\mathbf{F}}_{1}^{\tau},\dot{\mathbf{U}}-\dot{\mathbf{A}}_{U}^{\tau}\rangle + \langle\dot{\mathbf{F}}_{3}^{\tau},\dot{\mathbf{X}}^{\tau}-\dot{\mathbf{U}}(\dot{\mathbf{V}}^{\tau})^{H}\rangle\\
&+\frac{\mu^{\tau}}{2}\left(\|\dot{\mathbf{U}}-\dot{\mathbf{A}}_{U}^{\tau}\|_{F}^{2}+\|\dot{\mathbf{X}}^{\tau}-\dot{\mathbf{U}}(\dot{\mathbf{V}}^{\tau})^{H}\|_{F}^{2}\right)\\
=&\mathop{{\rm{arg\, min}}}\limits_{\dot{\mathbf{U}}}\ \|\dot{\mathbf{U}}-\dot{\mathbf{A}}_{U}^{\tau}+\dot{\mathbf{F}}_{1}^{\tau}/\mu^{\tau}\|_{F}^{2}\\
&+\|\dot{\mathbf{X}}^{\tau}-\dot{\mathbf{U}}(\dot{\mathbf{V}}^{\tau})^{H}+\dot{\mathbf{F}}_{3}^{\tau}/\mu^{\tau}\|_{F}^{2},
\end{split}
\end{equation}

\begin{equation}\small
\label{e17}
\begin{split}
\dot{\mathbf{V}}^{\tau+1}=&\mathop{{\rm{arg\, min}}}\limits_{\dot{\mathbf{U}}}\ \langle\dot{\mathbf{F}}_{2}^{\tau},\dot{\mathbf{V}}-\dot{\mathbf{A}}_{V}^{\tau}\rangle + \langle\dot{\mathbf{F}}_{3}^{\tau},\dot{\mathbf{X}}^{\tau}-\dot{\mathbf{U}}^{\tau+1}\dot{\mathbf{V}}^{H}\rangle\\
&+\frac{\mu^{\tau}}{2}\left(\|\dot{\mathbf{V}}-\dot{\mathbf{A}}_{V}^{\tau}\|_{F}^{2}+\|\dot{\mathbf{X}}^{\tau}-\dot{\mathbf{U}}^{\tau+1}\dot{\mathbf{V}}^{H}\|_{F}^{2}\right)\\
=&\mathop{{\rm{arg\, min}}}\limits_{\dot{\mathbf{U}}}\ \|\dot{\mathbf{V}}-\dot{\mathbf{A}}_{V}^{\tau}+\dot{\mathbf{F}}_{2}^{\tau}/\mu^{\tau}\|_{F}^{2}\\
&+\|\dot{\mathbf{X}}^{\tau}-\dot{\mathbf{U}}^{\tau+1}\dot{\mathbf{V}}^{H}+\dot{\mathbf{F}}_{3}^{\tau}/\mu^{\tau}\|_{F}^{2}.
\end{split}
\end{equation}
Let $\mathcal{A}(\dot{\mathbf{U}}):=\|\dot{\mathbf{U}}-\dot{\mathbf{A}}_{U}^{\tau}+\dot{\mathbf{F}}_{1}^{\tau}/\mu^{\tau}\|_{F}^{2}+\|\dot{\mathbf{X}}^{\tau}-\dot{\mathbf{U}}(\dot{\mathbf{V}}^{\tau})^{H}+\dot{\mathbf{F}}_{3}^{\tau}/\mu^{\tau}\|_{F}^{2}$, and $\mathcal{B}(\dot{\mathbf{V}}):=\|\dot{\mathbf{V}}-\dot{\mathbf{A}}_{V}^{\tau}+\dot{\mathbf{F}}_{2}^{\tau}/\mu^{\tau}\|_{F}^{2}+\|\dot{\mathbf{X}}^{\tau}-\dot{\mathbf{U}}^{\tau+1}\dot{\mathbf{V}}^{H}+\dot{\mathbf{F}}_{3}^{\tau}/\mu^{\tau}\|_{F}^{2}$. The gradient of $\mathcal{A}(\dot{\mathbf{U}})$ can be calculated as
\begin{equation}\small
\label{e18}
\begin{split}
\frac{\partial\mathcal{A}(\dot{\mathbf{U}})}{\partial\dot{\mathbf{U}}^{*}}=&\frac{\partial {\rm{Tr}}\{(\dot{\mathbf{U}}-\dot{\mathbf{S}}_{1})^{H}(\dot{\mathbf{U}}-\dot{\mathbf{S}}_{1})\} }{\partial\dot{\mathbf{U}}^{*}}\\
&+\frac{\partial {\rm{Tr}}\{(\dot{\mathbf{S}}_{2}-\dot{\mathbf{U}}\dot{\mathbf{S}}_{3})^{H}(\dot{\mathbf{S}}_{2}-\dot{\mathbf{U}}\dot{\mathbf{S}}_{3})\} }{\partial\dot{\mathbf{U}}^{*}}\\
=&\mathfrak{R}(\dot{\mathbf{U}})-\frac{1}{2}\dot{\mathbf{U}}^{*}-\mathfrak{R}(\dot{\mathbf{S}}_{1})+\frac{1}{2}\dot{\mathbf{S}}_{1}^{*}\\
&+\frac{1}{2}\dot{\mathbf{S}}_{2}^{*}\dot{\mathbf{S}}_{3}^{H}-\mathfrak{R}(\dot{\mathbf{S}}_{2})\dot{\mathbf{S}}_{3}^{H}+\mathfrak{R}(\dot{\mathbf{U}}\dot{\mathbf{S}}_{3})\dot{\mathbf{S}}_{3}^{H}\\
&-\frac{1}{2}\dot{\mathbf{U}}^{*}\dot{\mathbf{S}}_{3}^{*}\dot{\mathbf{S}}_{3}^{H}\\
=&\frac{1}{2}(\dot{\mathbf{U}}-\dot{\mathbf{S}}_{1})+\frac{1}{2}(\dot{\mathbf{U}}\dot{\mathbf{S}}_{3}\dot{\mathbf{S}}_{3}^{H}-\dot{\mathbf{S}}_{2}\dot{\mathbf{S}}_{3}^{H}),
\end{split}
\end{equation}
where $\dot{\mathbf{S}}_{1}=\dot{\mathbf{A}}_{U}^{\tau}-\dot{\mathbf{F}}_{1}^{\tau}/\mu^{\tau}$, $\dot{\mathbf{S}}_{2}=\dot{\mathbf{X}}^{\tau}+\dot{\mathbf{F}}_{3}^{\tau}/\mu^{\tau}$ and $\dot{\mathbf{S}}_{3}=(\dot{\mathbf{V}}^{\tau})^{H}$. Setting (\ref{e18}) to zero, we can obtain a unique solution 
\begin{equation}\small
\label{e19}
\begin{split}
\dot{\mathbf{U}}^{\tau+1}=&(\dot{\mathbf{S}}_{1}+\dot{\mathbf{S}}_{2}\dot{\mathbf{S}}_{3}^{H})(\mathbf{I}+\dot{\mathbf{S}}_{3}\dot{\mathbf{S}}_{3}^{H})^{-1}\\
=&\left(\dot{\mathbf{A}}_{U}^{\tau}-\dot{\mathbf{F}}_{1}^{\tau}/\mu^{\tau}+(\dot{\mathbf{X}}^{\tau}\right.\\
&+\left.\dot{\mathbf{F}}_{3}^{\tau}/\mu^{\tau})\dot{\mathbf{V}}^{\tau}\right)\left(\mathbf{I}+(\dot{\mathbf{V}}^{\tau})^{H}\dot{\mathbf{V}}^{\tau}\right)^{-1}.\\
\end{split}
\end{equation}
By the similar way, we can obtain the optimal solution of $\dot{\mathbf{V}}^{\tau+1}$ as
\begin{equation}\small
\label{e20}
\begin{split}
\dot{\mathbf{V}}^{\tau+1}=&\left(\dot{\mathbf{A}}_{V}^{\tau}-\dot{\mathbf{F}}_{2}^{\tau}/\mu^{\tau}+(\dot{\mathbf{X}}^{\tau}\right.\\
&\left.+\dot{\mathbf{F}}_{3}^{\tau}/\mu^{\tau})^{H}\dot{\mathbf{U}}^{\tau+1}\right)\left(\mathbf{I}+(\dot{\mathbf{U}}^{\tau+1})^{H}\dot{\mathbf{U}}^{\tau+1}\right)^{-1}.\\
\end{split}
\end{equation}

\textbf{Updating $\dot{\mathbf{A}}_{U}$ and $\dot{\mathbf{A}}_{V}$}: In the $\tau+1$-th iteration, fixing the other variables at their latest values, $\dot{\mathbf{A}}_{U}^{\tau+1}$ and $\dot{\mathbf{A}}_{V}^{\tau+1}$ are respectively the optimal solutions of the following problems
\begin{equation}\small
\label{e21}
\begin{split}
\dot{\mathbf{A}}_{U}^{\tau+1}=&\mathop{{\rm{arg\, min}}}\limits_{\dot{\mathbf{A}}_{U}}\ \frac{\lambda}{2}\|\dot{\mathbf{A}}_{U}\|_{\ast}+\langle\dot{\mathbf{F}}_{1}^{\tau},\dot{\mathbf{U}}^{\tau+1}-\dot{\mathbf{A}}_{U}\rangle\\
&+\frac{\mu^{\tau}}{2}\|\dot{\mathbf{U}}^{\tau+1}-\dot{\mathbf{A}}_{U}\|_{F}^{2}\\
=&\mathop{{\rm{arg\, min}}}\limits_{\dot{\mathbf{A}}_{U}}\ \frac{1}{2}\|\dot{\mathbf{A}}_{U}-(\dot{\mathbf{U}}^{\tau+1}+\dot{\mathbf{F}}_{1}^{\tau}/\mu^{\tau})\|_{F}^{2}\\
&+\frac{\lambda}{2\mu^{\tau}}\|\dot{\mathbf{A}}_{U}\|_{\ast},
\end{split}
\end{equation}
\begin{equation}\small
\label{e22}
\begin{split}
\dot{\mathbf{A}}_{V}^{\tau+1}=&\mathop{{\rm{arg\, min}}}\limits_{\dot{\mathbf{A}}_{V}}\ \frac{\lambda}{2}\|\dot{\mathbf{A}}_{V}\|_{\ast}+\langle\dot{\mathbf{F}}_{2}^{\tau},\dot{\mathbf{V}}^{\tau+1}-\dot{\mathbf{A}}_{V}\rangle\\
&+\frac{\mu^{\tau}}{2}\|\dot{\mathbf{V}}^{\tau+1}-\dot{\mathbf{A}}_{V}\|_{F}^{2}\\
=&\mathop{{\rm{arg\, min}}}\limits_{\dot{\mathbf{A}}_{V}}\ \frac{1}{2}\|\dot{\mathbf{A}}_{V}-(\dot{\mathbf{V}}^{\tau+1}+\dot{\mathbf{F}}_{2}^{\tau}/\mu^{\tau})\|_{F}^{2}\\
&+\frac{\lambda}{2\mu^{\tau}}\|\dot{\mathbf{A}}_{V}\|_{\ast}.
\end{split}
\end{equation}
The closed-form solutions of (\ref{e21}) and (\ref{e22}) can be obtained by the quaternion singular value thresholding (QSVT) \cite{DBLP:journals/tip/ChenXZ20}, \emph{i.e.},
\begin{equation}\small
\label{e23}
\dot{\mathbf{A}}_{U}^{\tau+1}=\mathcal{D}_{\frac{\lambda}{2\mu^{\tau}}}\left(\dot{\mathbf{U}}^{\tau+1}+\dot{\mathbf{F}}_{1}^{\tau}/\mu^{\tau}\right),
\end{equation}
\begin{equation}\small
\label{e24}
\dot{\mathbf{A}}_{U}^{\tau+1}=\mathcal{D}_{\frac{\lambda}{2\mu^{\tau}}}\left(\dot{\mathbf{V}}^{\tau+1}+\dot{\mathbf{F}}_{2}^{\tau}/\mu^{\tau}\right).
\end{equation}
The QSVT operator $\mathcal{D}_{\delta}(\dot{\mathbf{M}})$ is defined as $\mathcal{D}_{\delta}(\dot{\mathbf{M}})=\dot{\mathbf{P}}{\rm{diag}}\{{\rm{max}}(\sigma_{i}(\dot{\mathbf{M}})-\delta, 0)\}\dot{\mathbf{Q}}^{H}$, and $\dot{\mathbf{M}}=\dot{\mathbf{P}}{\rm{diag}}\{\sigma_{i}(\dot{\mathbf{M}})\}\dot{\mathbf{Q}}^{H}$ is the QSVD of quaternion matrix $\dot{\mathbf{M}}$.

\textbf{Updating $\dot{\mathbf{X}}$}: In the $\tau+1$-th iteration, fixing the other variables at their latest values,  $\dot{\mathbf{X}}^{\tau+1}$ is the optimal solution of the following problem
\begin{equation}\small
\begin{split}
\dot{\mathbf{X}}^{\tau+1}=&\mathop{{\rm{arg\, min}}}\limits_{\dot{\mathbf{X}}}\ \frac{1}{2}\|\mathcal{P}_{\Omega}(\dot{\mathbf{X}}-\dot{\mathbf{T}})\|_{F}^{2}\\
&+\frac{\mu^{\tau}}{2}\|\dot{\mathbf{X}}-\big(\dot{\mathbf{U}}^{\tau+1}(\dot{\mathbf{V}}^{\tau+1})^{H}-\frac{\dot{\mathbf{F}}_{3}^{\tau}}{\mu^{\tau}}\big)\|_{F}^{2}.
\end{split}
\end{equation}
Then, we can directly obtain the optimal $\dot{\mathbf{X}}^{\tau+1}$ as
\begin{equation}\small
\label{e25}
\begin{split}
\dot{\mathbf{X}}^{\tau+1}=&\mathcal{P}_{\Omega^{c}}\big(\dot{\mathbf{U}}^{\tau+1}(\dot{\mathbf{V}}^{\tau+1})^{H}-\dot{\mathbf{F}}_{3}^{\tau}/\mu^{\tau}\big)\\
&+\mathcal{P}_{\Omega}\bigg(\frac{\mu^{\tau}\dot{\mathbf{U}}^{\tau+1}(\dot{\mathbf{V}}^{\tau+1})^{H}-\dot{\mathbf{F}}_{3}^{\tau}+\dot{\mathbf{T}}}{1+\mu^{\tau}}\bigg),
\end{split}
\end{equation}
where $\Omega^{c}$ is the complement of $\Omega$, and we have used the
fact that $\mathcal{P}_{\Omega^{c}}(\dot{\mathbf{T}})=\mathbf{0}$ in (\ref{e25}).

\textbf{Updating $\dot{\mathbf{F}}_{1}, \dot{\mathbf{F}}_{2},\dot{\mathbf{F}}_{3}$ and $\mu$ }:
The multipliers $\dot{\mathbf{F}}_{1}^{\tau+1}, \dot{\mathbf{F}}_{2}^{\tau+1}$ and $\dot{\mathbf{F}}_{3}^{\tau+1}$ are directly obtained by
\begin{equation}\small
\label{e26}
\begin{split}
\dot{\mathbf{F}}_{1}^{\tau+1}&=\dot{\mathbf{F}}_{1}^{\tau}+\mu^{\tau}(\dot{\mathbf{U}}^{\tau+1}-\dot{\mathbf{A}}_{U}^{\tau+1}),\\
\dot{\mathbf{F}}_{2}^{\tau+1}&=\dot{\mathbf{F}}_{2}^{\tau}+\mu^{\tau}(\dot{\mathbf{V}}^{\tau+1}-\dot{\mathbf{A}}_{V}^{\tau+1}),\\
\dot{\mathbf{F}}_{3}^{\tau+1}&=\dot{\mathbf{F}}_{3}^{\tau}+\mu^{\tau}(\dot{\mathbf{X}}^{\tau+1}-\dot{\mathbf{U}}^{\tau+1}(\dot{\mathbf{V}}^{\tau+1})^{H}).
\end{split}
\end{equation}
And the penalty parameter $\mu^{\tau+1}$ is
\begin{equation}\small
\label{e27}
\mu^{\tau+1}={\rm{min}}(\beta\mu^{\tau}, \mu_{max}).
\end{equation}

\subsection{Rank estimation}
\label{r_est}
In most cases, we do not know the true rank, $d$, of quaternion matrix data. However, it is essential for the success of the proposed methods. Thus, in this section,
we develop a method for estimating the rank of quaternion matrix data.
This method starts from an input overestimated rank $d$ of quaternion matrix $\dot{\mathbf{X}}$. Assume that the rank of $\dot{\mathbf{X}}^{\tau}$ is $d^{\tau}$. We compute the singular values of $( \dot{\mathbf{U}}^{\tau})^{H}\dot{\mathbf{U}}^{\tau}$, and ordered non-increasing, \emph{i.e.}, $\sigma^{\tau}_{1}\geq\sigma^{\tau}_{2}\geq, \ldots, \geq \sigma^{\tau}_{{d^{\tau}}}$. Then, we compute the quotient sequnce $\hat{\sigma}^{\tau}_{m}=\sigma^{\tau}_{m}/\sigma^{\tau}_{m+1},\:(m=1, \ldots, d^{\tau}-1)$. Assume that
\begin{equation*}\small
p^{\tau}=\mathop{{\rm{arg\, max}}}\limits_{1\leq m\leq d^{\tau}-1}\hat{\sigma}^{\tau}_{m},
\end{equation*}
and define
\begin{equation}\small
\label{geprank}
\delta^{\tau}=\frac{(d^{\tau}-1)\hat{\sigma}^{\tau}_{p^{\tau}}}{\sum_{m\neq p^{\tau}}\hat{\sigma}^{\tau}_{m}},
\end{equation}
where the value of $\delta^{\tau}$ represents how many times the maximum drop (occurring at the $ \sigma^{\tau}_{p^{\tau}}$) is larger than the average of the rest of drops \cite{DBLP:journals/oms/ShenWZ14}. Once $\delta^{\tau}>20$, \emph{i.e.}, there is a large drop in the estimated rank, we reduce ${d}^{\tau}$ to $p^{\tau}$. This adjustment is done only one time and works sufficiently well in our test.

Finally, as described above, the \textbf{Q-DFN}-based quaternion matrix completion method (\emph{i.e.}, the solutions of the problem (\ref{en3})) can be summarized in TABLE \ref{tab_algorithm0}, the \textbf{Q-DNN}-based quaternion matrix completion method (\emph{i.e.}, the solutions of the problem (\ref{e11})) can be summarized in TABLE \ref{tab_algorithm1}.  Similarly, we also present the \textbf{Q-FNN}-based quaternion matrix completion method (\emph{i.e.}, the solutions of the problem (\ref{e12})) in TABLE \ref{tab_algorithm2}, and provide the details in the Appendix \ref{a_sec4}.
\begin{table}[htbp]
	\caption{The \textbf{Q-DFN}-based quaternion matrix completion method.}
	\hrule
	\label{tab_algorithm0}
	\begin{algorithmic}[1]
		\REQUIRE The observed quaternion matrix data $\dot{\mathbf{T}}\in\mathbb{H}^{M\times N}$ with $\mathcal{P}_{\Omega^{c}}(\dot{\mathbf{T}})=\mathbf{0}$, $\lambda$, $\mu_{\max}$ and $\beta$.
		\STATE \textbf{Initialize} $\mu^{0}$, $d^{0}$ and $\dot{\mathbf{X}}^{0}=\dot{\mathbf{T}}$.
		\STATE \textbf{Repeat}
		\STATE Update $\dot{\mathbf{U}}^{\tau+1}$ and $\dot{\mathbf{V}}^{\tau+1}$  by (\ref{en9}) and (\ref{en10}). 
		\STATE Update $\dot{\mathbf{X}}^{\tau+1}$ by (\ref{en11}).
		\STATE Update $\dot{\mathbf{F}}^{\tau+1}$ by (\ref{en12}).
		\STATE Update  $\mu^{\tau+1}$ by (\ref{en13}).
		\IF {$\delta^{\tau}> 20$ in (\ref{geprank})}
		\STATE  Re-estimate rank (\emph{see} \ref{r_est}) and adjust sizes of the iterates.
		\ENDIF
		\STATE  $\tau\longleftarrow \tau+1$.
		\STATE \textbf{Until convergence}
		\ENSURE   $\dot{\mathbf{U}}^{\tau+1}$, $\dot{\mathbf{V}}^{\tau+1}$ and $\dot{\mathbf{X}}^{\tau+1}$.
	\end{algorithmic}
	\hrule
\end{table}

\begin{table}[htbp]
	\caption{The \textbf{Q-DNN}-based quaternion matrix completion method.}
	\hrule
	\label{tab_algorithm1}
	\begin{algorithmic}[1]
		\REQUIRE The observed quaternion matrix data $\dot{\mathbf{T}}\in\mathbb{H}^{M\times N}$ with $\mathcal{P}_{\Omega^{c}}(\dot{\mathbf{T}})=\mathbf{0}$, $\lambda$, $\mu_{\max}$ and $\beta$.
		\STATE \textbf{Initialize} $\mu^{0}$, $d^{0}$, $\dot{\mathbf{A}}_{U}^{0}$, $\dot{\mathbf{A}}_{V}^{0}$ and $\dot{\mathbf{X}}^{0}=\dot{\mathbf{T}}$.
		\STATE \textbf{Repeat}
		\STATE Update $\dot{\mathbf{U}}^{\tau+1}$ and $\dot{\mathbf{V}}^{\tau+1}$  by (\ref{e19}) and (\ref{e20}). 
		\STATE Update $\dot{\mathbf{A}}_{U}^{\tau+1}$ and $\dot{\mathbf{A}}_{V}^{\tau+1}$  by (\ref{e23}) and (\ref{e24}). 
		\STATE Update $\dot{\mathbf{X}}^{\tau+1}$ by (\ref{e25}).
		\STATE Update $\dot{\mathbf{F}}_{1}^{\tau+1}, \dot{\mathbf{F}}_{2}^{\tau+1}$ and $\dot{\mathbf{F}}_{3}^{\tau+1}$ by (\ref{e26}).
		\STATE Update  $\mu^{\tau+1}$ by (\ref{e27}).
		\IF {$\delta^{\tau}> 20$ in (\ref{geprank})}
		\STATE  Re-estimate rank (\emph{see} \ref{r_est}) and adjust sizes of the iterates.
		\ENDIF
		\STATE  $\tau\longleftarrow \tau+1$.
		\STATE \textbf{Until convergence}
		\ENSURE   $\dot{\mathbf{U}}^{\tau+1}$, $\dot{\mathbf{V}}^{\tau+1}$ and $\dot{\mathbf{X}}^{\tau+1}$.
	\end{algorithmic}
	\hrule
\end{table}

\begin{table}[htbp]
	\caption{The \textbf{Q-FNN}-based quaternion matrix completion method.}
	\hrule
	\label{tab_algorithm2}
	\begin{algorithmic}[1]
		\REQUIRE The observed quaternion matrix data $\dot{\mathbf{T}}\in\mathbb{H}^{M\times N}$ with $\mathcal{P}_{\Omega^{c}}(\dot{\mathbf{T}})=\mathbf{0}$, $\lambda$, $\mu_{\max}$ and $\beta$.
		\STATE \textbf{Initialize} $\mu^{0}$, $d^{0}$, $\dot{\mathbf{A}}_{V}^{0}$ and $\dot{\mathbf{X}}^{0}=\dot{\mathbf{T}}$.
		\STATE \textbf{Repeat}
		\STATE $\dot{\mathbf{U}}^{\tau+1}=\left((\mu^{\tau}\dot{\mathbf{X}}^{\tau}
		+\dot{\mathbf{F}}_{2}^{\tau})\dot{\mathbf{V}}^{\tau}\right)\left(\frac{2\lambda}{3}\mathbf{I}+\mu^{\tau}(\dot{\mathbf{V}}^{\tau})^{H}\dot{\mathbf{V}}^{\tau}\right)^{-1}$. 
		\STATE		
		$\dot{\mathbf{V}}^{\tau+1}=\left(\dot{\mathbf{A}}_{V}^{\tau}-\dot{\mathbf{F}}_{1}^{\tau}/\mu^{\tau}+(\dot{\mathbf{X}}^{\tau}\right.$
		\STATE
		$\qquad \qquad\left.+\dot{\mathbf{F}}_{2}^{\tau}/\mu^{\tau})^{H}\dot{\mathbf{U}}^{\tau+1}\right)\left(\mathbf{I}+(\dot{\mathbf{U}}^{\tau+1})^{H}\dot{\mathbf{U}}^{\tau+1}\right)^{-1}$. 	
		\STATE $\dot{\mathbf{A}}_{V}^{\tau+1}=\mathcal{D}_{\frac{2\lambda}{3\mu^{\tau}}}\left(\dot{\mathbf{V}}^{\tau+1}+\dot{\mathbf{F}}_{1}^{\tau}/\mu^{\tau}\right)$.
		\STATE $\dot{\mathbf{X}}^{\tau+1}=\mathcal{P}_{\Omega^{c}}\big(\dot{\mathbf{U}}^{\tau+1}(\dot{\mathbf{V}}^{\tau+1})^{H}-\dot{\mathbf{F}}_{2}^{\tau}/\mu^{\tau}\big)$
		\STATE \qquad\qquad$+\mathcal{P}_{\Omega}\bigg(\frac{\mu^{\tau}\dot{\mathbf{U}}^{\tau+1}(\dot{\mathbf{V}}^{\tau+1})^{H}-\dot{\mathbf{F}}_{2}^{\tau}+\dot{\mathbf{T}}}{1+\mu^{\tau}}\bigg)$.
		\STATE $\dot{\mathbf{F}}_{1}^{\tau+1}=\dot{\mathbf{F}}_{1}^{\tau}+\mu^{\tau}(\dot{\mathbf{V}}^{\tau+1}-\dot{\mathbf{A}}_{V}^{\tau+1})$.
		\STATE $\dot{\mathbf{F}}_{2}^{\tau+1}=\dot{\mathbf{F}}_{2}^{\tau}+\mu^{\tau}(\dot{\mathbf{X}}^{\tau+1}-\dot{\mathbf{U}}^{\tau+1}(\dot{\mathbf{V}}^{\tau+1})^{H})$
		\STATE $\mu^{\tau+1}={\rm{min}}(\beta\mu^{\tau}, \mu_{max})$.
		\IF {$\delta^{\tau}>20$ in (\ref{geprank})}
		\STATE  Re-estimate rank (\emph{see} \ref{r_est}) and adjust sizes of the iterates.
		\ENDIF
		\STATE  $\tau\longleftarrow \tau+1$.
		\STATE \textbf{Until convergence}
		\ENSURE   $\dot{\mathbf{U}}^{\tau+1}$, $\dot{\mathbf{V}}^{\tau+1}$ and $\dot{\mathbf{X}}^{\tau+1}$.
	\end{algorithmic}
	\hrule
\end{table}

The stopping criterion of the three algorithms is defined as following relative  error 
\begin{equation*}
{\rm{RE}}:=\frac{\|\dot{\mathbf{U}}^{\tau}(\dot{\mathbf{V}}^{\tau})^{H}-\dot{\mathbf{X}}^{\tau}\|_{F}}{\|\dot{\mathbf{T}}\|_{F}}\leq tol,
\end{equation*}
where, $tol>0$ (we set $tol=1e-4$ in the experiments) is the stopping tolerance. 

\section{Experimental results}
\label{sec4}
In this section, several experiments on some natural color images are conducted to evaluate the effectiveness of the proposed Q-DFN-based, Q-DNN-based and Q-FNN-based quaternion matrix completion methods.
All the experiments are run in MATLAB $2014b$ under Windows $10$ on a personal computer with $1.60$GHz CPU and $8$GB memory.

\textbf{Compared methods:} In the simulations, we 
compare the proposed Q-DNN and Q-FNN with six state-of-the-art low-rank matrix completion methods including:
\begin{itemize}
	\item \textbf{D-N}\footnote{\url{http://doi.ieeecomputersociety.org/10.1109/TPAMI.2017.2748590}\label{web1}} \cite{DBLP:journals/pami/ShangCLLL18}: a BF-based real matrix completion method which minimize the double nuclear norm. 
	\item \textbf{F-N}
	\cite{DBLP:journals/pami/ShangCLLL18}: a BF-based real matrix completion method which minimize the Frobenius/nuclear norm. 
   \item \textbf{WNNM}\footnote{\url{http://www4.comp.polyu.edu.hk/cslzhang/code}} \cite{DBLP:journals/ijcv/GuXMZFZ17}:  a real matrix completion method  which  uses  the weighted nuclear norm regularization term.
   \item \textbf{MC-NC}\footnote{\url{https://github.com/sudalvxin/MC-NC.git}} \cite{DBLP:journals/tip/NieHL19}: a real matrix completion method based  on  Non-Convex relaxation.
  \item
  \textbf{LRQA-2} \cite{DBLP:journals/tip/ChenXZ20}: a quaternion matrix completion method based a nonconvex  rank surrogate (laplace function).
  \item
  \textbf{RegL1-ALM}\footnote{\url{https://sites.google.com/site/yinqiangzheng/}}  \cite{DBLP:conf/cvpr/ZhengLSYO12}: a robust BF-based real matrix completion method which utilizes the L1-norm rather than Frobenius norm to build loss function.
\end{itemize}

\textbf{Parameter and initialization setting:} For Q-DFN, Q-DNN and Q-FNN, we set $\lambda=0.05\sqrt{{\rm{max}}(M,N)}$, $\mu_{\max}=10^{20}$ and $\beta=1.03$. Let
$\mu^{0}=10^{-3}$ ($\mu^{0}=10^{-2}$ for  Q-DNN), $\dot{\mathbf{A}}_{U}^{0}= \mathbf{I}$, $\dot{\mathbf{A}}_{V}^{0}=\mathbf{I}$, and the appropriate $d^{0}$ is chosen from $\{40,60,80,100,120\}$. In addition, all compared methods are from the source codes and the parameter settings are based on the suggestions in the orginal papers.

\textbf{Quantitative assessment:} To evaluate the performance of proposed methods, except visual quality, we employ two widely used quantitative quality indexes, including the peak signal-to-noise ratio (PSNR) and the structure similarity (SSIM) \cite{DBLP:journals/tip/WangBSS04}. 

\begin{figure}[htbp]
	\centering
	\includegraphics[width=7.5cm,height=1cm]{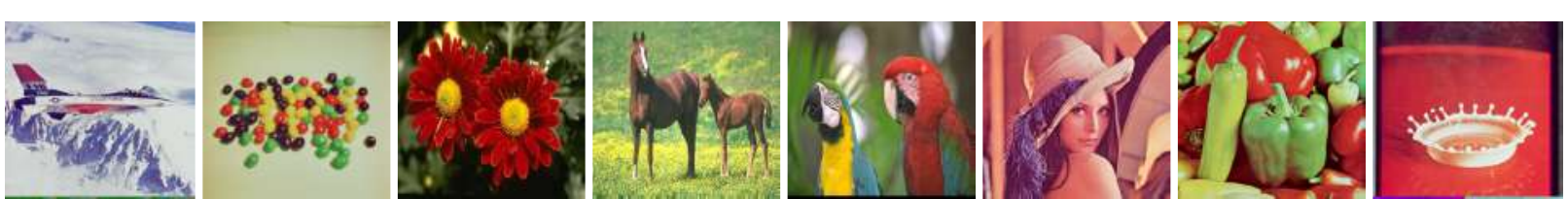}
	\caption{The $8$ color images (from left to right, Image(1) $256\times 256\times 3$,
		Image(2)  $256\times 256\times 3$, Image(3)  $256\times 256\times 3$, Image(4)  $256\times 256\times 3$, Image(5)  $256\times 256\times 3$, Image(6)  $256\times 256\times 3$, Image(7)  $256\times 256\times 3$ and Image
		(8)  $256\times 256\times 3$).}
	\label{Ytu}
\end{figure}

The widely used $8$ color images with different sizes are selected as the test samples shown in Fig.\ref{Ytu}. Similar to the settings in \cite{DBLP:journals/pami/ShangCLLL18,DBLP:journals/tgrs/ChenGWWPH17,DBLP:journals/tip/ChenXZ20}, all real matrix completion methods among the compared methods are performed on each channel of the test images individually.

\begin{figure}[!htb]
	\centering
	\subfigure[]{
		\includegraphics[width=2.7cm,height=3cm]{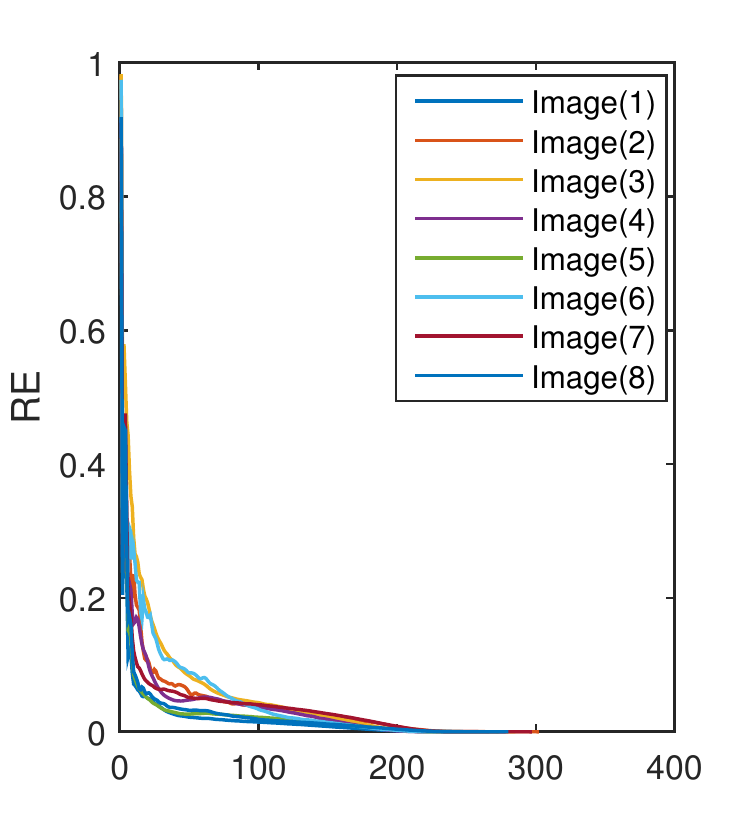}
	}
	\hspace{-0.25in}
	\subfigure[]{
		\includegraphics[width=2.7cm,height=3cm]{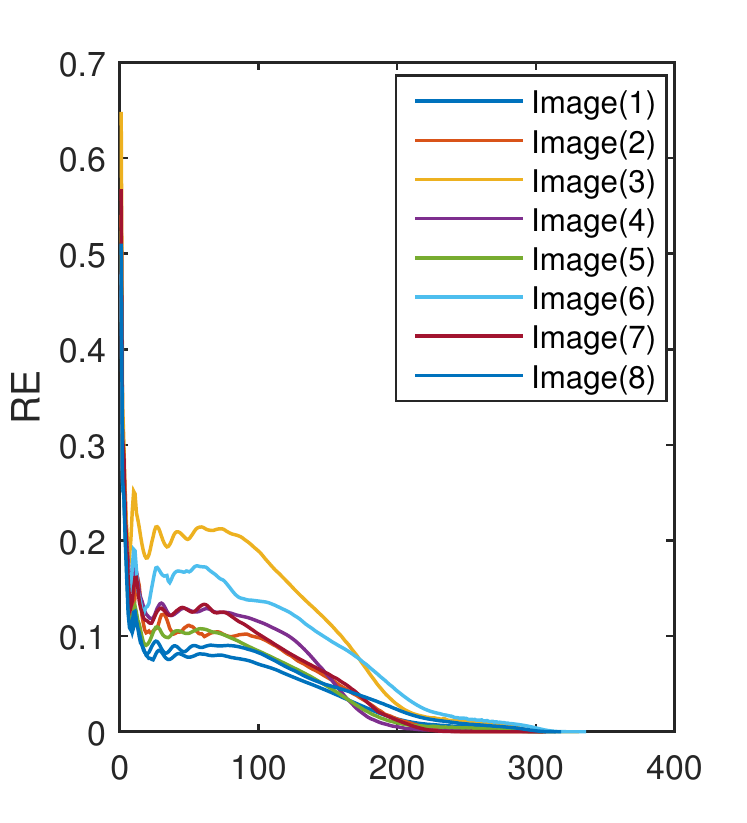}
	}
	\hspace{-0.25in}
	\subfigure[]{
		\includegraphics[width=2.7cm,height=3cm]{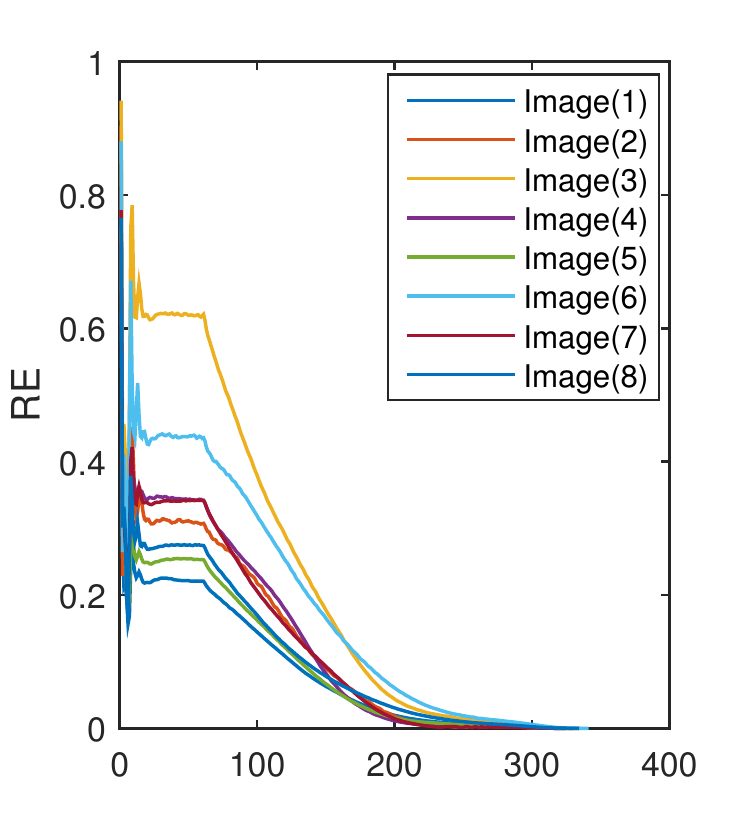}
	}
	\caption{Relative error (RE) versus iteration on all images (from left to right: (a) Q-DFN, (b) Q-DNN and (c) Q-FNN).}
	\label{fig0}
\end{figure}

Fig.\ref{fig0} shows the empirical convergence of the proposed three methods for color image inpainting on all images with ${\rm{MR}}=0.50$, where ${\rm{MR}}$ denotes the missing ratio of pixels (the larger the value of ${\rm{MR}}$, the more pixels are lost).

\begin{figure*}[htbp]
	\centering
\subfigure[]{
	\includegraphics[width=1.39cm,height=11cm]{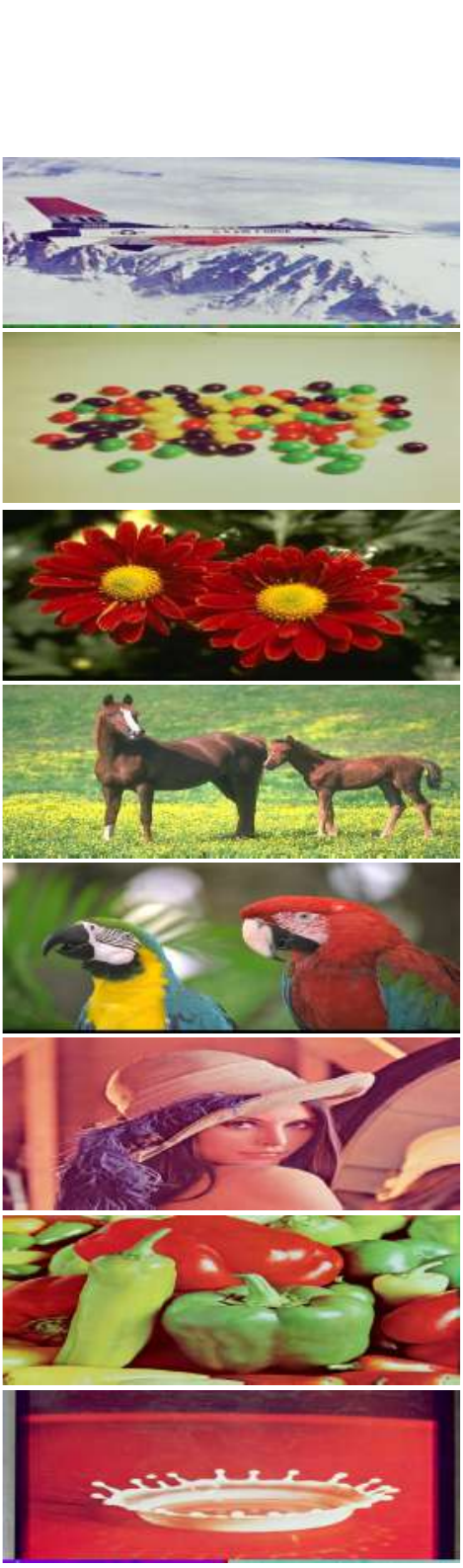}
	}
\hspace{-0.18in}
\subfigure[]{
	\includegraphics[width=1.39cm,height=11cm]{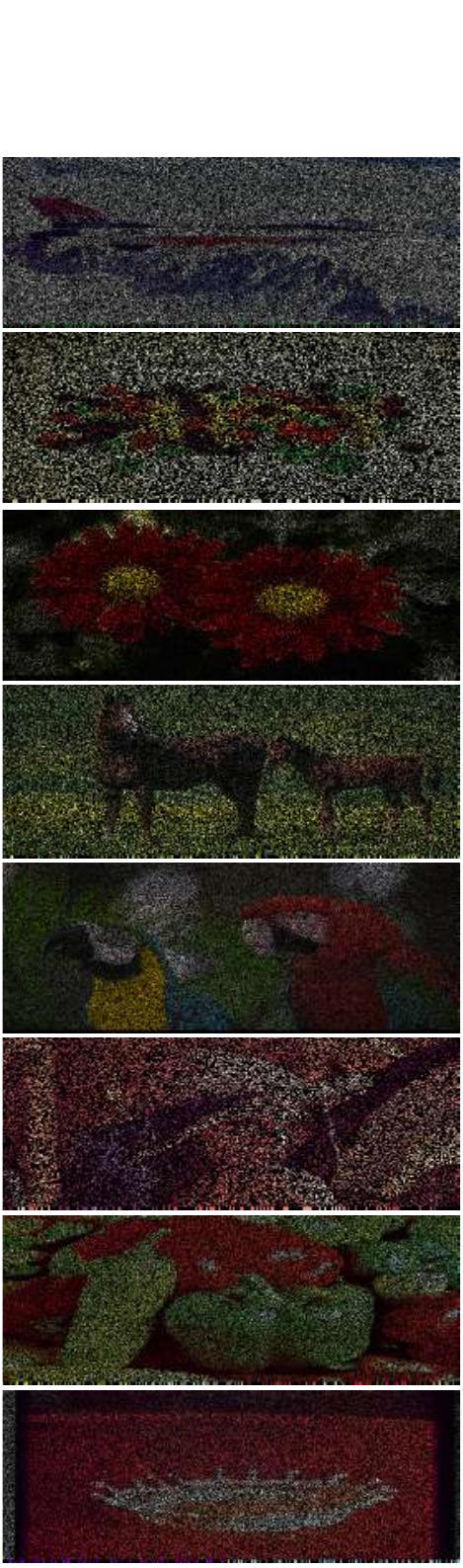}
}
\hspace{-0.18in}
\subfigure[]{
	\includegraphics[width=1.39cm,height=11cm]{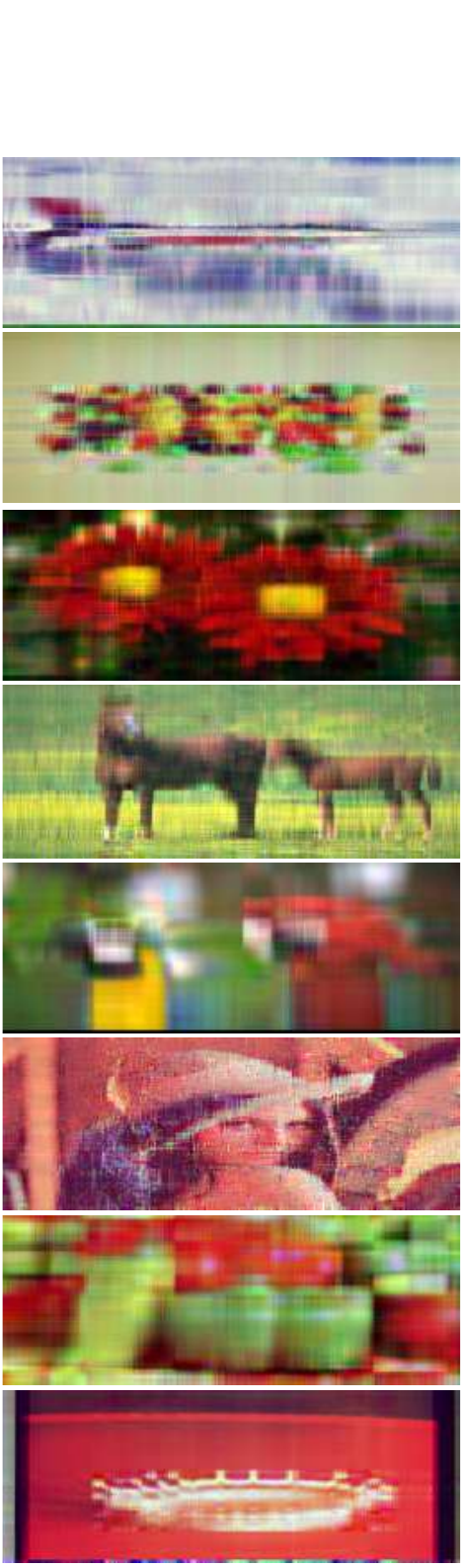}
}
\hspace{-0.18in}
\subfigure[]{
	\includegraphics[width=1.39cm,height=11cm]{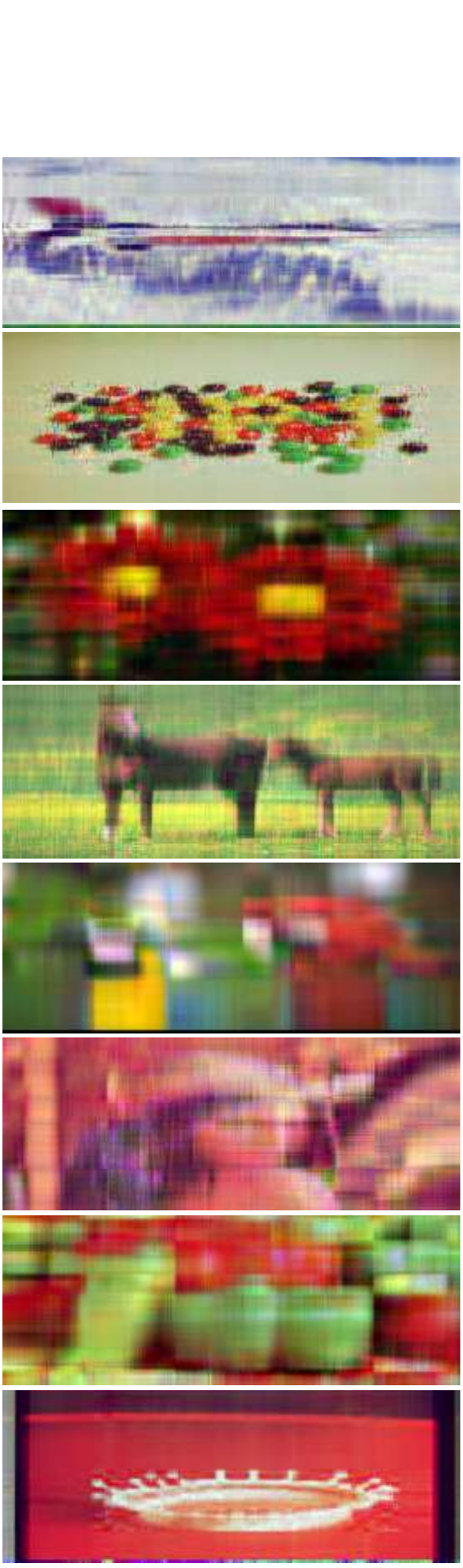}
}
\hspace{-0.18in}
\subfigure[]{
	\includegraphics[width=1.39cm,height=11cm]{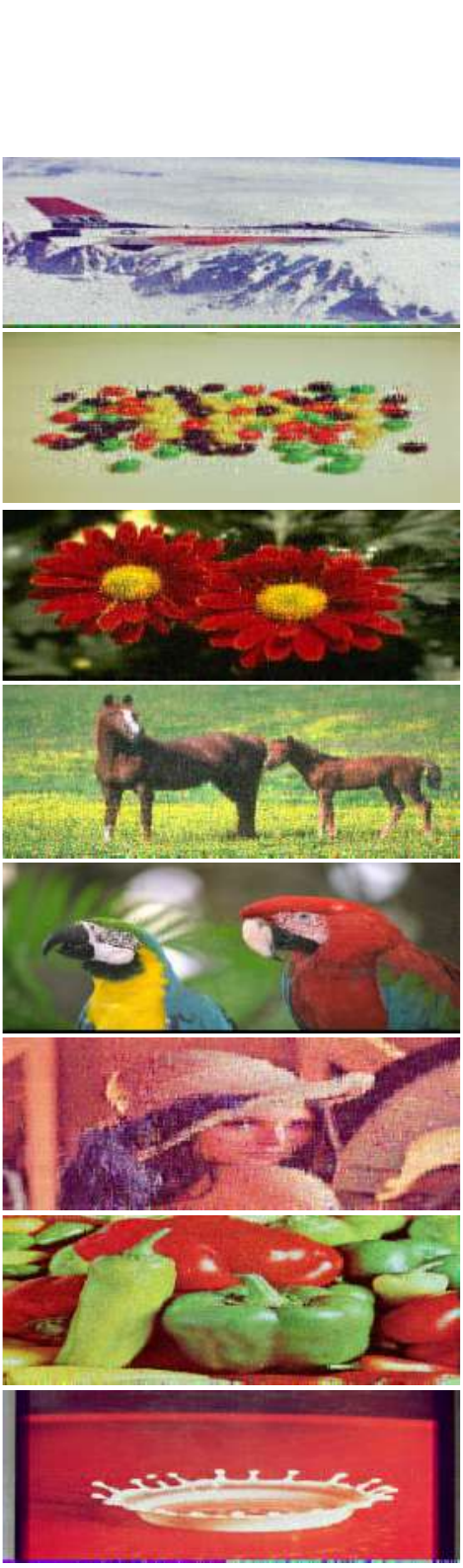}
}
\hspace{-0.18in}
\subfigure[]{
	\includegraphics[width=1.39cm,height=11cm]{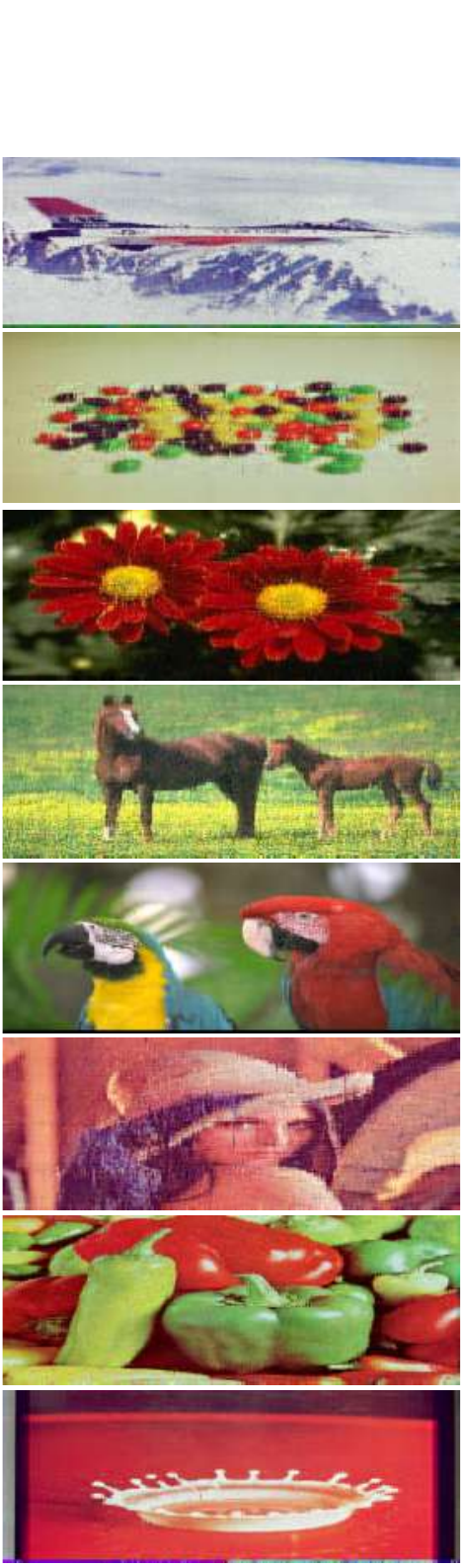}
}
\hspace{-0.18in}
\subfigure[]{
	\includegraphics[width=1.39cm,height=11cm]{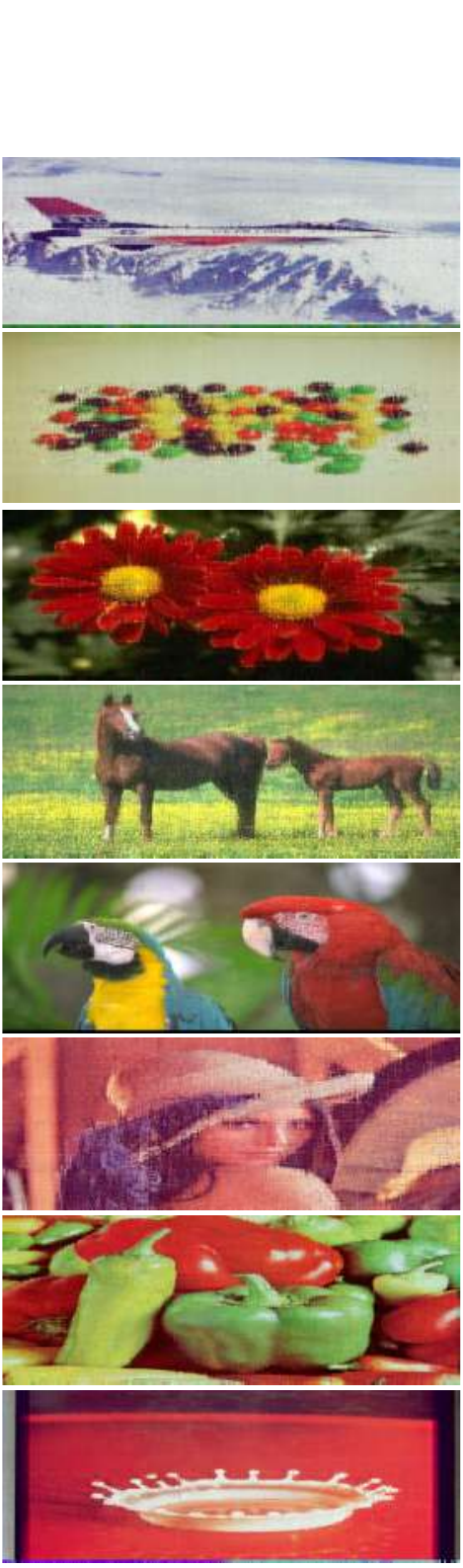}
}
\hspace{-0.18in}
\subfigure[]{
	\includegraphics[width=1.39cm,height=11cm]{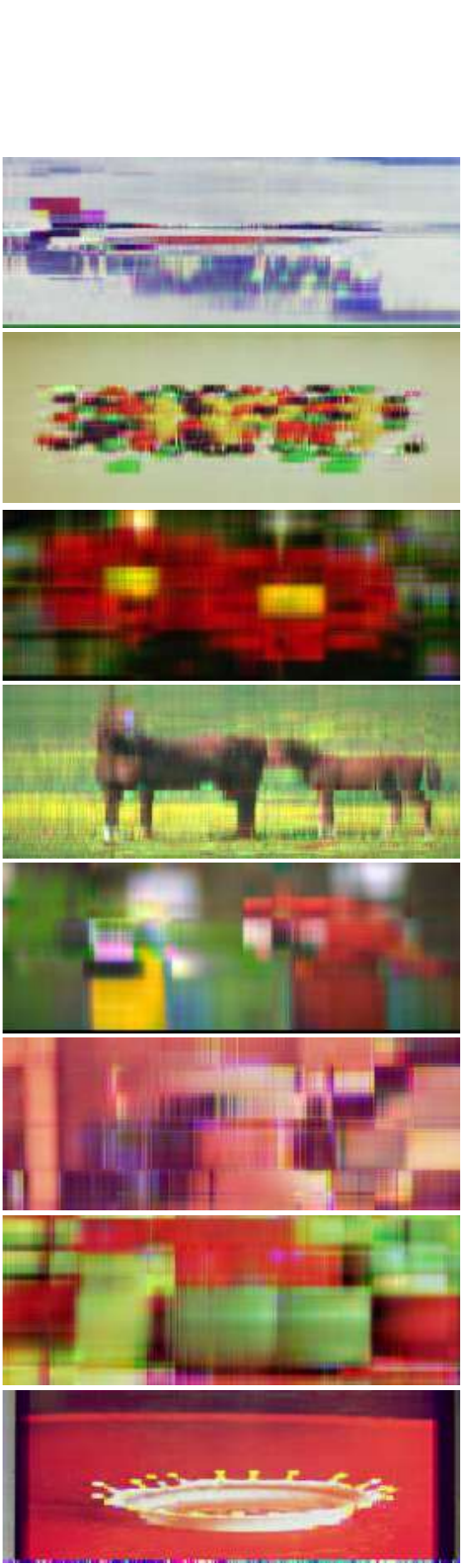}
}
\hspace{-0.18in}
\subfigure[]{
	\includegraphics[width=1.39cm,height=11cm]{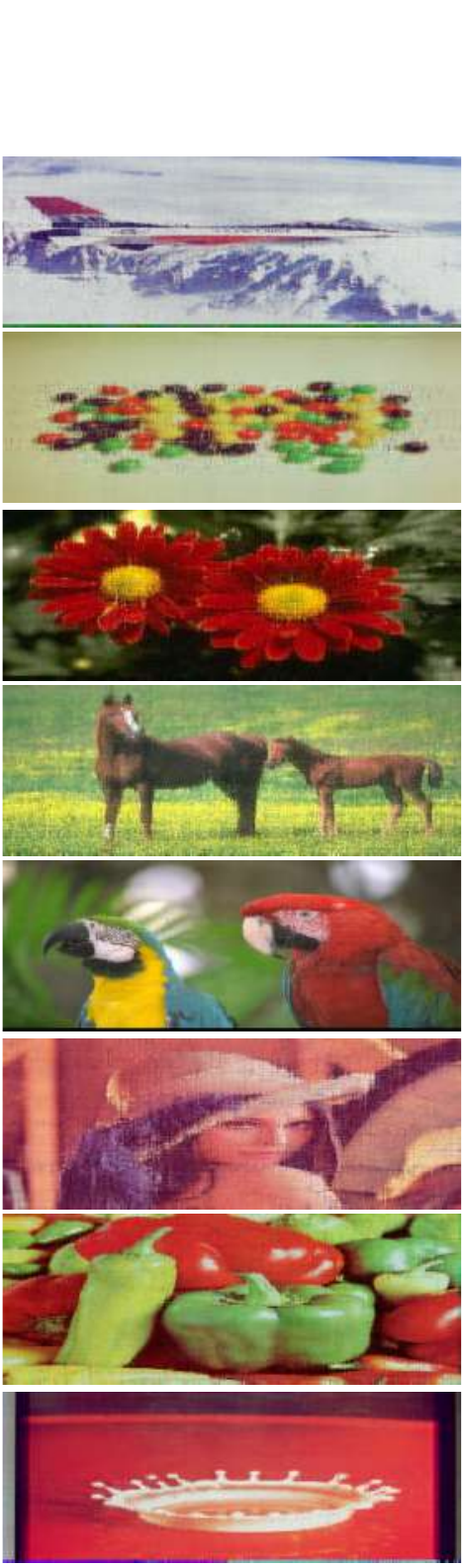}
}
\hspace{-0.18in}
\subfigure[]{
	\includegraphics[width=1.39cm,height=11cm]{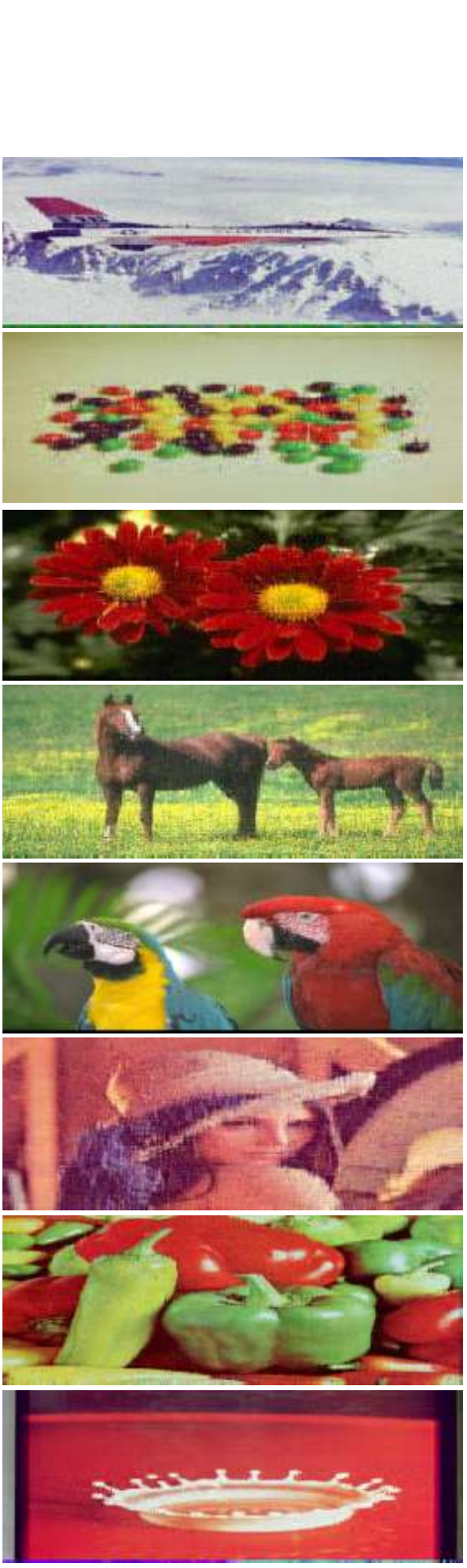}
}
\hspace{-0.18in}
\subfigure[]{
	\includegraphics[width=1.39cm,height=11cm]{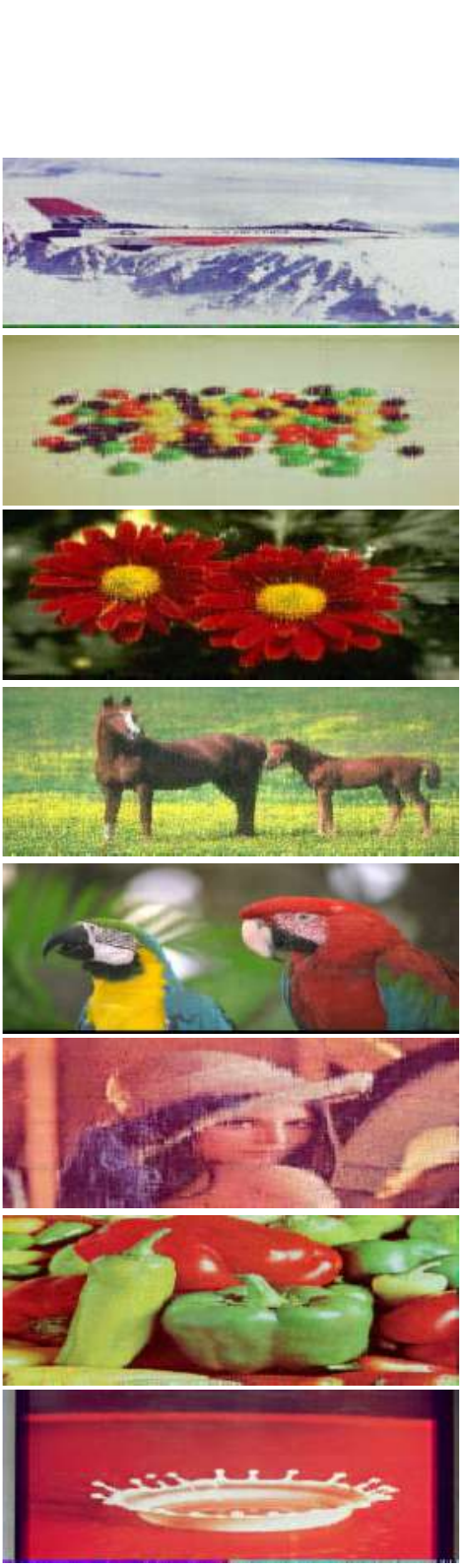}
}\\
\vspace{-0.1in}
\subfigure[\textbf{Bold} fonts denote the best performance; \underline{underline} ones represent the second-best results.]{
		\resizebox{13.5cm}{4.2cm}{
		\begin{tabular}{|c|c|c|c|c|c|c|c|c|c|c|}
			\hline 
			Images & Indexes  & D-N \cite{DBLP:journals/pami/ShangCLLL18} & F-N \cite{DBLP:journals/pami/ShangCLLL18} & WNNM \cite{DBLP:journals/ijcv/GuXMZFZ17} & MC-NC \cite{DBLP:journals/tip/NieHL19} & LRQA-2 \cite{DBLP:journals/tip/ChenXZ20} & RegL1-ALM \cite{DBLP:conf/cvpr/ZhengLSYO12} & \textbf{Q-DFN} & \textbf{Q-DNN} &\textbf{Q-FNN}\\ \toprule
			\hline
			
			\multirow{3}{*}{Image(1)}
			&PSNR  &20.956  &21.313  &23.549&26.002 &25.767  &20.467 &25.332 &\underline{26.085}    &\textbf{26.170}  \\ 
			\cline{2-11}
			&SSIM  &0.404  &0.413  &0.430  &0.607  &0.735    &0.508 &\underline{0.761}&0.747    &\textbf{0.769}  \\ 
			\cline{2-11} 
			&time(s)  &25.08  &20.05  &142.27  &93.56   &743.07  &636.51 &94.31&129.04   &137.67 \\ 
			\hline 
			
			\multirow{3}{*}{Image(2)}	
			&PSNR  &21.067  &23.259  &23.648  &25.604   &25.679  &21.777&25.291 &\underline{25.901} &\textbf{26.017}  \\ 
			\cline{2-11}
			&SSIM  &0.834 &0.877  &0.884  &0.926  &0.929  &0.877 &0.928&\underline{0.931}    &\textbf{0.937}  \\ 
			\cline{2-11} 
			&time(s)  &5.54  &5.06  &41.80  &28.32  &92.44  &154.19 &60.10&66.05&73.79  \\ 
			\hline 
			
			\multirow{3}{*}{Image(3)}	
			&PSNR  &21.383 &21.350  &23.554  &25.121  &25.764  &19.367 &25.672&\underline{25.985}   &\textbf{26.007} \\ 
			\cline{2-11}
			&SSIM  &0.741  &0.745  &0.789  &0.847  &0.913   &0.742 &0.915&\underline{0.916}    &\textbf{0.920} \\ 
			\cline{2-11} 
			&time(s) &18.02  &14.08  &89.17  &281.25  &407.69   &368.69&64.63 &99.22    &121.30 \\ 
			\hline 
			
			\multirow{3}{*}{Image(4)}	
			&PSNR  &19.722  &19.774  &18.945  &20.013  &21.665 &19.423 &\textbf{21.780}&\underline{21.686} &21.636  \\ 
			\cline{2-11}
			&SSIM  &0.764  &0.762  &0.718  &0.775  &0.850   &0.764 &\textbf{0.856}&\underline{0.852}    &0.848  \\ 
			\cline{2-11} 
			&time(s) &17.95  &14.24  &87.36  &216.34  &337.88  &498.33 &77.68&109.30  &126.18  \\ 
			\hline 
			
			\multirow{3}{*}{Image(5)}	
			&PSNR  &22.804 &22.847  &25.899  &27.453  &27.795 &21.212  &27.721&\underline{27.969} &\textbf{28.040}\\ 
			\cline{2-11}
			&SSIM  &0.860  &0.854 &0.857  &0.913  &0.928   &0.776  &0.929&\underline{0.930}    &\textbf{0.932} \\ 
			\cline{2-11} 
			&time(s)  &31.52 &27.86  &264.36  &160.73  &1694.73  &825.09   &122.23&179.18   &200.70\\ 
			\hline 	
			
			\multirow{3}{*}{Image(6)}	
			&PSNR  &21.359  &20.637  &21.265  &23.426  &\underline{23.878}  &19.198  &23.482 &23.724 &\textbf{24.065}\\ 
			\cline{2-11}
			&SSIM  &0.865  &0.852  &0.853  &0.912  &\underline{0.924}   &0.845 &\underline{0.924}&0.919    &\textbf{0.925} \\ 
			\cline{2-11} 
			&time(s)  &7.10  &6.27  &40.71  &24.77  &85.99  &123.76   &61.54 &61.47   &71.98\\ 
			\hline 	
			
			\multirow{3}{*}{Image(7)}	
			&PSNR  &18.998  &18.745  &23.012  &25.957  &\underline{25.909}  &17.358 &25.801  &25.735 &\textbf{26.403}\\ 
			\cline{2-11}
			&SSIM  &0.834  &0.833  &0.895  &0.941  &\underline{0.945}   &0.809 &0.937 &0.942    &\textbf{0.951} \\ 
			\cline{2-11} 
			&time(s)  &25.00  &20.92  &151.91  &122.56  &796.07  &548.46   &110.09 &126.56   &132.61\\ 
			\hline 	
			
			\multirow{3}{*}{Image(8)}	
			&PSNR  &24.607  &25.823  &28.377  &30.422  &30.040   &24.799 &29.850 &\underline{30.475} &\textbf{30.573}\\ 
			\cline{2-11}
			&SSIM  &0.918  &0.924  &0.935  &0.956  &0.965   &0.930  &0.961&\underline{0.967}    &\textbf{0.969} \\ 
			\cline{2-11} 
			&time(s)  &24.55  &19.49  &148.67  &70.87  &880.53  &852.91  &107.51&128.68  &130.61\\ 
			\hline 					
		\end{tabular}}
	}%
	\caption{(a) is the original image. (b) is the observed image (${\rm{MR}}=0.70$). (c)-(k) are the inpainting results of D-N, F-N, WNNM, MC-NC, LRQA-2, RegL1-ALM,  Q-DFN, Q-DNN  and Q-FNN,  respectively.  (l)  summaries  the  PSNR  values, SSIM values and the running  time of all methods. \textbf{The figure is viewed better in zoomed PDF}.}
	\label{fig2}
\end{figure*}
\begin{figure}[htbp]
	\centering
	\includegraphics[width=8cm,height=5cm]{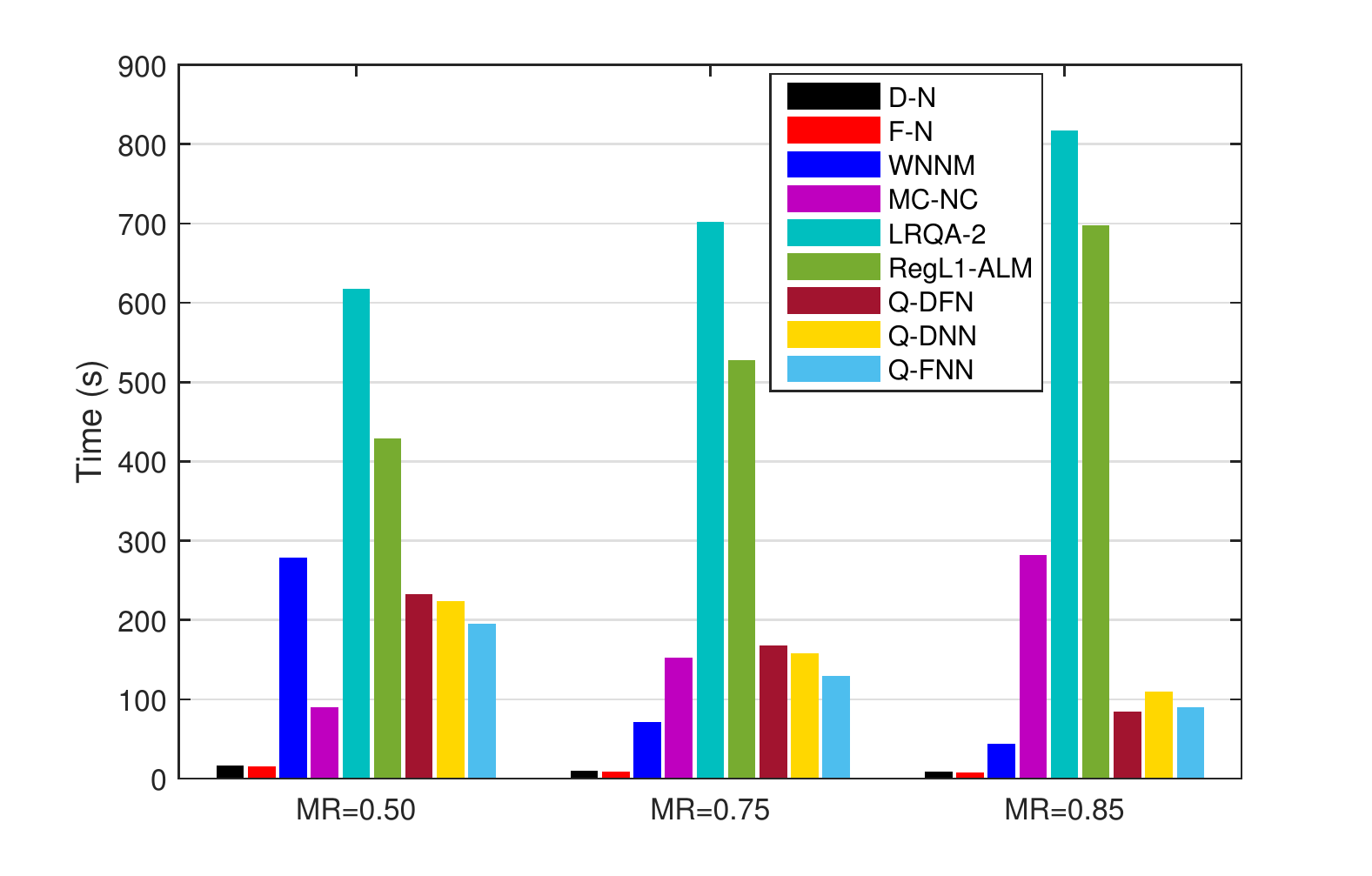}
	\caption{Average runtime (second) of all methods for color image inpainting on the eight testing color images with different  ${\rm{MRs}}$.}
	\label{atime}
\end{figure}
\begin{table*}[htbp]
	\caption{Quantitative assessment indexes (PSNR/SSIM) of different methods on the eight color images (\textbf{Bold} fonts denote the best performance; \underline{underline} ones represent the second-best results).}
	\centering
	\resizebox{15cm}{6cm}{
	\begin{tabular}{|c|c|c|c|c|c|c|c|c|c|}		
		\hline
		  Methods:& D-N \cite{DBLP:journals/pami/ShangCLLL18} & F-N \cite{DBLP:journals/pami/ShangCLLL18} & WNNM \cite{DBLP:journals/ijcv/GuXMZFZ17} & MC-NC \cite{DBLP:journals/tip/NieHL19} & LRQA-2 \cite{DBLP:journals/tip/ChenXZ20} & RegL1-ALM \cite{DBLP:conf/cvpr/ZhengLSYO12} &\textbf{Q-DFN}& \textbf{Q-DNN} &\textbf{Q-FNN} \\ \toprule
		\hline
		Images:  &\multicolumn{9}{c|}{${\rm{MR}}=0.50$}\\
		\hline
		Image(1)&21.409/0.421&21.409/0.422&28.388/0.605&29.154/0.778&29.171/0.829&19.324/0.483&30.055/\textbf{0.885}&\textbf{30.702}/0.878&\underline{30.526}/\underline{0.883}\\
		Image(2)&22.990/0.858&22.977/0.857&29.350/0.957&\underline{30.865}/\underline{0.974}&30.275/0.970&20.881/0.880 &30.180/0.973   &\textbf{30.947}/\textbf{0.975}&30.173/0.971\\
		Image(3)&21.833/0.742&21.833/0.742&27.630/0.901&\underline{29.674}/0.944&29.556/0.964&20.084/0.789   &29.408/\textbf{0.968}   &\textbf{29.724}/\underline{0.967}&29.591/0.966\\
		Image(4)&21.048/0.780&21.048/0.780&21.631/0.835&23.888/0.900&\underline{24.316}/0.912&19.671/0.789   &\textbf{24.400}/\textbf{0.919}   &24.043/0.914&24.308/\underline{0.918}\\  
		Image(5)&23.670/0.801&23.670/0.802&29.858/0.921&31.562/0.957&30.645/0.952&20.897/0.771&31.539/\textbf{0.967} &\textbf{31.838}/\underline{0.966}&\underline{31.798}/\textbf{0.967}\\
		Image(6)&21.912/0.875&21.001/0.879&24.960/0.929&\textbf{27.760}/\textbf{0.967}&27.469/0.962&19.965/0.870&27.474/\underline{0.965} &\underline{27.475}/0.964&27.192/0.961\\
		Image(7)&19.636/0.842&19.638/0.844&26.805/0.950&29.247/0.972&29.277/0.972&17.317/0.816      &29.151/0.972&\underline{29.541}/\underline{0.973}&\textbf{29.836}/\textbf{0.976}\\
		Image(8)&25.144/0.924&25.146/0.926&32.508/0.962&\underline{34.203}/0.978&33.286/0.981&22.842/0.924   &34.069/0.984   &\textbf{34.686}/\textbf{0.987}&34.144/\underline{0.985} \\
		\hline
		Aver. &22.205&22.090&27.641&\underline{29.540}&29.249&20.122&29.534&29.869&\textbf{29.696} \\ \toprule
		\hline
		Images  &\multicolumn{9}{c|}{${\rm{MR}}=0.75$}\\
		\hline
		Image(1)&20.761/0.398&20.759/0.393&21.189/0.379&24.724/0.543&24.659/0.697&19.926/0.484&24.133/\underline{0.719}&\underline{24.841}/0.699&\textbf{25.134}/\textbf{0.725}\\
		Image(2)&20.871/0.827&20.876/0.827&21.850/0.841&23.799/0.891&24.111/0.905&19.854/0.842&23.778/0.905&\textbf{24.638}/\textbf{0.913}&\underline{24.520}/\underline{0.912} \\
		Image(3)&20.326/0.706&20.331/0.707&22.367/0.741&23.902/0.808&24.738/0.889&19.584/0.750&24.273/0.885&\underline{24.956}/\underline{0.894}&\textbf{25.069}/\textbf{0.897} \\
		Image(4)&19.358/0.751&19.355/0.751&18.107/0.671&19.237/0.737&\textbf{21.050}/\textbf{0.826}&18.975/0.753&20.963/0.830&20.719/0.812&\underline{20.858}/\underline{0.819} \\  
		Image(5)&22.388/0.769&22.387/0.769&24.746/0.827&26.455/0.896&26.933/0.915&21.743/0.769&26.694/\underline{0.916}&\underline{26.949}/0.915&\textbf{27.023}/\textbf{0.919} \\
		Image(6)&19.833/0.813&19.906/0.815&20.014/0.816&22.151/0.886&22.743/0.904&18.425/0.773&22.334/\underline{0.907}&\underline{22.907}/0.904&\textbf{23.107}/\textbf{0.910} \\
		Image(7)&18.528/0.824&18.528/0.824&21.850/0.870&24.459/0.926&\underline{24.792}/0.932&17.287/0.813&23.869/0.920&24.727/\underline{0.930}&\textbf{25.076}/\textbf{0.936} \\
		Image(8)&24.508/0.910&24.491/0.911&27.182/0.923&28.981/0.946&28.666/0.957&23.232/0.921&28.001/0.953&\underline{29.144}/\underline{0.960}&\textbf{29.247}/\textbf{0.962} \\
		\hline
		Aver. &20.817&20.829&22.163&24.213&24.711&19.878  &24.289&\underline{24.860}&\textbf{25.004}   \\ \toprule	
			\hline
		Images  &\multicolumn{9}{c|}{${\rm{MR}}=0.85$}\\
		\hline
		Image(1)&19.847/0.317 &19.859/0.317&18.877/0.251&20.651/0.352&21.604/0.567&18.821/0.339&21.391/\underline{0.584}&\underline{21.860}/0.570&\textbf{22.066}/\textbf{0.596}\\
		Image(2)&18.979/0.740 &19.073/0.747&18.277/0.739&19.285/0.776&20.924/0.832&17.533/0.691&21.120/\underline{0.853}&\underline{21.154}/0.844&\textbf{21.601}/\textbf{0.860} \\
		Image(3)&20.427/0.698 &20.436/0.698&19.548/0.619&20.488/0.674&\underline{22.201}/\underline{0.804}&19.660/0.718&21.386/0.790&22.007/0.796&\textbf{22.299}/\textbf{0.818} \\
		Image(4)&17.578/0.655 &17.605/0.658&16.295/0.570&17.353/0.633&19.316/\underline{0.766}&16.211/0.615&\underline{19.412}/\textbf{0.776}&19.226/0.749&\textbf{19.507}/\underline{0.766}\\  
		Image(5)&21.905/0.742 &21.922/0.743&22.089/0.740&23.603/0.823&24.512/0.853&21.498/0.757&24.244/0.876&\underline{24.532}/\underline{0.865}&\textbf{24.806}/\textbf{0.879}\\
		Image(6)&18.287/0.755 &18.169/0.750&17.218/0.709&18.297/0.769&20.334/0.855&17.157/0.708&19.662/0.853&\textbf{20.471}/\underline{0.859}&\underline{20.444}/\textbf{0.860} \\
		Image(7)&19.685/0.835 &19.674/0.836&18.556/0.773&20.017/0.829&\underline{21.823}/\underline{0.885}&19.282/0.837&20.817/0.869&21.355/0.869&\textbf{22.118}/\textbf{0.891} \\
		Image(8)&23.628/0.897 &23.683/0.898&23.636/0.885&24.857/0.909&25.863/0.935&21.815/0.876&24.842/0.928&\underline{25.933}/\underline{0.937}&\textbf{26.189}/\textbf{0.939}\\
		\hline
		Aver. &20.042&20.053&19.312&20.569&\underline{22.072}&18.997&21.606&22.067&\textbf{22.379} \\ \toprule	
	\end{tabular}}
  \label{Index4SR_2}
\end{table*}

\begin{figure}[htb]
	\centering
	\subfigure[Original]{
		\includegraphics[width=2.3cm,height=2cm]{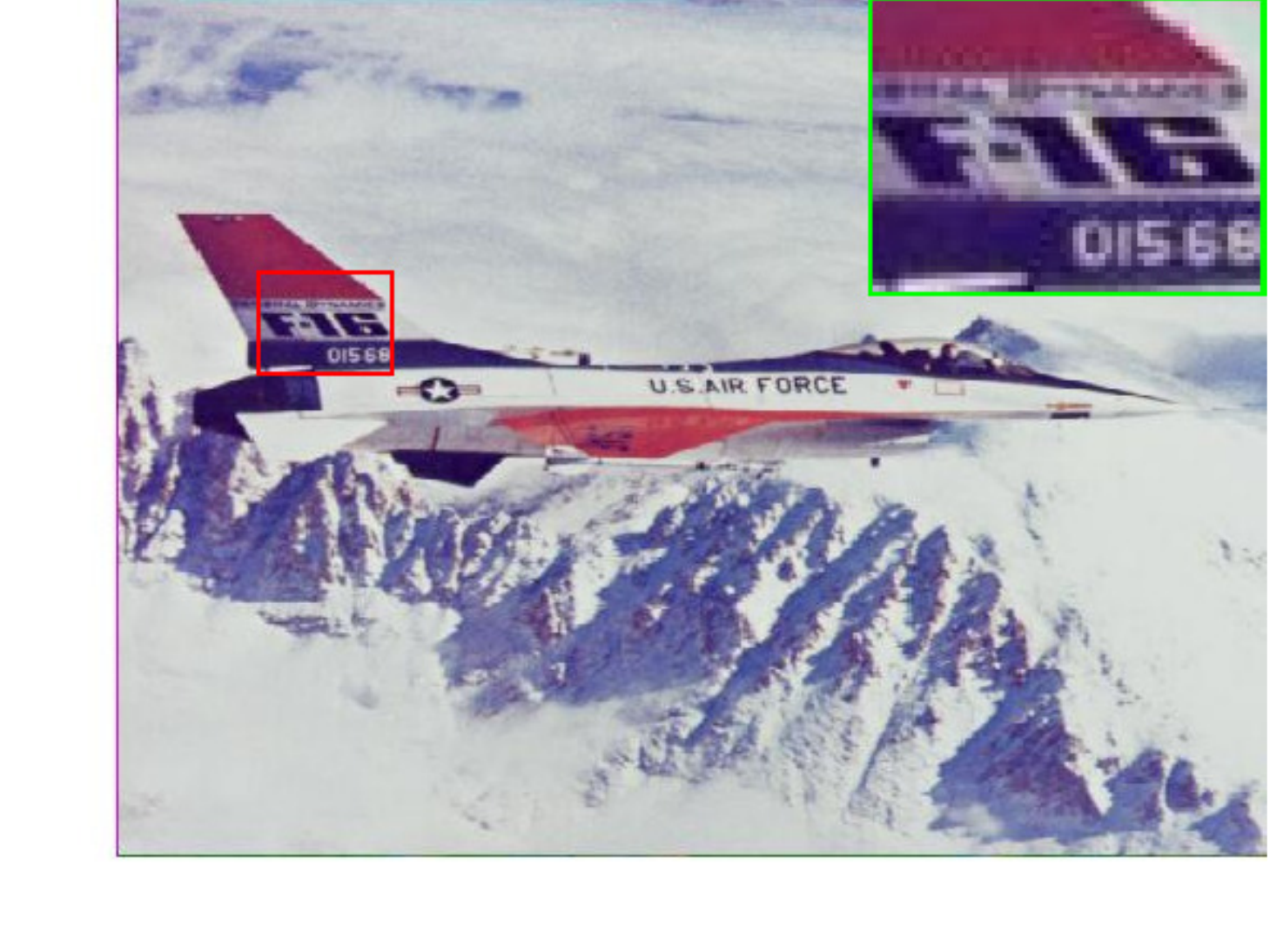}
	}
	\hspace{-0.18in}
	\subfigure[Observed]{
		\includegraphics[width=2.3cm,height=2cm]{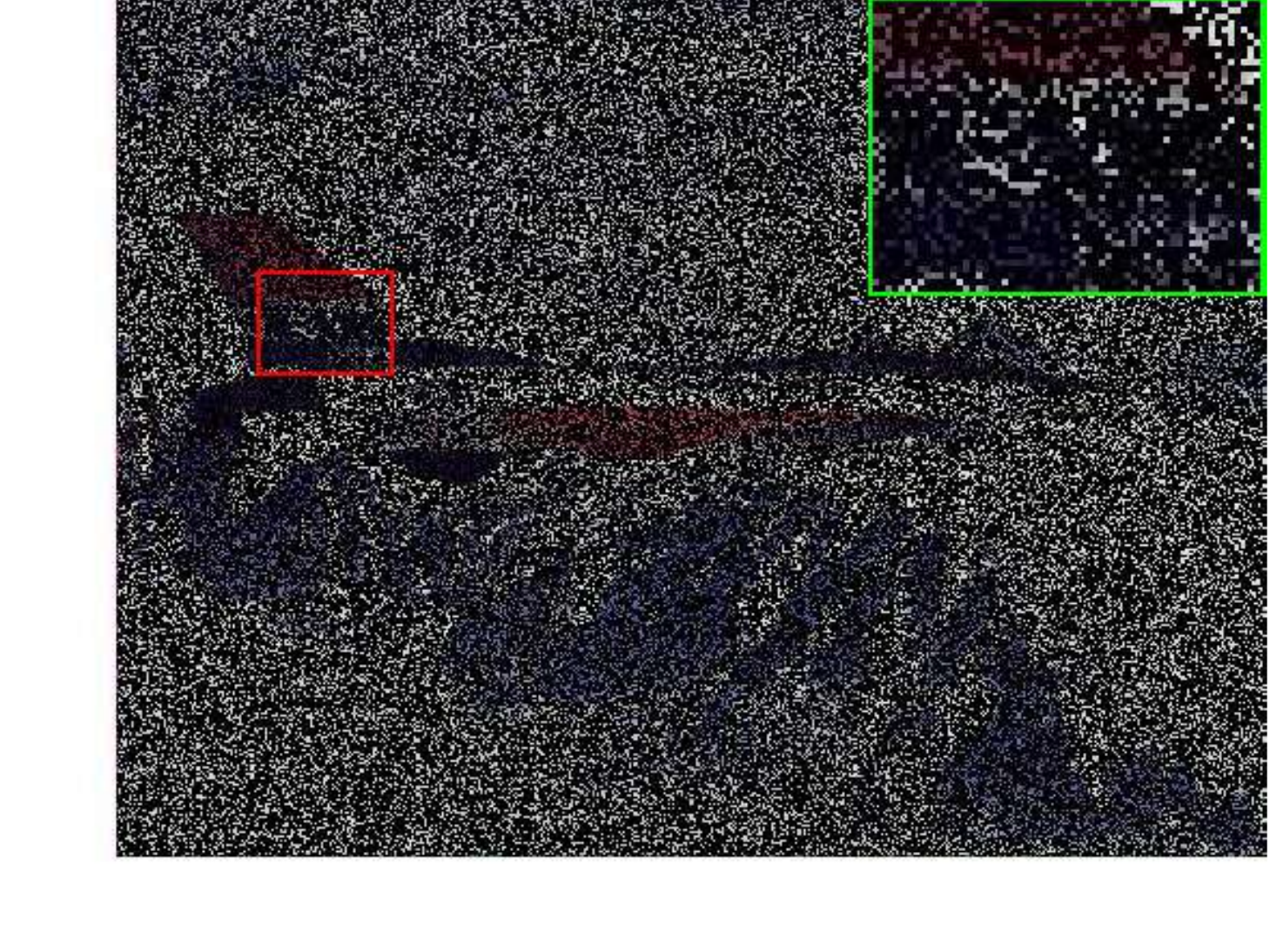}
	}\\
	\hspace{-0.18in}
	\subfigure[D-N \cite{DBLP:journals/pami/ShangCLLL18}]{
		\includegraphics[width=2.3cm,height=2cm]{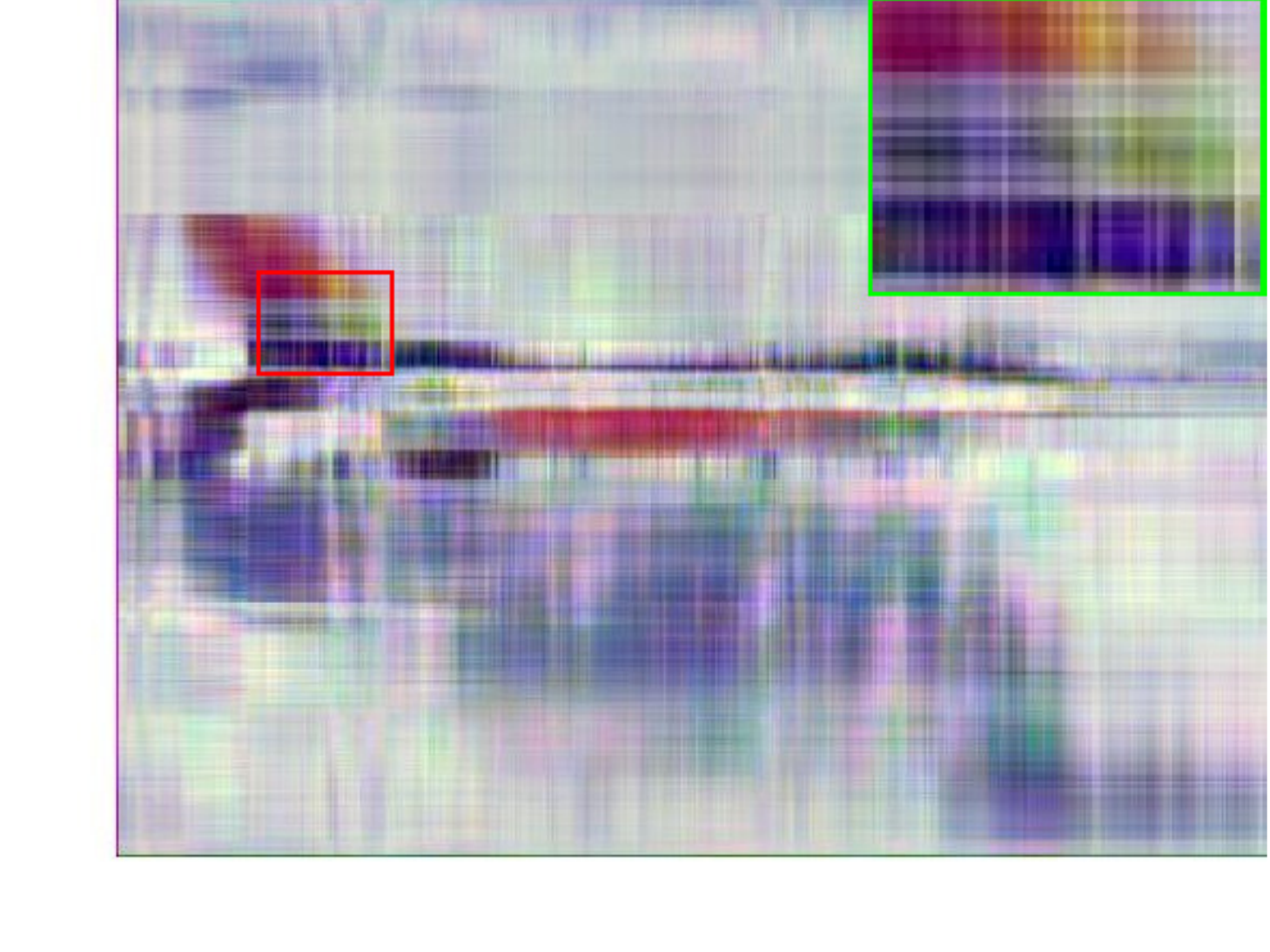}
	}
	\hspace{-0.18in}
\subfigure[F-N \cite{DBLP:journals/pami/ShangCLLL18}]{
	\includegraphics[width=2.3cm,height=2cm]{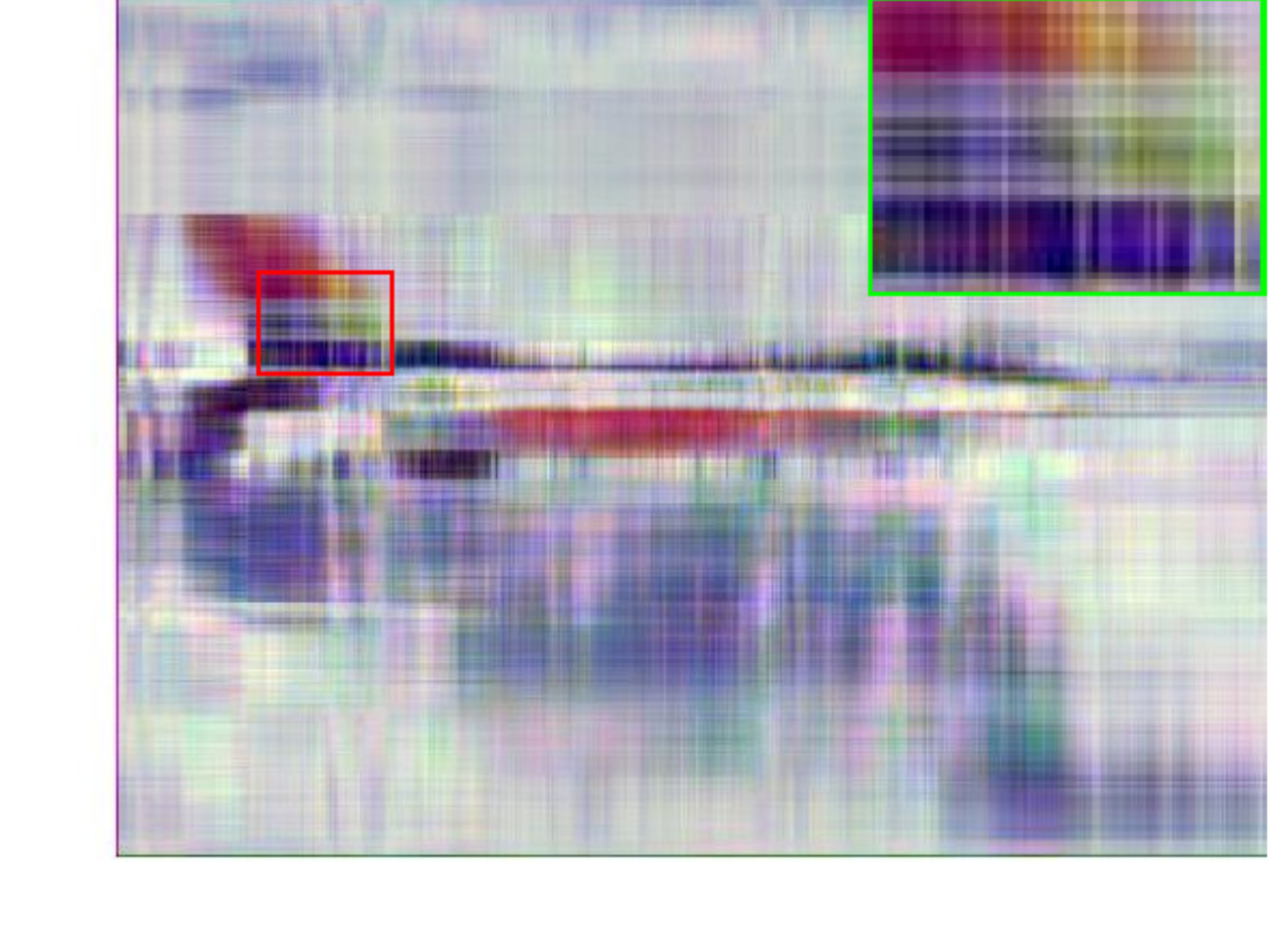}
}
	\hspace{-0.18in}
\subfigure[WNNM \cite{DBLP:journals/ijcv/GuXMZFZ17}]{
	\includegraphics[width=2.3cm,height=2cm]{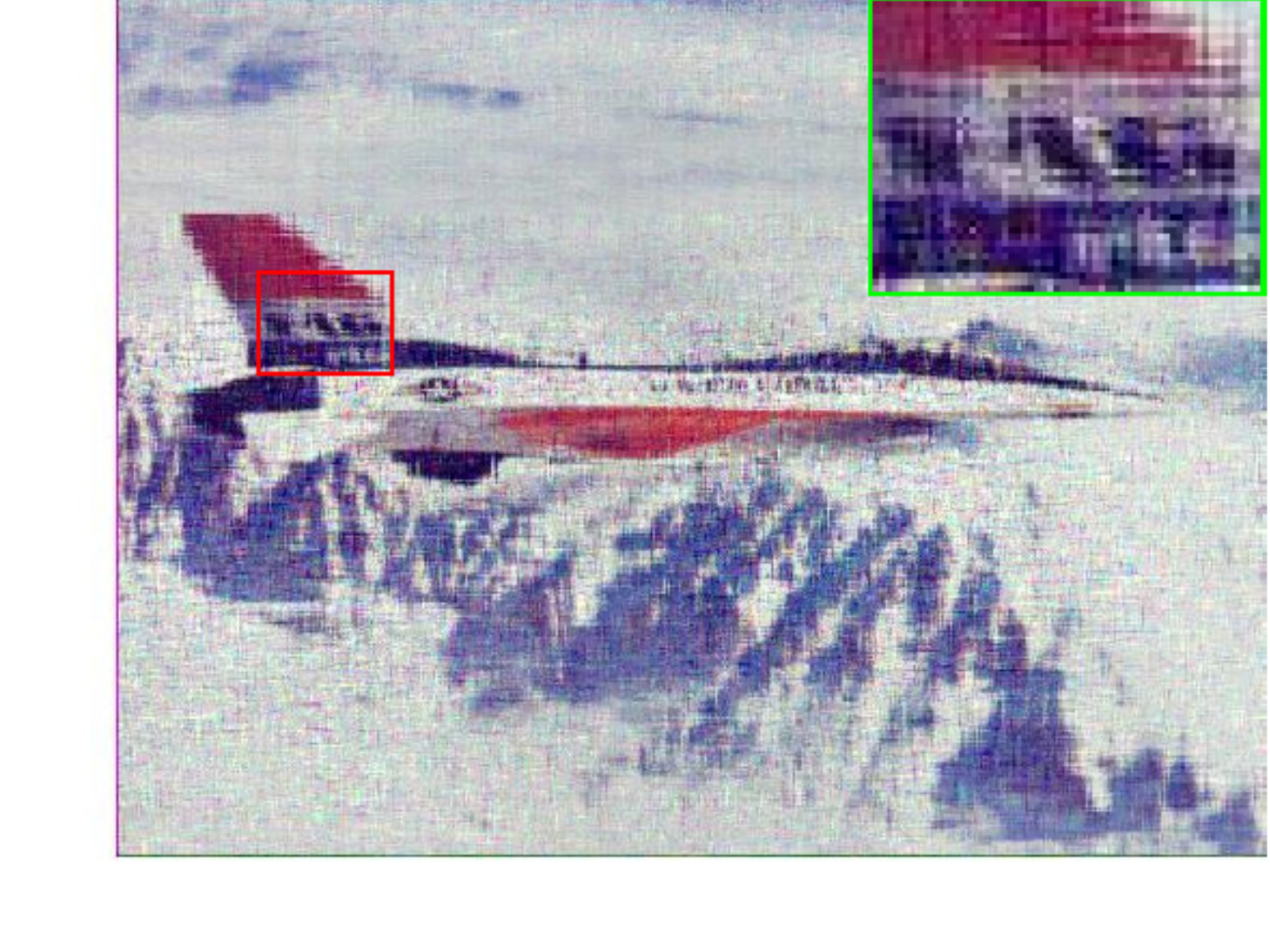}
}\\
	\hspace{-0.18in}
\subfigure[MC-NC \cite{DBLP:journals/tip/NieHL19}]{
	\includegraphics[width=2.3cm,height=2cm]{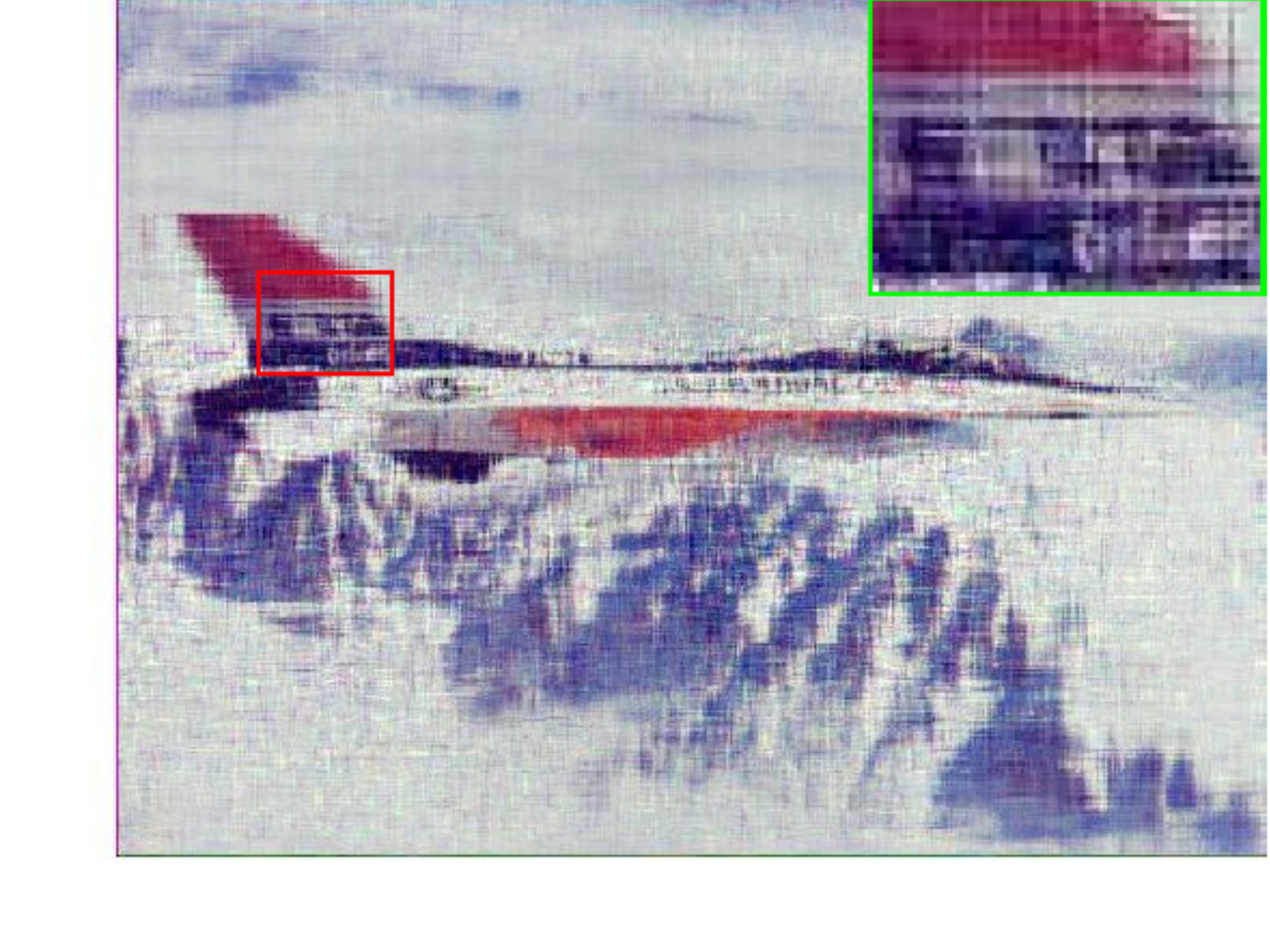}
}
	\hspace{-0.18in}
\subfigure[LRQA-2 \cite{DBLP:journals/tip/ChenXZ20}]{
	\includegraphics[width=2.3cm,height=2cm]{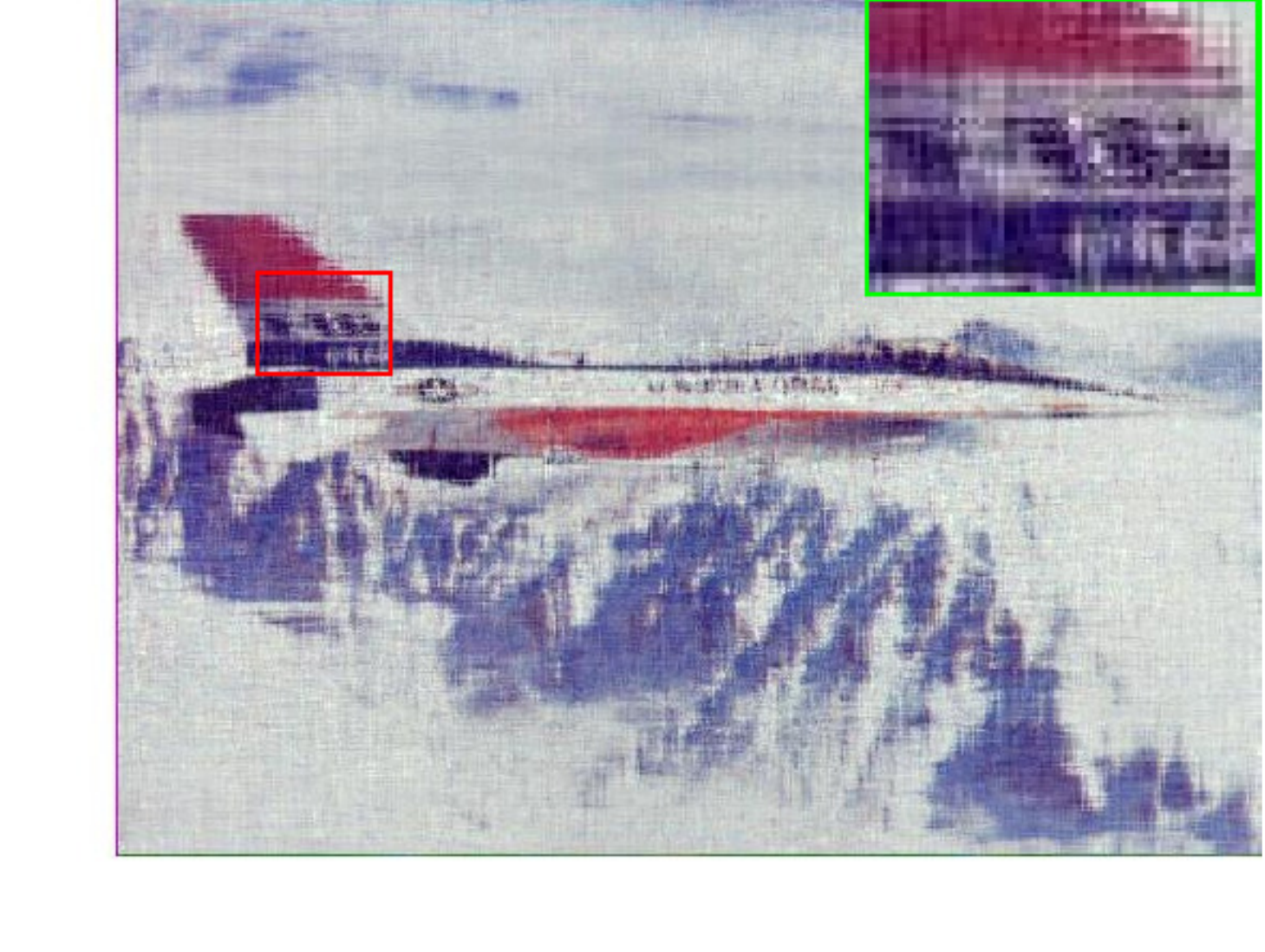}
}
	\hspace{-0.18in}
\subfigure[RegL1-ALM \cite{DBLP:conf/cvpr/ZhengLSYO12}]{
	\includegraphics[width=2.3cm,height=2cm]{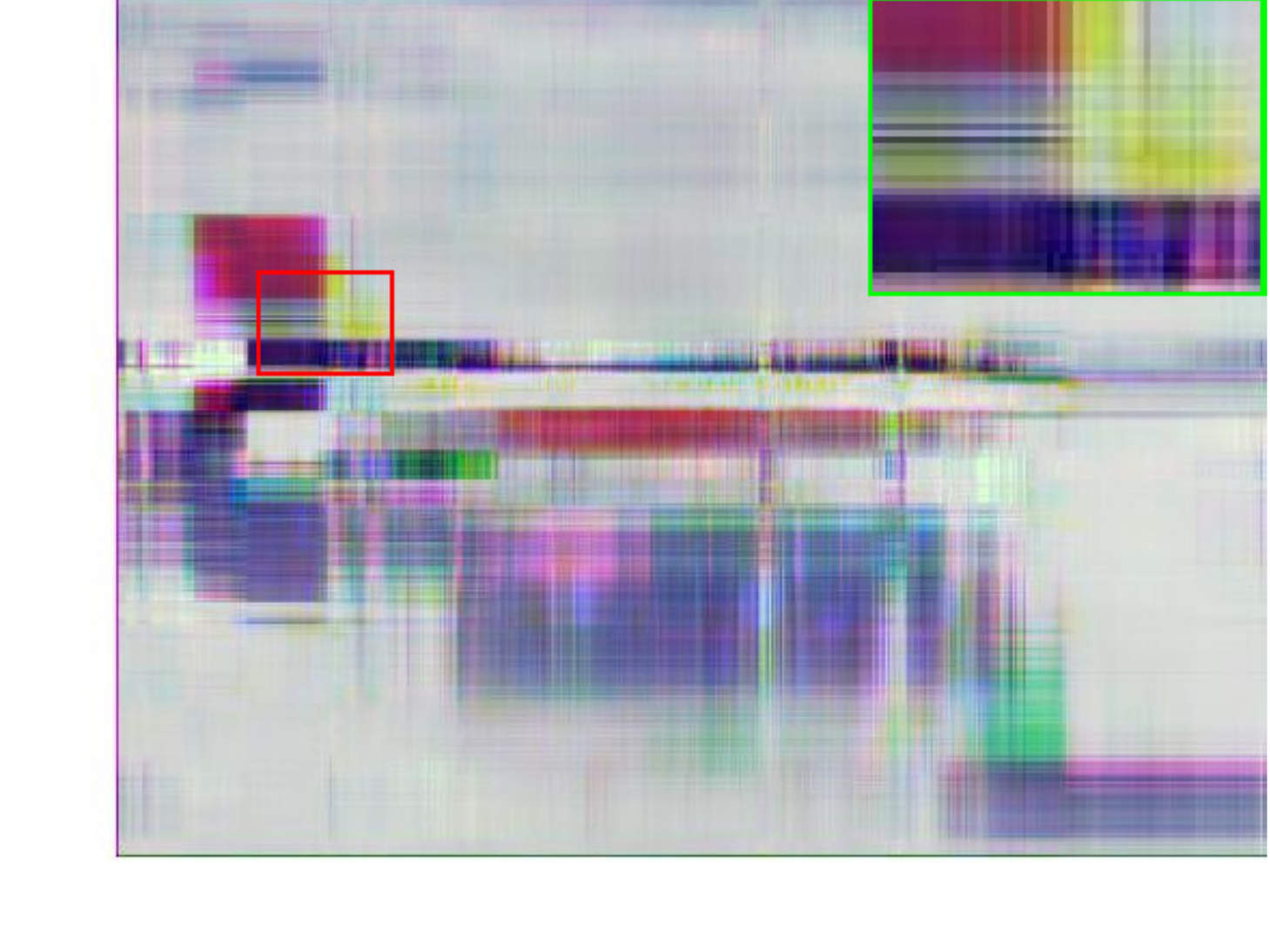}
}\\
	\hspace{-0.18in}
\subfigure[Q-DFN]{
	\includegraphics[width=2.3cm,height=2cm]{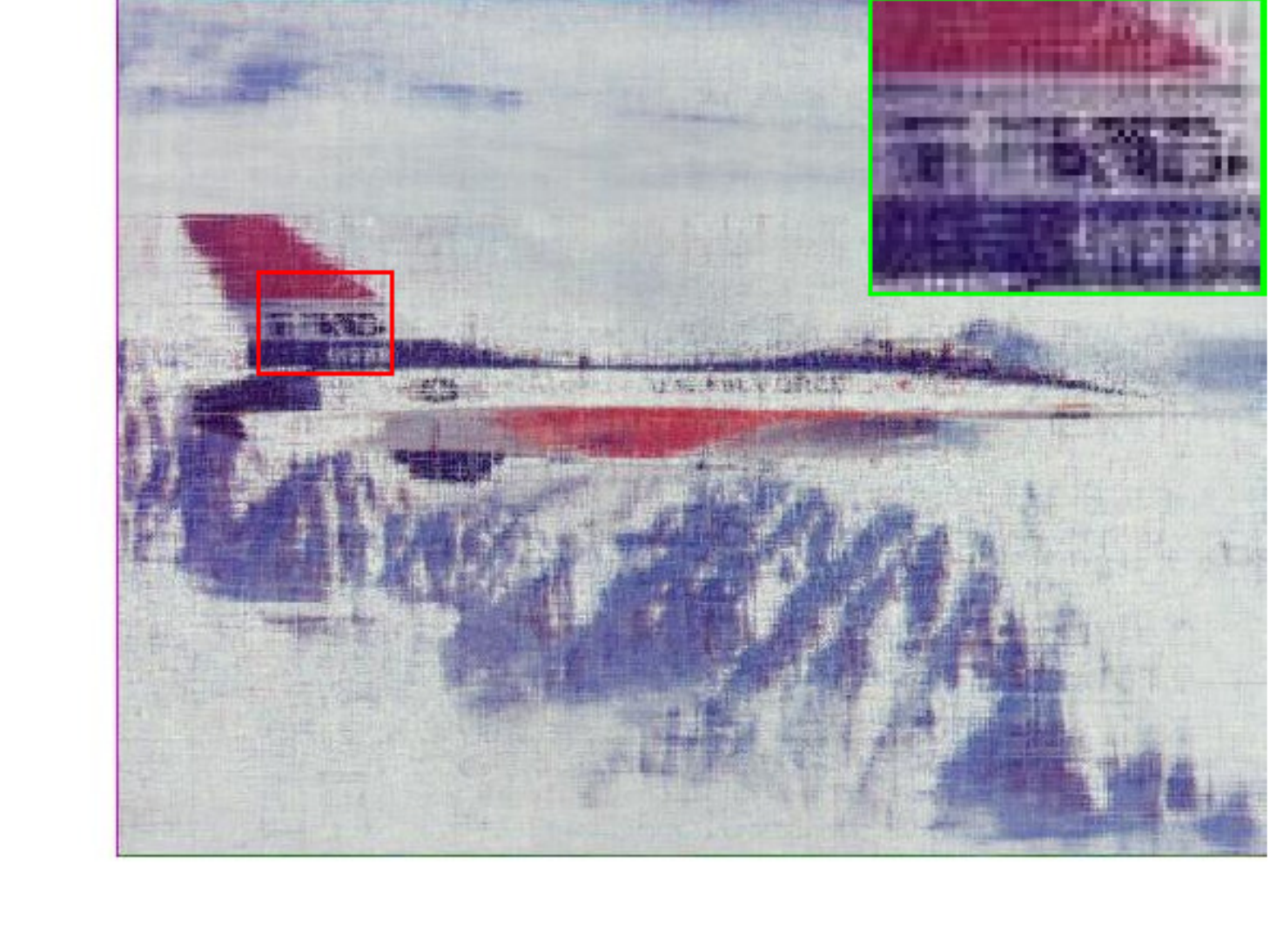}
}
\hspace{-0.18in}
\subfigure[Q-DNN]{
	\includegraphics[width=2.3cm,height=2cm]{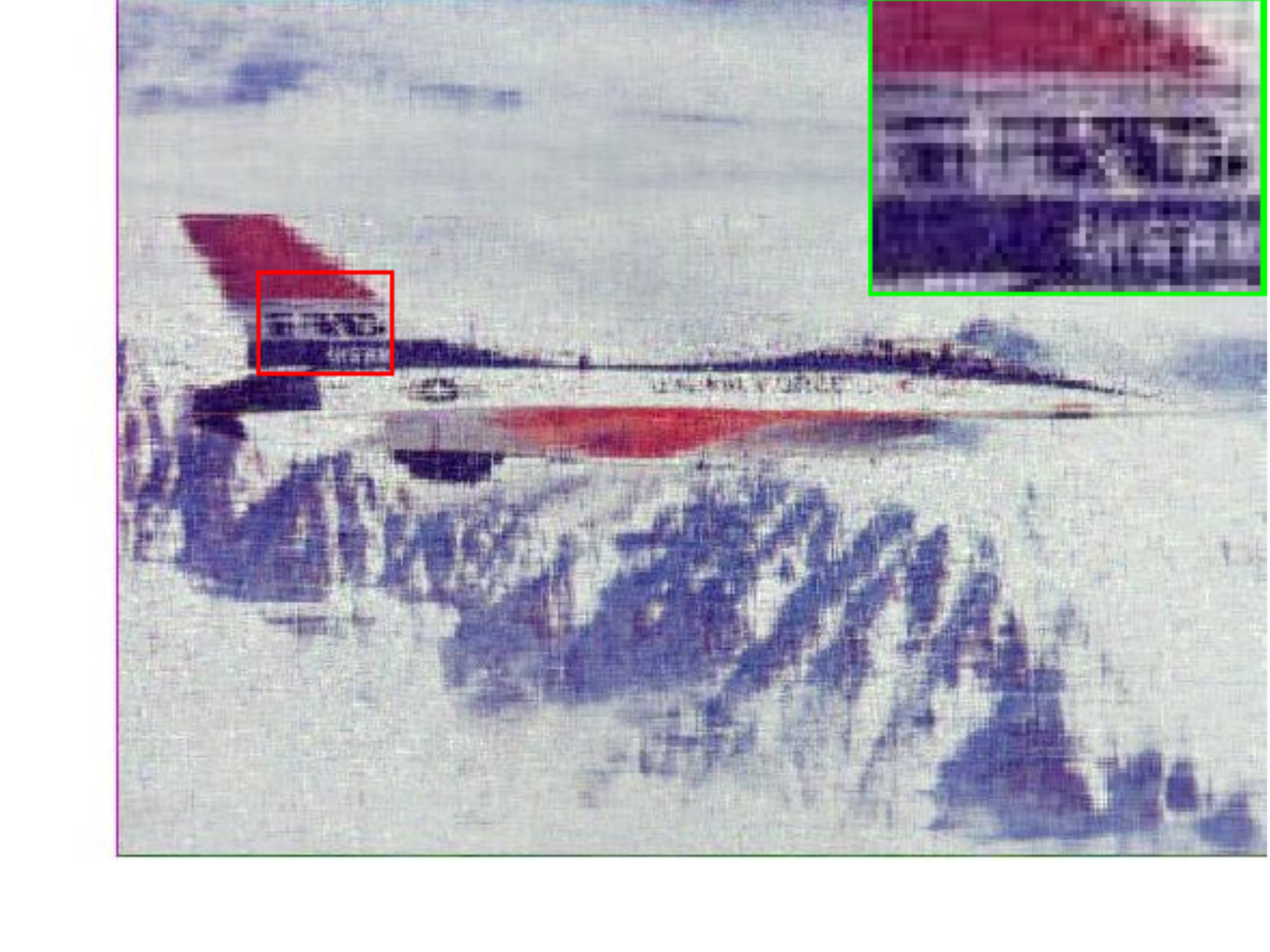}
}
\hspace{-0.18in}
\subfigure[Q-FNN]{
	\includegraphics[width=2.3cm,height=2cm]{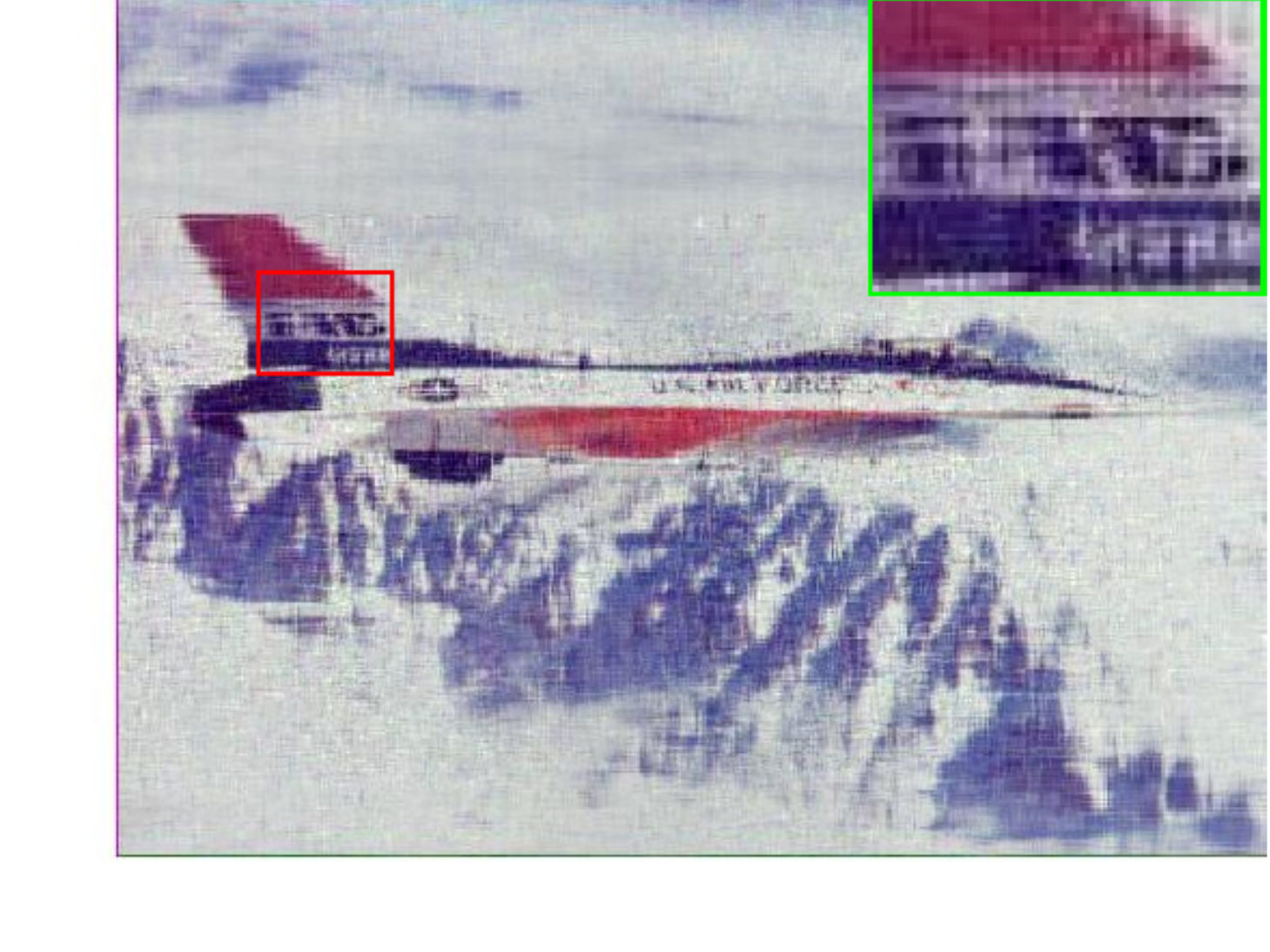}
}
	\caption{Color image inpainting  results on Image(1). (a) is the original image. (b) is the observed image (${\rm{MR}}=0.75$). (c)-(k) are the inpainting results of D-N, F-N, WNNM, MC-NC, LRQA-2, RegL1-ALM,  Q-DFN, Q-DNN  and Q-FNN,  respectively. }
	\label{fig01}
\end{figure}

Fig.\ref{fig2} displays the visual and quantitative comparisons between the proposed three methods and all compared inpainting methods on the eight testing color images with  ${\rm{MR}}=0.70$. TABLE \ref{Index4SR_2}
reports the quantitative PSNR and SSIM values (and the average values of PSNR) of all inpainting methods on the eight testing color images with different  ${\rm{MRs}}$. Fig.\ref{atime} shows the average runtime of all methods for color image inpainting on the eight testing color images with different  ${\rm{MRs}}$. From all the experimental results, we can observe and summarize the following points.
\begin{itemize}
	 \item The proposed three algorithms are indeed convergent (\emph{see} Fig.\ref{fig0}).
	  Although the convergence curves fluctuate at the beginning of the iteration (especially Q-DNN and Q-FNN), the three algorithms converge as the number of iterations increases.
	 \item Compared with LRMC-based methods, LRQMC-based methods (including our proposed three methods and  LRQA-2) , in the overwhelming majority of cases, show better performance both visually (\emph{see} Fig.\ref{fig2} and Fig.\ref{fig01}) and quantitatively (\emph{see} TABLE \ref{Index4SR_2}). It is mainly due to the advantage of quaternion representation of color pixel values. As the latest LRMC-based method, MC-NC shows the comparable performance with the LRQMC-based methods for lower  ${\rm{MR}}$ values, \emph{e.g.}, ${\rm{MR}}=0.50$.
	 \item As LRQMC-based methods, the performance of LRQA-2, and the proposed Q-DFN, Q-DNN and Q-FNN are very close. Nonetheless, the runtime of the proposed Q-DFN, Q-DNN and Q-FNN is much shorter than that of LRQA-2 (\emph{see} Fig.\ref{atime}). This is mainly because our proposed methods only need to handle two much smaller factor quaternion matrices rather than calculating the QSVD of original big size quaternion matrices in each iteration as LRQA-2 must do.  
\end{itemize}

\section{Conclusion}
\label{sec5}
We propose three novel low-rank quaternion matrix completion (LRQMC) methods, named Q-DFN, Q-DNN, and Q-FFN to recover missing data of the color image, which can holistically handle the three RGB channels. 
Based on three quaternion-based bilinear factor (QBF) matrix norm minimization models, the proposed methods only need to process two much smaller factor quaternion matrices, thus it can effectively reduce the time consumption caused by the calculation of quaternion singular value decompositions (QSVD). The alternating direction method of multipliers (ADMM) framework is applied to solve the models, which indeed guarantees the convergence (as we empirically show in the experiments ) of the proposed algorithms. Furthermore, experimental results on real color image inpainting also demonstrate the effectiveness of the developed methods.

One of our future work aims to extend the proposed models to some other low-rank quaternion matrix approximate problems such as color face recognition, color image superresolution, and color image denoising. Recently, some convolutional neural networks (CNN)-based approaches can complete images with very large missing blocks based on numerous training samples \cite{DBLP:conf/bmvc/TsengLL17}. We also would like to extend our ideas to the quaternion-based convolutional neural networks (QCNN)  \cite{DBLP:conf/eccv/ZhuXXC18} framework in the next works.

\appendices
\section{Basic knowledge of quaternion algebras}
\label{a_sec1}
Quaternion space $\mathbb{H}$ was first introduced by W. Hamilton \cite{articleHamilton84} in 1843, which is an extension of the complex space $\mathbb{C}$.  A quaternion $\dot{q}\in\mathbb{H}$ with a real component and three imaginary components is defined as
\begin{equation}\footnotesize
\label{equ2}
\dot{q}=q_{0}+q_{1}i+q_{2}j+q_{3}k,
\end{equation}
where $q_{l}\in\mathbb{R}\: (l=0,1,2,3)$, and $i, j, k$ are
imaginary number units and obey the quaternion rules that
\begin{align}\footnotesize
\left\{
\begin{array}{lc}
i^{2}=j^{2}=k^{2}=ijk=-1,\\
ij=-ji=k, jk=-kj=i, ki=-ik=j.
\end{array}
\right.
\end{align}
$\dot{q}$ can be decomposed into a real part $\mathfrak{R}(\dot{q}):=q_{0}$ and an imaginary part $\mathfrak{I}(\dot{q}):=q_{1}i+q_{2}j+q_{3}k$ such that $\dot{q}=\mathfrak{R}(\dot{q})+\mathfrak{I}(\dot{q})$.
If the real part $\mathfrak{R}(\dot{q})=0$, $\dot{q}$ is named a pure quaternion. Given two quaternions $\dot{p}$ and $\dot{q}\in\mathbb{H}$, the sum and multiplication of them are respectively
\begin{equation*}\footnotesize
\label{eqn1}
\dot{p}+\dot{q}=(p_{0}+q_{0})+(p_{1}+q_{1})i+(p_{2}+q_{2})j+(p_{3}+q_{3})k
\end{equation*}
and
\begin{align*}\footnotesize
\label{eqn2}
\dot{p}\dot{q}=&(p_{0}q_{0}-p_{1}q_{1}-p_{2}q_{2}-p_{3}q_{3})\\
&+(p_{0}q_{1}+p_{1}q_{0}+p_{2}q_{3}-p_{3}q_{2})i\\
&+(p_{0}q_{2}-p_{1}q_{3}+p_{2}q_{0}+p_{3}q_{1})j\\
&+(p_{0}q_{3}+p_{1}q_{2}-p_{2}q_{1}+p_{3}q_{0})k.
\end{align*}
It is noticeable that the multiplication of two quaternions is not
commutative so that in general $\dot{p}\dot{q}\neq\dot{q}\dot{p}$.
The conjugate and the modulus of a quaternion $\dot{q}$ are,
respectively, defined as follows
\begin{align*}\footnotesize
\dot{q}^{\ast}&=q_{0}-q_{1}i-q_{2}j-q_{3}k,\\
|\dot{q}|&=\sqrt{\dot{q}\dot{q}^{\ast}}=\sqrt{q_{0}^{2}+q_{1}^{2}+q_{2}^{2}+q_{3}^{2}}.
\end{align*}

Analogously, a quaternion matrix $\dot{\mathbf{Q}}=(\dot{q}_{mn})\in\mathbb{H}^{M\times N}$ is written
as $\dot{\mathbf{Q}}=\mathbf{Q}_{0}+\mathbf{Q}_{1}i+\mathbf{Q}_{2}j+\mathbf{Q}_{3}k$, where $\mathbf{Q}_{l}\in\mathbb{R}^{M\times N}\: (l=0,1,2,3)$, $\dot{\mathbf{Q}}$ is named a pure quaternion matrix when $\mathfrak{R}(\dot{\mathbf{Q}})=\mathbf{Q}_{0}=\mathbf{0}$. The quaternion matrix Frobenius norm is defined as $\|\dot{\mathbf{Q}}\|_{F}=\sqrt{\sum_{m=1}^{M}\sum_{n=1}^{N}|\ddot{q}_{mn}|^{2}}=\sqrt{{\rm{tr}}\{(\dot{\mathbf{Q}})^{H}\dot{\mathbf{Q}}\}}$. Using Cayley-Dickson notation \cite{DBLP:journals/sigpro/BihanM04}, $\dot{\mathbf{Q}}$ can be expressed as $\dot{\mathbf{Q}}=\mathbf{Q}_{a}+\mathbf{Q}_{b}j$, where $\mathbf{Q}_{a}=\mathbf{Q}_{0}+\mathbf{Q}_{1}i\in\mathbb{C}^{M\times N}$, $\mathbf{Q}_{b}=\mathbf{Q}_{2}+\mathbf{Q}_{3}i\in\mathbb{C}^{M\times N}$. Then the quaternion matrix $\dot{\mathbf{Q}}$ can be denoted as an equivalent complex matrix 
\begin{equation}\footnotesize
\label{qtorm}
\mathbf{Q}_{c}=\left(\begin{array}{cc}
\mathbf{Q}_{a}&  \mathbf{Q}_{b} \\ 
-\mathbf{Q}_{b}^{\ast}&  \mathbf{Q}_{a}^{\ast}  
\end{array} \right)\in\mathbb{C}^{2M\times 2N}.
\end{equation}
Based on (\ref{qtorm}), the QSVD of quaternion matrix $\dot{\mathbf{Q}}$ can be obtained by applying the classical complex SVD algorithm to $\mathbf{Q}_{c}$. The relation between the QSVD of quaternion matrix  $\dot{\mathbf{Q}}\in\mathbb{H}^{M\times N}$ and the SVD of its equivalent complex matrix $\mathbf{Q}_{c}\in\mathbb{C}^{2M\times 2N}$ ($\mathbf{Q}_{c}=\mathbf{U}\check{\mathbf{D}}\mathbf{V}^{H}$) is defined as \cite{DBLP:journals/tip/XuYXZN15}
\begin{align}\footnotesize
\label{qsvd}
\left\{
\begin{array}{lc}
\mathbf{D}={\rm{row}}_{odd}({\rm{col}}_{odd}(\check{\mathbf{D}})),\\
\dot{\mathbf{U}}={\rm{col}}_{odd}(\mathbf{U}_{1})+{\rm{col}}_{odd}(-(\mathbf{U}_{2})^{\ast})j,\\
\dot{\mathbf{V}}={\rm{col}}_{odd}(\mathbf{V}_{1})+{\rm{col}}_{odd}(-(\mathbf{V}_{2})^{\ast})j,
\end{array}
\right.
\end{align}
such that $\dot{\mathbf{Q}}=\dot{\mathbf{U}}\mathbf{D}\dot{\mathbf{V}}^{H}$, where
\begin{align*}\footnotesize
\mathbf{U}=\left(\begin{array}{c}
(\mathbf{U}_{1})_{M\times 2M} \\
(\mathbf{U}_{2})_{M\times 2M}
\end{array} \right),\quad
\mathbf{V}=\left(\begin{array}{c}
(\mathbf{V}_{1})_{N\times 2N} 	\\
(\mathbf{V}_{2})_{N\times 2N}
\end{array} \right),
\end{align*}
and ${\rm{row}}_{odd}(\mathbf{M})$, ${\rm{col}}_{odd}(\mathbf{M})$ respectively extract the odd rows and odd columns of matrix $\mathbf{M}$. 

Readers can find more details on quaternion algebra in \cite{10029950538, Girard2007Quaternions, Altmann1986Rotations}.

\section{The proof of \textbf{Theorem \ref{Th2}}}
\label{a_sec2}
Proof: According to the QSVD in \textbf{Theorem \ref{Th1}}, there exist unitary quaternion matrices $\dot{\mathbf{A}}\in\mathbb{H}^{M\times M}$ and  $\dot{\mathbf{B}}\in\mathbb{H}^{N\times N}$ such that
\begin{equation}\footnotesize
\label{enn1}
\dot{\mathbf{X}}=\dot{\mathbf{A}}\left(\begin{array}{cc}
\mathbf{D}_{d}	&  \mathbf{0}\\
\mathbf{0}	& \mathbf{0}
\end{array}\right)\dot{\mathbf{B}}^{H},
\end{equation}
where $\mathbf{D}_{d}={\rm{diag}}\{d_{1},\ldots,d_{r},d_{r+1},\ldots,d_{d}\}\in\mathbb{R}^{d\times d}$ and the $d_{1},\ldots,d_{r}$ are the positive singular values of $\dot{\mathbf{X}}$, $d_{r+1},\ldots,d_{d}$ are equal to $0$. Then, (\ref{enn1}) can be rewritten as
\begin{equation*}\footnotesize
\dot{\mathbf{X}}=\dot{\mathbf{A}}\left( \begin{array}{c}
\mathbf{D}_{d}^{\frac{1}{2}}\\
\mathbf{0}
\end{array}\right)_{M\times d}\left( \begin{array}{cc}
\mathbf{D}_{d}^{\frac{1}{2}}& \mathbf{0}
\end{array} \right)_{d\times N}\dot{\mathbf{B}}^{H}.
\end{equation*}
Let $\dot{\mathbf{U}}=\dot{\mathbf{A}}\left( \begin{array}{c}
\mathbf{D}_{d}^{\frac{1}{2}}\\
\mathbf{0}
\end{array}\right)\in\mathbb{H}^{M\times d}$ and $\dot{\mathbf{V}}=\dot{\mathbf{B}}\left( \begin{array}{c}
\mathbf{D}_{d}^{\frac{1}{2}}\\
\mathbf{0}
\end{array}\right)\in\mathbb{H}^{N\times d}$, they obviously meet ${\rm{rank}}(\dot{\mathbf{U}})={\rm{rank}}(\dot{\mathbf{V}})=r$, and $\dot{\mathbf{X}}=\dot{\mathbf{U}}\dot{\mathbf{V}}^{H}$.

\section{The proofs of \textbf{Theorem \ref{th0}}, \textbf{Theorem \ref{th1}}  and \textbf{Theorem \ref{th2}}}
\label{a_sec3}
To prove the \textbf{Theorem \ref{th0}}, \textbf{Theorem \ref{th1}} and \textbf{Theorem \ref{th2}}, we first give the following lemma.
\begin{lemma}
	\label{lem1}	
	Let two quaternion matrices $\dot{\mathbf{U}}\in\mathbb{H}^{M\times d}$ and $\dot{\mathbf{V}}\in\mathbb{H}^{N\times d}$ be given and let $p>0$, denote $K={\rm{min}}(M,N,d)$, the following inequality holds for the decreasingly ordered singular values of $\dot{\mathbf{U}}\dot{\mathbf{V}}^{H}$, $\dot{\mathbf{U}}$ and $\dot{\mathbf{V}}$. 
	\begin{equation}\footnotesize
	\sum_{k}^{K}\sigma_{k}^{p}(\dot{\mathbf{U}}\dot{\mathbf{V}}^{H})\leq\sum_{k}^{K}\sigma_{k}^{p}(\dot{\mathbf{U}})\sigma_{k}^{p}(\dot{\mathbf{V}}).
	\end{equation}
\end{lemma}
\textit{Proof:}
\begin{align*}\footnotesize
\sum_{k}^{K}\sigma_{k}^{p}(\dot{\mathbf{U}}\dot{\mathbf{V}}^{H})&=\frac{1}{2}\sum_{k}^{K}\sigma_{k}^{p}(\mathcal{P}(\dot{\mathbf{U}}\dot{\mathbf{V}}^{H}))\\
&=\frac{1}{2}\sum_{k}^{K}\sigma_{k}^{p}(\mathcal{P}(\dot{\mathbf{U}})\mathcal{P}(\dot{\mathbf{V}}^{H}))\\
&\leq \frac{1}{2}\sum_{k}^{K}\sigma_{k}^{p}(\mathcal{P}(\dot{\mathbf{U}})) \sigma_{k}^{p}(\mathcal{P}(\dot{\mathbf{V}}^{H}))\\
&=\sum_{k}^{K}\sigma_{k}^{p}(\dot{\mathbf{U}})\sigma_{k}^{p}(\dot{\mathbf{V}}),
\end{align*}
where the inequality follows from \cite{DBLP:books/daglib/0019186} (Theorem 3.3.14).

Proof of \textbf{Theorem \ref{th0}}:
Since $\dot{\mathbf{X}}=\dot{\mathbf{U}}\dot{\mathbf{V}}^{H}$, where
$\dot{\mathbf{U}}\in\mathbb{H}^{M\times d}$ and $\dot{\mathbf{V}}\in\mathbb{H}^{N\times d}$, denote $K={\rm{min}}(M,N,d)$, we have 
\begin{align*}\footnotesize
\|\dot{\mathbf{X}}\|_{\ast}&=\sum_{k}^{K}\sigma_{k}(\dot{\mathbf{X}})
=	\sum_{k}^{K}\sigma_{k}(\dot{\mathbf{U}}\dot{\mathbf{V}}^{H})
\leq \sum_{k}^{K}\sigma_{k}(\dot{\mathbf{U}})\sigma_{k}(\dot{\mathbf{V}})\\
&\leq \left(\sum_{k}^{K}\sigma_{k}^{2}(\dot{\mathbf{U}})\right)^{1/2} \left(\sum_{k}^{K}\sigma_{k}^{2}(\dot{\mathbf{V}})\right)^{1/2} \\
&\leq \frac{1}{2}\left(\sum_{k}^{K}\sigma_{k}^{2}(\dot{\mathbf{U}})\right)+\frac{1}{2}\left(\sum_{k}^{K}\sigma_{k}^{2}(\dot{\mathbf{V}})\right)\\
&\leq \frac{1}{2}\left(\sum_{k}^{{\rm{min}}(M,d)}\!\!\!\!\sigma_{k}^{2}(\dot{\mathbf{U}})\right)+\frac{1}{2}\left(\sum_{k}^{{\rm{min}}(N,d)}\!\!\!\!\sigma_{k}^{2}(\dot{\mathbf{V}})\right)\\
&=\frac{1}{2}\left(\|\dot{\mathbf{U}}\|_{F}^{2}+\|\dot{\mathbf{V}}\|_{F}^{2}\right),
\end{align*}
where the first inequality  follows from \textbf{Lemma \ref{lem1}} as $p=1$, the second inequality follows from the well-known Holder's inequality \cite{DBLP:journals/appml/Yang03a}, the third inequality holds due to the Jensen's inequality \cite{NeedhamA}, and since we always have $K={\rm{min}}(M,N,d)\leq{\rm{min}}(M,d), K={\rm{min}}(M,N,d)\leq{\rm{min}}(N,d)$, thus the last inequality holds. On the other hand, let $\dot{\mathbf{U}}_{\star}=\dot{\mathbf{A}}_{\dot{\mathbf{X}}}\mathbf{D}_{\dot{\mathbf{X}}}^{\frac{1}{2}}$ and $\dot{\mathbf{V}}_{\star}=\dot{\mathbf{B}}_{\dot{\mathbf{X}}}\mathbf{D}_{\dot{\mathbf{X}}}^{\frac{1}{2}}$, where $\dot{\mathbf{X}}=\dot{\mathbf{A}}_{\dot{\mathbf{X}}}\mathbf{D}_{\dot{\mathbf{X}}}\dot{\mathbf{B}}_{\dot{\mathbf{X}}}^{H}$ is the QSVD of $\dot{\mathbf{X}}$. Then, we have $\dot{\mathbf{X}}=\dot{\mathbf{U}}_{\star}\dot{\mathbf{V}}_{\star}^{H}$ and $\|\dot{\mathbf{X}}\|_{\ast}=\frac{1}{2}\left(\|\dot{\mathbf{U}}_{\star}\|_{F}^{2}+\|\dot{\mathbf{V}}_{\star}\|_{F}^{2}\right)$.

Hence, from above, we have
 \begin{equation*}\footnotesize
  \mathop{{\rm{min}}}\limits_{\dot{\mathbf{U}},\dot{\mathbf{V}}\atop \dot{\mathbf{X}}=\dot{\mathbf{U}}\dot{\mathbf{V}}^{H}} \frac{1}{2}\|\dot{\mathbf{U}}\|_{F}^{2}+\frac{1}{2}\|\dot{\mathbf{V}}\|_{F}^{2}=\|\dot{\mathbf{X}}\|_{\ast}.
 \end{equation*}

Proof of \textbf{Theorem \ref{th1}}:
Since $\dot{\mathbf{X}}=\dot{\mathbf{U}}\dot{\mathbf{V}}^{H}$, where
$\dot{\mathbf{U}}\in\mathbb{H}^{M\times d}$ and $\dot{\mathbf{V}}\in\mathbb{H}^{N\times d}$, denote $K={\rm{min}}(M,N,d)$, we have 
\begin{align*}\footnotesize
\|\dot{\mathbf{X}}\|_{Q-S_{1/2}}^{1/2}&=\sum_{k}^{K}\sigma_{k}^{1/2}(\dot{\mathbf{X}})
=	\sum_{k}^{K}\sigma_{k}^{1/2}(\dot{\mathbf{U}}\dot{\mathbf{V}}^{H})\\
&\leq \sum_{k}^{K}\sigma_{k}^{1/2}(\dot{\mathbf{U}})\sigma_{k}^{1/2}(\dot{\mathbf{V}})\\
&\leq \left(\sum_{k}^{K}\sigma_{k}(\dot{\mathbf{U}})\right)^{1/2} \left(\sum_{k}^{K}\sigma_{k}(\dot{\mathbf{V}})\right)^{1/2} \\
&\leq \frac{1}{2}\left(\sum_{k}^{K}\sigma_{k}(\dot{\mathbf{U}})\right)+\frac{1}{2}\left(\sum_{k}^{K}\sigma_{k}(\dot{\mathbf{V}})\right)\\
&\leq \frac{1}{2}\left(\sum_{k}^{{\rm{min}}(M,d)}\!\!\!\!\sigma_{k}(\dot{\mathbf{U}})\!\right)+\frac{1}{2}\left(\sum_{k}^{{\rm{min}}(N,d)}\!\!\!\!\sigma_{k}(\dot{\mathbf{V}})\!\right)\\
&=\frac{1}{2}\left(\|\dot{\mathbf{U}}\|_{\ast}+\|\dot{\mathbf{V}}\|_{\ast}\right),
\end{align*}
where the first inequality follows from \textbf{Lemma \ref{lem1}} as $p=\frac{1}{2}$, the second inequality follows from the well-known Holder's inequality \cite{DBLP:journals/appml/Yang03a}, the third inequality holds due to the Jensen's inequality \cite{NeedhamA}, and since we always have $K={\rm{min}}(M,N,d)\leq{\rm{min}}(M,d), K={\rm{min}}(M,N,d)\leq{\rm{min}}(N,d)$, thus the last inequality holds. On the other hand, let $\dot{\mathbf{U}}_{\star}=\dot{\mathbf{A}}_{\dot{\mathbf{X}}}\mathbf{D}_{\dot{\mathbf{X}}}^{\frac{1}{2}}$ and $\dot{\mathbf{V}}_{\star}=\dot{\mathbf{B}}_{\dot{\mathbf{X}}}\mathbf{D}_{\dot{\mathbf{X}}}^{\frac{1}{2}}$, where $\dot{\mathbf{X}}=\dot{\mathbf{A}}_{\dot{\mathbf{X}}}\mathbf{D}_{\dot{\mathbf{X}}}\dot{\mathbf{B}}_{\dot{\mathbf{X}}}^{H}$ is the QSVD of $\dot{\mathbf{X}}$. Then, we have $\dot{\mathbf{X}}=\dot{\mathbf{U}}_{\star}\dot{\mathbf{V}}_{\star}^{H}$ and $\|\dot{\mathbf{X}}\|_{Q-S_{1/2}}=\frac{1}{4}\left(\|\dot{\mathbf{U}}_{\star}\|_{\ast}+\|\dot{\mathbf{V}}_{\star}\|_{\ast}\right)^{2}$.

Hence, from above, we have
\begin{equation*}\footnotesize
\mathop{{\rm{min}}}\limits_{\dot{\mathbf{U}},\dot{\mathbf{V}}\atop \dot{\mathbf{X}}=\dot{\mathbf{U}}\dot{\mathbf{V}}^{H}} \frac{1}{4}\left(\|\dot{\mathbf{U}}\|_{\ast}+\|\dot{\mathbf{V}}\|_{\ast}\right)^{2}=\|\dot{\mathbf{X}}\|_{Q-S_{1/2}}.
\end{equation*}

Proof of \textbf{Theorem \ref{th2}}: 
Since $\dot{\mathbf{X}}=\dot{\mathbf{U}}\dot{\mathbf{V}}^{H}$, where
$\dot{\mathbf{U}}\in\mathbb{H}^{M\times d}$ and $\dot{\mathbf{V}}\in\mathbb{H}^{N\times d}$, denote $K={\rm{min}}(M,N,d)$, we have 
\begin{align*}\footnotesize
\|\dot{\mathbf{X}}\|_{Q-S_{2/3}}^{2/3}&=\sum_{k}^{K}\sigma_{k}^{2/3}(\dot{\mathbf{X}})
=	\sum_{k}^{K}\sigma_{k}^{2/3}(\dot{\mathbf{U}}\dot{\mathbf{V}}^{H})\\
&\leq \sum_{k}^{K}\sigma_{k}^{2/3}(\dot{\mathbf{U}})\sigma_{k}^{2/3}(\dot{\mathbf{V}})\\
&\leq \left(\sum_{k}^{K}\sigma_{k}^{\frac{2}{3}*3}(\dot{\mathbf{U}})\right)^{1/3} \left(\sum_{k}^{K}\sigma_{k}^{\frac{2}{3}*\frac{3}{2}}(\dot{\mathbf{V}})\right)^{2/3} \\
&\leq \frac{1}{3}\left(\sum_{k}^{K}\sigma_{k}^{2}(\dot{\mathbf{U}})\right)+\frac{2}{3}\left(\sum_{k}^{K}\sigma_{k}(\dot{\mathbf{V}})\right)\\
&\leq \frac{1}{3}\left(\sum_{k}^{{\rm{min}}(M,d)}\!\!\!\!\sigma_{k}^{2}(\dot{\mathbf{U}})\right)+\frac{2}{3}\left(\sum_{k}^{{\rm{min}}(N,d)}\!\!\!\!\sigma_{k}(\dot{\mathbf{V}})\right)\\
&=\frac{1}{3}\left(\|\dot{\mathbf{U}}\|_{F}^{2}+2\|\dot{\mathbf{V}}\|_{\ast}\right),
\end{align*}
where the  first  inequality  follows  from \textbf{Lemma \ref{lem1}} as $p=\frac{2}{3}$, the second inequality follows from the well-known Holder's inequality \cite{DBLP:journals/appml/Yang03a}, the third inequality holds due to the Jensen's inequality \cite{NeedhamA}, and since we always have $K={\rm{min}}(M,N,d)\leq{\rm{min}}(M,d), K={\rm{min}}(M,N,d)\leq{\rm{min}}(N,d)$, thus the last inequality holds. On the other hand, let $\dot{\mathbf{U}}_{\star}=\dot{\mathbf{A}}_{\dot{\mathbf{X}}}\mathbf{D}_{\dot{\mathbf{X}}}^{\frac{1}{3}}$ and $\dot{\mathbf{V}}_{\star}=\dot{\mathbf{B}}_{\dot{\mathbf{X}}}\mathbf{D}_{\dot{\mathbf{X}}}^{\frac{2}{3}}$, where $\dot{\mathbf{X}}=\dot{\mathbf{A}}_{\dot{\mathbf{X}}}\mathbf{D}_{\dot{\mathbf{X}}}\dot{\mathbf{B}}_{\dot{\mathbf{X}}}^{H}$ is the QSVD of $\dot{\mathbf{X}}$. Then, we have $\dot{\mathbf{X}}=\dot{\mathbf{U}}_{\star}\dot{\mathbf{V}}_{\star}^{H}$ and $\|\dot{\mathbf{X}}\|_{Q-S_{2/3}}=\left((\|\dot{\mathbf{U}}_{\star}\|_{F}^{2}+2\|\dot{\mathbf{V}}_{\star}\|_{\ast})/3\right)^{3/2}$.

Hence, from above, we have
\begin{equation*}\footnotesize
\mathop{{\rm{min}}}\limits_{\dot{\mathbf{U}},\dot{\mathbf{V}}\atop \dot{\mathbf{X}}=\dot{\mathbf{U}}\dot{\mathbf{V}}^{H}} \left(\frac{\|\dot{\mathbf{U}}\|_{F}^{2}+2\|\dot{\mathbf{V}}\|_{\ast}}{3}\right)^{3/2}=\|\dot{\mathbf{X}}\|_{Q-S_{2/3}}.
\end{equation*}

\section{Solving the problem (\ref{e14})}
\label{a_sec4}
Similar to the problem (\ref{e13}), the problem (\ref{e14}) is solved by minimizing the following augmented Lagrangian function
\begin{equation}\footnotesize
\label{e28}
\begin{split}
&\mathcal{L}_{\mu}(\dot{\mathbf{U}},\dot{\mathbf{V}},\dot{\mathbf{A}}_{V},\dot{\mathbf{X}},\dot{\mathbf{F}}_{1},\dot{\mathbf{F}}_{2})\\
&=\frac{\lambda}{3}\big(\|\dot{\mathbf{U}}\|_{F}^{2}+2\|\dot{\mathbf{A}}_{V}\|_{\ast}\big)+\langle\dot{\mathbf{F}}_{1},\dot{\mathbf{V}}-\dot{\mathbf{A}}_{V}\rangle\\
&\quad +\langle\dot{\mathbf{F}}_{2},\dot{\mathbf{X}}-\dot{\mathbf{U}}\dot{\mathbf{V}}^{H}\rangle+
\frac{\mu}{2}\left(\|\dot{\mathbf{V}}-\dot{\mathbf{A}}_{V}\|_{F}^{2}\right.\\
&\quad\left.+\|\dot{\mathbf{X}}-\dot{\mathbf{U}}\dot{\mathbf{V}}^{H}\|_{F}^{2}\right)+\frac{1}{2}\|\mathcal{P}_{\Omega}(\dot{\mathbf{X}}-\dot{\mathbf{T}})\|_{F}^{2},
\end{split}
\end{equation}
where $\mu>0$ is the penalty parameter, $\dot{\mathbf{F}}_{1}$ and $\dot{\mathbf{F}}_{2}$ are Lagrange multipliers.

\textbf{Updating $\dot{\mathbf{U}}$ and $\dot{\mathbf{V}}$}:
\begin{equation}\footnotesize
\label{e29}
\begin{split}
\dot{\mathbf{U}}^{\tau+1}=&\mathop{{\rm{arg\, min}}}\limits_{\dot{\mathbf{U}}}\ 
\frac{1}{2}\|\dot{\mathbf{X}}^{\tau}-\dot{\mathbf{U}}(\dot{\mathbf{V}}^{\tau})^{H}+\dot{\mathbf{F}}_{2}^{\tau}/\mu^{\tau}\|_{F}^{2}\\
&+\frac{\lambda}{3\mu^{\tau}}\|\dot{\mathbf{U}}\|_{F}^{2}
\end{split}
\end{equation}

\begin{equation}\footnotesize
\label{e30}
\begin{split}
\dot{\mathbf{V}}^{\tau+1}=&\mathop{{\rm{arg\, min}}}\limits_{\dot{\mathbf{V}}}\ \|\dot{\mathbf{V}}-\dot{\mathbf{A}}_{V}^{\tau}+\dot{\mathbf{F}}_{1}^{\tau}/\mu^{\tau}\|_{F}^{2}\\
&+\|\dot{\mathbf{X}}^{\tau}-\dot{\mathbf{U}}^{\tau+1}\dot{\mathbf{V}}^{H}+\dot{\mathbf{F}}_{2}^{\tau}/\mu^{\tau}\|_{F}^{2}.
\end{split}
\end{equation}
By the similar way as (\ref{e18}), we can obtain the optimal solution of $\dot{\mathbf{U}}^{\tau+1}$ and $\dot{\mathbf{V}}^{\tau+1}$ as
\begin{equation}\footnotesize
\label{e31}
\dot{\mathbf{U}}^{\tau+1}=\left((\mu^{\tau}\dot{\mathbf{X}}^{\tau}
+\dot{\mathbf{F}}_{2}^{\tau})\dot{\mathbf{V}}^{\tau}\right)\left(\frac{2\lambda}{3}\mathbf{I}+\mu^{\tau}(\dot{\mathbf{V}}^{\tau})^{H}\dot{\mathbf{V}}^{\tau}\right)^{-1}.
\end{equation}
\begin{equation}\footnotesize
\label{e32}
\begin{split}
\dot{\mathbf{V}}^{\tau+1}=&\left(\dot{\mathbf{A}}_{V}^{\tau}-\dot{\mathbf{F}}_{1}^{\tau}/\mu^{\tau}+(\dot{\mathbf{X}}^{\tau}\right.\\
&\left.+\dot{\mathbf{F}}_{2}^{\tau}/\mu^{\tau})^{H}\dot{\mathbf{U}}^{\tau+1}\right)\left(\mathbf{I}+(\dot{\mathbf{U}}^{\tau+1})^{H}\dot{\mathbf{U}}^{\tau+1}\right)^{-1}.\\
\end{split}
\end{equation}

\textbf{Updating $\dot{\mathbf{A}}_{V}$}:
\begin{equation}\footnotesize
\label{e33}
\begin{split}
\dot{\mathbf{A}}_{V}^{\tau+1}=&\mathop{{\rm{arg\, min}}}\limits_{\dot{\mathbf{A}}_{V}}\ \frac{1}{2}\|\dot{\mathbf{A}}_{V}-(\dot{\mathbf{V}}^{\tau+1}+\dot{\mathbf{F}}_{1}^{\tau}/\mu^{\tau})\|_{F}^{2}\\
&+\frac{2\lambda}{3\mu^{\tau}}\|\dot{\mathbf{A}}_{V}\|_{\ast}.
\end{split}
\end{equation}
The closed-form solution of (\ref{e33}) can be obtained by the QSVT  \emph{i.e.},
\begin{equation}\footnotesize
\label{e34}
\dot{\mathbf{A}}_{V}^{\tau+1}=\mathcal{D}_{\frac{2\lambda}{3\mu^{\tau}}}\left(\dot{\mathbf{V}}^{\tau+1}+\dot{\mathbf{F}}_{1}^{\tau}/\mu^{\tau}\right).
\end{equation}

\textbf{Updating $\dot{\mathbf{X}}$}:
\begin{equation}\footnotesize
\begin{split}
\dot{\mathbf{X}}^{\tau+1}=&\mathop{{\rm{arg\, min}}}\limits_{\dot{\mathbf{X}}}\ \frac{1}{2}\|\mathcal{P}_{\Omega}(\dot{\mathbf{X}}-\dot{\mathbf{T}})\|_{F}^{2}\\
&+\frac{\mu^{\tau}}{2}\|\dot{\mathbf{X}}-\big(\dot{\mathbf{U}}^{\tau+1}(\dot{\mathbf{V}}^{\tau+1})^{H}-\frac{\dot{\mathbf{F}}_{2}^{\tau}}{\mu^{\tau}}\big)\|_{F}^{2}.
\end{split}
\end{equation}
Then, we can directly obtain the optimal $\dot{\mathbf{X}}^{\tau+1}$ as
\begin{equation}\footnotesize
\label{e36}
\begin{split}
\dot{\mathbf{X}}^{\tau+1}=&\mathcal{P}_{\Omega^{c}}\big(\dot{\mathbf{U}}^{\tau+1}(\dot{\mathbf{V}}^{\tau+1})^{H}-\dot{\mathbf{F}}_{2}^{\tau}/\mu^{\tau}\big)\\
&+\mathcal{P}_{\Omega}\bigg(\frac{\mu^{\tau}\dot{\mathbf{U}}^{\tau+1}(\dot{\mathbf{V}}^{\tau+1})^{H}-\dot{\mathbf{F}}_{2}^{\tau}+\dot{\mathbf{T}}}{1+\mu^{\tau}}\bigg),
\end{split}
\end{equation}
where $\Omega^{c}$ is the complement of $\Omega$, and we have used the
fact that $\mathcal{P}_{\Omega^{c}}(\dot{\mathbf{T}})=\mathbf{0}$ in (\ref{e36}).

\textbf{Updating $\dot{\mathbf{F}}_{1}, \dot{\mathbf{F}}_{2}$ and $\mu$ }: 
\begin{equation}\footnotesize
\label{e35}
\begin{split}
\dot{\mathbf{F}}_{1}^{\tau+1}&=\dot{\mathbf{F}}_{1}^{\tau}+\mu^{\tau}(\dot{\mathbf{V}}^{\tau+1}-\dot{\mathbf{A}}_{V}^{\tau+1}),\\
\dot{\mathbf{F}}_{2}^{\tau+1}&=\dot{\mathbf{F}}_{2}^{\tau}+\mu^{\tau}(\dot{\mathbf{X}}^{\tau+1}-\dot{\mathbf{U}}^{\tau+1}(\dot{\mathbf{V}}^{\tau+1})^{H}),\\
\mu^{\tau+1}&={\rm{min}}(\beta\mu^{\tau}, \mu_{max}).
\end{split}
\end{equation}

\section*{Acknowledgment}
This work was supported by The Science and Technology Development Fund, Macau SAR (File no. FDCT/085/2018/A2) and University of Macau (File no. MYRG2019-00039-FST).

\ifCLASSOPTIONcaptionsoff
  \newpage
\fi



%

\bibliographystyle{IEEEtran}
\bibliography{Myreference}

\begin{thebibliography}{10}
\providecommand{\url}[1]{#1}
\csname url@samestyle\endcsname
\providecommand{\newblock}{\relax}
\providecommand{\bibinfo}[2]{#2}
\providecommand{\BIBentrySTDinterwordspacing}{\spaceskip=0pt\relax}
\providecommand{\BIBentryALTinterwordstretchfactor}{4}
\providecommand{\BIBentryALTinterwordspacing}{\spaceskip=\fontdimen2\font plus
\BIBentryALTinterwordstretchfactor\fontdimen3\font minus
  \fontdimen4\font\relax}
\providecommand{\BIBforeignlanguage}[2]{{%
\expandafter\ifx\csname l@#1\endcsname\relax
\typeout{** WARNING: IEEEtran.bst: No hyphenation pattern has been}%
\typeout{** loaded for the language `#1'. Using the pattern for}%
\typeout{** the default language instead.}%
\else
\language=\csname l@#1\endcsname
\fi
#2}}
\providecommand{\BIBdecl}{\relax}
\BIBdecl

\bibitem{DBLP:journals/csur/ZhouYZY14}
X.~Zhou, C.~Yang, H.~Zhao, and W.~Yu, ``Low-rank modeling and its applications
  in image analysis,'' \emph{{ACM} Comput. Surv.}, vol.~47, no.~2, pp.
  36:1--36:33, 2014.

\bibitem{DBLP:journals/jacm/CandesLMW11}
E.~J. Cand{\`{e}}s, X.~Li, Y.~Ma, and J.~Wright, ``Robust principal component
  analysis?'' \emph{J. {ACM}}, vol.~58, no.~3, pp. 11:1--11:37, 2011.

\bibitem{DBLP:journals/siamjo/CaiCS10}
J.~Cai, E.~J. Cand{\`{e}}s, and Z.~Shen, ``A singular value thresholding
  algorithm for matrix completion,'' \emph{{SIAM} Journal on Optimization},
  vol.~20, no.~4, pp. 1956--1982, 2010.

\bibitem{DBLP:journals/focm/CandesR09}
E.~J. Cand{\`{e}}s and B.~Recht, ``Exact matrix completion via convex
  optimization,'' \emph{Foundations of Computational Mathematics}, vol.~9,
  no.~6, pp. 717--772, 2009.

\bibitem{DBLP:conf/cvpr/GuZZF14}
S.~Gu, L.~Zhang, W.~Zuo, and X.~Feng, ``Weighted nuclear norm minimization with
  application to image denoising,'' in \emph{2014 {IEEE} Conference on Computer
  Vision and Pattern Recognition, {CVPR} 2014, Columbus, OH, USA, June 23-28,
  2014}, 2014, pp. 2862--2869.

\bibitem{DBLP:journals/ijcv/GuXMZFZ17}
S.~Gu, Q.~Xie, D.~Meng, W.~Zuo, X.~Feng, and L.~Zhang, ``Weighted nuclear norm
  minimization and its applications to low level vision,'' \emph{Int. J.
  Comput. Vis.}, vol. 121, no.~2, pp. 183--208, 2017.

\bibitem{DBLP:conf/aaai/NieHD12}
F.~Nie, H.~Huang, and C.~H.~Q. Ding, ``Low-rank matrix recovery via efficient
  schatten p-norm minimization,'' in \emph{Proceedings of the Twenty-Sixth
  {AAAI} Conference on Artificial Intelligence, July 22-26, 2012, Toronto,
  Ontario, Canada}, 2012.

\bibitem{DBLP:journals/jcam/LiuHC14}
L.~Liu, W.~Huang, and D.~Chen, ``Exact minimum rank approximation via schatten
  p-norm minimization,'' \emph{J. Comput. Appl. Math.}, vol. 267, pp. 218--227,
  2014.

\bibitem{DBLP:journals/tip/XieGLZZZ16}
Y.~Xie, S.~Gu, Y.~Liu, W.~Zuo, W.~Zhang, and L.~Zhang, ``Weighted schatten
  p-norm minimization for image denoising and background subtraction,''
  \emph{{IEEE} Trans. Image Processing}, vol.~25, no.~10, pp. 4842--4857, 2016.

\bibitem{DBLP:journals/mpc/WenYZ12}
Z.~Wen, W.~Yin, and Y.~Zhang, ``Solving a low-rank factorization model for
  matrix completion by a nonlinear successive over-relaxation algorithm,''
  \emph{Math. Program. Comput.}, vol.~4, no.~4, pp. 333--361, 2012.

\bibitem{DBLP:conf/cvpr/KimLO15}
E.~Kim, M.~Lee, and S.~Oh, ``Elastic-net regularization of singular values for
  robust subspace learning,'' in \emph{{IEEE} Conference on Computer Vision and
  Pattern Recognition, {CVPR} 2015, Boston, MA, USA, June 7-12, 2015}, 2015,
  pp. 915--923.

\bibitem{DBLP:conf/cvpr/ZhengLSYO12}
Y.~Zheng, G.~Liu, S.~Sugimoto, S.~Yan, and M.~Okutomi, ``Practical low-rank
  matrix approximation under robust l1-norm,'' in \emph{2012 {IEEE} Conference
  on Computer Vision and Pattern Recognition, Providence, RI, USA, June 16-21,
  2012}, 2012, pp. 1410--1417.

\bibitem{DBLP:journals/pami/ShangCLLL18}
F.~Shang, J.~Cheng, Y.~Liu, Z.~Luo, and Z.~Lin, ``Bilinear factor matrix norm
  minimization for robust {PCA:} algorithms and applications,'' \emph{{IEEE}
  Trans. Pattern Anal. Mach. Intell.}, vol.~40, no.~9, pp. 2066--2080, 2018.

\bibitem{DBLP:conf/iccv/CabralTCB13}
R.~S. Cabral, F.~D. la~Torre, J.~P. Costeira, and A.~Bernardino, ``Unifying
  nuclear norm and bilinear factorization approaches for low-rank matrix
  decomposition,'' in \emph{{IEEE} International Conference on Computer Vision,
  {ICCV} 2013, Sydney, Australia, December 1-8, 2013}, 2013, pp. 2488--2495.

\bibitem{DBLP:journals/sigpro/FanLYLN19}
H.~Fan, J.~Li, Q.~Yuan, X.~Liu, and M.~Ng, ``Hyperspectral image denoising with
  bilinear low rank matrix factorization,'' \emph{Signal Process.}, vol. 163,
  pp. 132--152, 2019.

\bibitem{DBLP:journals/tip/MairalES08}
J.~Mairal, M.~Elad, and G.~Sapiro, ``Sparse representation for color image
  restoration,'' \emph{{IEEE} Trans. Image Processing}, vol.~17, no.~1, pp.
  53--69, 2008.

\bibitem{DBLP:conf/iccv/XuZ0F17}
J.~Xu, L.~Zhang, D.~Zhang, and X.~Feng, ``Multi-channel weighted nuclear norm
  minimization for real color image denoising,'' in \emph{{IEEE} International
  Conference on Computer Vision, {ICCV} 2017, Venice, Italy, October 22-29,
  2017}, 2017, pp. 1105--1113.

\bibitem{DBLP:journals/iet-ipr/ChenLSLS14}
B.~Chen, Q.~Liu, X.~Sun, X.~Li, and H.~Shu, ``Removing gaussian noise for
  colour images by quaternion representation and optimisation of weights in
  non-local means filter,'' \emph{{IET} Image Processing}, vol.~8, no.~10, pp.
  591--600, 2014.

\bibitem{DBLP:journals/imst/XuYL10}
J.~Xu, L.~Ye, and W.~Luo, ``Color edge detection using multiscale quaternion
  convolution,'' \emph{Int. J. Imaging Systems and Technology}, vol.~20, no.~4,
  pp. 354--358, 2010.

\bibitem{DBLP:journals/mssp/GaiYW015}
S.~Gai, G.~Yang, M.~Wan, and L.~Wang, ``Denoising color images by reduced
  quaternion matrix singular value decomposition,'' \emph{Multidim. Syst. Sign.
  Process.}, vol.~26, no.~1, pp. 307--320, 2015.

\bibitem{DBLP:journals/ijon/YuZY19}
Y.~Yu, Y.~Zhang, and S.~Yuan, ``Quaternion-based weighted nuclear norm
  minimization for color image denoising,'' \emph{Neurocomputing}, vol. 332,
  pp. 283--297, 2019.

\bibitem{1199526_P2003}
P.~{Bas}, N.~{Le Bihan}, and J.~. {Chassery}, ``Color image watermarking using
  quaternion fourier transform,'' in \emph{2003 IEEE International Conference
  on Acoustics, Speech, and Signal Processing, 2003. Proceedings. (ICASSP
  '03).}, vol.~3, April 2003, pp. III--521.

\bibitem{DBLP:journals/tip/ZouKW16}
C.~Zou, K.~I. Kou, and Y.~Wang, ``Quaternion collaborative and sparse
  representation with application to color face recognition,'' \emph{{IEEE}
  Trans. Image Processing}, vol.~25, no.~7, pp. 3287--3302, 2016.

\bibitem{DBLP:journals/tip/ChenXZ20}
Y.~Chen, X.~Xiao, and Y.~Zhou, ``Low-rank quaternion approximation for color
  image processing,'' \emph{{IEEE} Trans. Image Processing}, vol.~29, pp.
  1426--1439, 2020.

\bibitem{DBLP:journals/ijcv/SubakanV11}
{\"{O}}.~N. Subakan and B.~C. Vemuri, ``A quaternion framework for color image
  smoothing and~segmentation,'' \emph{International Journal of Computer
  Vision}, vol.~91, no.~3, pp. 233--250, 2011.

\bibitem{DBLP:journals/sigpro/ChenSZCTDL12}
B.~Chen, H.~Shu, H.~Zhang, G.~Chen, C.~Toumoulin, J.~Dillenseger, and L.~Luo,
  ``Quaternion zernike moments and their invariants for color image analysis
  and object recognition,'' \emph{Signal Processing}, vol.~92, no.~2, pp.
  308--318, 2012.

\bibitem{8611251}
J.~L. {Contreras-Hernandez}, D.~L. {Almanza-Ojeda}, S.~{Ledesma-Orozco},
  A.~{Garcia-Perez}, R.~J. {Romero-Troncoso}, and M.~A. {Ibarra-Manzano},
  ``Quaternion signal analysis algorithm for induction motor fault detection,''
  \emph{IEEE Transactions on Industrial Electronics}, vol.~66, no.~11, pp.
  8843--8850, 2019.

\bibitem{10029950538}
\BIBentryALTinterwordspacing
F.~ZHANG, ``Quaternions and matrices of quaternions,'' \emph{Linear akgebra abd
  its applications}, vol. 251, pp. 21--57, 1997. [Online]. Available:
  \url{https://ci.nii.ac.jp/naid/10029950538/en/}
\BIBentrySTDinterwordspacing

\bibitem{Cand2009Exact}
E.~J. Candès and B.~Recht, ``Exact matrix completion via convex
  optimization,'' \emph{Foundations of Computational Mathematics}, vol.~9,
  no.~6, p. 717, 2009.

\bibitem{DBLP:journals/siammax/GillisG11}
N.~Gillis and F.~Glineur, ``Low-rank matrix approximation with weights or
  missing data is np-hard,'' \emph{{SIAM} J. Matrix Analysis Applications},
  vol.~32, no.~4, pp. 1149--1165, 2011.

\bibitem{DBLP:journals/pami/HuZYLH13}
Y.~Hu, D.~Zhang, J.~Ye, X.~Li, and X.~He, ``Fast and accurate matrix completion
  via truncated nuclear norm regularization,'' \emph{{IEEE} Trans. Pattern
  Anal. Mach. Intell.}, vol.~35, no.~9, pp. 2117--2130, 2013.

\bibitem{DBLP:conf/cvpr/LuTYL14}
C.~Lu, J.~Tang, S.~Yan, and Z.~Lin, ``Generalized nonconvex nonsmooth low-rank
  minimization,'' in \emph{2014 {IEEE} Conference on Computer Vision and
  Pattern Recognition, {CVPR} 2014, Columbus, OH, USA, June 23-28, 2014}, 2014,
  pp. 4130--4137.

\bibitem{DBLP:journals/tit/ZhangHZ13}
M.~Zhang, Z.~Huang, and Y.~Zhang, ``Restricted p-isometry properties of
  nonconvex matrix recovery,'' \emph{{IEEE} Trans. Inf. Theory}, vol.~59,
  no.~7, pp. 4316--4323, 2013.

\bibitem{DBLP:journals/tip/ZhangQZYGW20}
H.~Zhang, J.~Qian, B.~Zhang, J.~Yang, C.~Gong, and Y.~Wei, ``Low-rank matrix
  recovery via modified schatten-p norm minimization with convergence
  guarantees,'' \emph{{IEEE} Trans. Image Processing}, vol.~29, pp. 3132--3142,
  2020.

\bibitem{DBLP:journals/tsp/XuM15}
D.~Xu and D.~P. Mandic, ``The theory of quaternion matrix derivatives,''
  \emph{{IEEE} Trans. Signal Processing}, vol.~63, no.~6, pp. 1543--1556, 2015.

\bibitem{DBLP:journals/oms/ShenWZ14}
Y.~Shen, Z.~Wen, and Y.~Zhang, ``Augmented lagrangian alternating direction
  method for matrix separation based on low-rank factorization,''
  \emph{Optimization Methods and Software}, vol.~29, no.~2, pp. 239--263, 2014.

\bibitem{DBLP:journals/tip/NieHL19}
F.~Nie, Z.~Hu, and X.~Li, ``Matrix completion based on non-convex low-rank
  approximation,'' \emph{{IEEE} Trans. Image Processing}, vol.~28, no.~5, pp.
  2378--2388, 2019.

\bibitem{DBLP:journals/tip/WangBSS04}
Z.~Wang, A.~C. Bovik, H.~R. Sheikh, and E.~P. Simoncelli, ``Image quality
  assessment: from error visibility to structural similarity,'' \emph{{IEEE}
  Trans. Image Processing}, vol.~13, no.~4, pp. 600--612, 2004.

\bibitem{DBLP:journals/tgrs/ChenGWWPH17}
Y.~Chen, Y.~Guo, Y.~Wang, D.~Wang, C.~Peng, and G.~He, ``Denoising of
  hyperspectral images using nonconvex low rank matrix approximation,''
  \emph{{IEEE} Trans. Geoscience and Remote Sensing}, vol.~55, no.~9, pp.
  5366--5380, 2017.

\bibitem{DBLP:conf/bmvc/TsengLL17}
C.~W. Tseng, H.~Lin, and S.~Lai, ``General deep image completion with
  lightweight conditional generative adversarial networks,'' in \emph{British
  Machine Vision Conference 2017, {BMVC} 2017, London, UK, September 4-7,
  2017}, 2017.

\bibitem{DBLP:conf/eccv/ZhuXXC18}
X.~Zhu, Y.~Xu, H.~Xu, and C.~Chen, ``Quaternion convolutional neural
  networks,'' in \emph{Computer Vision - {ECCV} 2018 - 15th European
  Conference, Munich, Germany, September 8-14, 2018, Proceedings, Part {VIII}},
  2018, pp. 645--661.

\bibitem{articleHamilton84}
W.~Rowan~Hamilton, ``Ii. on quaternions; or on a new system of imaginaries in
  algebra,'' \emph{Phil. Mag., 3rd Ser.}, vol.~25, 01 1844.

\bibitem{DBLP:journals/sigpro/BihanM04}
N.~L. Bihan and J.~I. Mars, ``Singular value decomposition of quaternion
  matrices: a new tool for vector-sensor signal processing,'' \emph{Signal
  Processing}, vol.~84, no.~7, pp. 1177--1199, 2004.

\bibitem{DBLP:journals/tip/XuYXZN15}
Y.~Xu, L.~Yu, H.~Xu, H.~Zhang, and T.~Nguyen, ``Vector sparse representation of
  color image using quaternion matrix analysis,'' \emph{{IEEE} Trans. Image
  Processing}, vol.~24, no.~4, pp. 1315--1329, 2015.

\bibitem{Girard2007Quaternions}
P.~R. Girard, \emph{Quaternions, Clifford Algebras and Relativistic Physics},
  2007.

\bibitem{Altmann1986Rotations}
S.~L. Altmann, ``Rotations, quaternions, and double groups,'' \emph{Acta
  Crystallographica}, vol.~44, no.~4, 1986.

\bibitem{DBLP:books/daglib/0019186}
R.~A. Horn and C.~R. Johnson, \emph{Topics in Matrix Analysis}.\hskip 1em plus
  0.5em minus 0.4em\relax Cambridge University Press, 1991.

\bibitem{DBLP:journals/appml/Yang03a}
X.~Yang, ``H{\"{o}}lder's inequality,'' \emph{Appl. Math. Lett.}, vol.~16,
  no.~6, pp. 897--903, 2003.

\bibitem{NeedhamA}
T.~Needham, ``A visual explanation of jensen's inequality,'' \emph{American
  Mathematical Monthly}, 1993.

\end{thebibliography}

%

%
%
%
%
%
%
%
%
%
\end{document}